\documentclass[
 superscriptaddress,
 amsmath,amssymb,
 aps, physrev,
]{revtex4-2}

\usepackage{newtxtext}
\usepackage{newtxmath}
\usepackage{natbib}
\usepackage[percent]{overpic}
\usepackage{tikz}
\usepackage{subcaption}

\usepackage{graphicx}
\usepackage{bm}
\usepackage{hyperref}

\hypersetup{
    colorlinks = true,
    urlcolor   = blue,
    citecolor  = black,
}

\usepackage{color}
\usepackage{xcolor}

\newcommand{\Aarnes}{\citet{aarnes2025}}
\newcommand{\Babiker}{{\citet{babiker2023}}}
\newcommand{\Semati}{\citet{semati2025}}

\renewcommand{\rm}[1]{\mathrm{#1}}
\newcommand{\Turbscript}{\mathrm{T}}
\newcommand{\Weber}{\mathrm{We}}
\newcommand{\Froude}{\mathrm{Fr}}

\newcommand{\Taylor}{\lambda}

\newcommand{\Lint}{L_\infty}
\newcommand{\Tint}{T_\infty}
\newcommand{\ReynoldsNumber}{\mathrm{Re}}
\newcommand{\Reint}{\ReynoldsNumber_\Turbscript}
\newcommand{\Rel}{\ReynoldsNumber_\lambda}
\newcommand{\ReL}{\ReynoldsNumber_L}
\newcommand{\Wet}{\Weber_\Turbscript}
\newcommand{\Frt}{\Froude_\Turbscript}
\newcommand{\WeL}{\Weber_L}
\newcommand{\FrL}{\Froude_L}
\newcommand{\meanT}[1]{\langle #1 \rangle}
\newcommand{\meanH}[1]{\overline{#1}}
\newcommand{\mean}[1]{\meanT{\meanH{#1}}}
\newcommand{\MSbeta}{\meanH{\beta^2}}
\newcommand{\MSbetas}{\meanH{\beta_s^2}}
\newcommand{\Area}{S}
\newcommand{\pp}{\mathrm{p.p.}}
\newcommand{\Cbb}{\mathit{C}_{\beta_s^2, \beta^2}}
\newcommand{\CAb}{\mathit{C}_{\Area,\beta^2}}
\newcommand{\CNb}{\mathit{C}_{N,\beta_s^2}}
\newcommand{\Cbbpp}{\mathit{C}_{\beta_s^2, \beta^2}^\pp}
\newcommand{\zref}{z_\text{ref}}
\newcommand{\hide}[1]{}
\newcommand{\rf}{_{\mathrm{ref}}}

\begin{document}
\preprint{APS-1}

\title{
Experimental investigation relating free-surface features to sub-surface turbulence
}

\author{Omer M. Babiker}
\affiliation{Department of Energy and Process Engineering, Norwegian University of Science and Technology, 7491 Trondheim, Norway}

\author{J{\o}rgen R. Aarnes}
\affiliation{Department of Energy and Process Engineering, Norwegian University of Science and Technology, 7491 Trondheim, Norway}

\author{Ali Semati}
\affiliation{Department of Energy and Process Engineering, Norwegian University of Science and Technology, 7491 Trondheim, Norway}

\author{Am\'{e}lie Ferran}
\affiliation{Department of Energy and Process Engineering, Norwegian University of Science and Technology, 7491 Trondheim, Norway}

\author{Yi Hui Tee}
\affiliation{Department of Energy and Process Engineering, Norwegian University of Science and Technology, 7491 Trondheim, Norway}
\affiliation{Department of Mechanical Engineering, University of Ottawa, 
Ottawa, ON, K1N 6N5, Canada}

\author{R. Jason Hearst}
\affiliation{Department of Energy and Process Engineering, Norwegian University of Science and Technology, 7491 Trondheim, Norway}

\author{Simen {\AA}. Ellingsen}
\affiliation{Department of Energy and Process Engineering, Norwegian University of Science and Technology, 7491 Trondheim, Norway}
\email{simen.a.ellingsen@ntnu.no}

\begin{abstract}
Turbulent flows beneath a free surface play a central role in the Earth system, yet their coupling to observable surface features remains incompletely understood. Recent studies using Direct Numerical Simulations (DNS) have reported strong correlation between observable surface features and surface divergence as well as velocity statistics directly beneath, but were limited to Reynolds numbers ($Re$) far below those typical of natural flows, and do not carry the inherent challenges of measurement and flow fidelity that real flows present. We present a laboratory study in which free-surface topology and sub-surface turbulent velocity are measured simultaneously in a jet-stirred tank, extending these numerical results to the physical domain. Using a novel combination of particle-image velocimetry (PIV) and free-surface profilometry, we access $Re$ up to two orders of magnitude higher than in the DNS. A computer vision method developed for identifying turbulent imprints on the free surface is successfully applied to experimental data, enabling direct comparison with the DNS. The correlation between time series of mean-square surface divergence and surface features is found to persist as strongly at higher Reynolds numbers, despite the increased disparity of turbulent scales. Beyond the thin viscous layer, all surface-to-bulk correlations scale with the integral length scale across both experimental and numerical cases. The normalized cross-correlation between mean-square horizontal velocity divergence and surface area covered by structures decreases linearly with depth and remains significant even two integral scales beneath the surface, unlike point-to-point correlations which decay fast, illustrating how correlations are near-instantaneous but spatially non-local. These results demonstrate that visible surface features provide considerable quantitative information about energetic flow events even below the surface-influenced layer. 
\end{abstract}

\maketitle


\section{Introduction}
\label{sec:intro}

The surface of a flowing river displays a myriad of recognizable free-surface features, conventionally referred to as `dimples', `scars' and `boils' (a few other terms have also been used in the past, to the same effect, see e.g.\ \cite{banerjee1994,brocchini2001a}). Dimples may be described as near-circular indentations in the free surface due to surface-attached `bathtub' vortices, boils are smooth surface elevations created by strong upwelling events, and scars are long, narrow indentations in the surface which are typically found near boils and are imprints of horizontally aligned vortices beneath the surface~\cite{babiker2023,aarnes2025}.
Recent studies have indicated that a key sub-surface flow property---the mean-square surface divergence (defined below)---can be inferred from observations of these readily observable surface features \citep{muraroFreesurfaceBehaviourShallow2021},
suggesting a potential avenue for remote sensing. 

A qualitative intuition for near-surface structures near strong upwelling regions has emerged gradually over a long time from observations, experiments and theory \citep[e.g.,][]{banerjee1994,longuet-higgins96, rashidi1997,kumar1998, brocchini2001a,babiker2023,aarnes2025}.
Progress has been made towards quantifying bottom topography from free-surface spectral properties \citep{dolcetti2016, dolcetti2019} and surface imprints \citep{mandel2019, gakhar2020, gakhar2022}, yet much remains to be known before it is possible to extract quantitative information like gas and heat flux, sub-surface mixing rate or the full flow field from, say, automatic surface-feature recognition by computer vision. Machine learning approaches that tackle this challenge are still in their relative infancy but have shown success in reconstructing sub-surface velocity fields based on surface information only \citep{xuan2023, moen2025}.

Our aim in this paper is to further the understanding of how and to what extent deformations of a free surface are correlated with the sub-surface flow field. We consider here the air--water interface atop turbulent water flow in the absence of a mean current in either phase. The viscosity and density of the air have negligible effects in this case, and the interface may be considered a free surface, described in theory and simulation as having constant pressure and no shear stress from the air side. The turbulence on the water side is weak enough to deform the surface relatively gently, yet strong enough to leave imprints on the free surface which are clearly visible by eye.

Recently, some of the present co-authors analysed data from direct numerical simulations (DNS) of said flow problem (see \Babiker{} and \Aarnes, hereafter BA and AA, respectively). In BA a computer vision method for automatic detection of dimples was developed, and it was discovered that the instantaneous number of dimples on the free surface in the computational domain is closely correlated with the mean-square of the surface divergence within the field of view.
In AA the analysis of dimples was significantly extended to include statistics of flow field properties beneath observed dimples and scars in the flow for different Reynolds numbers ($Re$) and Weber numbers ($We$); these parameters are defined below. The detection method was augmented to identify both dimples and scars. 
DNS allow the study of the free-surface flow in full detail with unmatched fidelity, providing the perfect test-bed for studying the interactions between surface and bulk in fine detail. On the downside, computational cost limits
the Reynolds numbers that DNS can access to flows which are far removed from those encountered in most natural flows. 

In the present study we move beyond these limitations by performing experiments of turbulent free-surface flows, using a novel set-up which allows continuous, high-accuracy data capture of the surface deformations and the sub-surface flow field simultaneously. With our experiments, we can test several of the observations and conclusions from the numerical studies at much higher Reynolds numbers, thus increasing the value of the simulations by extending the range where conclusions drawn from them can be expected to hold. 
While there are certain differences between the experiment and DNS due to the stated intention of deforming the surface (this contrasts with studies like that of \cite{ruth2024} where the specific intention was to observe free-surface turbulence without significant surface deformations), we show in Sec.~\ref{sub-sub-sec:experimental_flows} that the experiment and DNS are highly similar in the relevant aspects within a depth of $\mathcal{O}\left(\Lint\right)$, where $\Lint$ is the integral lengthscale of the turbulence, i.e., within the `blockage layer' which is where surface-to-bulk interactions are significant. It should be noted that the velocity fields in the experiments are measured slightly below the surface, allowing us to measure the depth-wise \emph{horizontal divergence} $\beta(z)$, defined as 
\begin{equation} 
    \label{eq:div_hor}
    \beta(x,y,z,t)=\partial_xu_x+\partial_yu_y = -\partial_zw_z\, ,
\end{equation} 
with velocity components $(u_x,u_y,u_z)$ in the $(x,y,z)$ directions, respectively, rather than surface divergence, which we choose to define as 
\begin{equation}\label{eq:div_surf}
    \beta_s(x,y,t)\equiv\lim_{z\to0^-}\beta(x,y,z,t)\, . 
\end{equation}
The definition \eqref{eq:div_surf} supposes that the surface is nearly flat ($|\nabla\eta|\ll1$, where $\eta(x,y,t)$ is the instantaneous surface elevation) and orthogonal to the $z$ axis, sufficiently well satisfied for all our present purposes. The surface divergence is only available for the DNS data, not the experimental data which were measured below the surface (subtleties regarding coordinate definitions near $z=0$ for the dataset are discussed in AA; they are of no practical consequence here). The divergences $\beta$ and $\beta_s$ are closely related for sufficiently shallow depths, much because the peaks in surface divergence are due to large turbulent events, which penetrate deeply. 

\begin{figure}
    \centering
    \begin{overpic}[width=\linewidth]{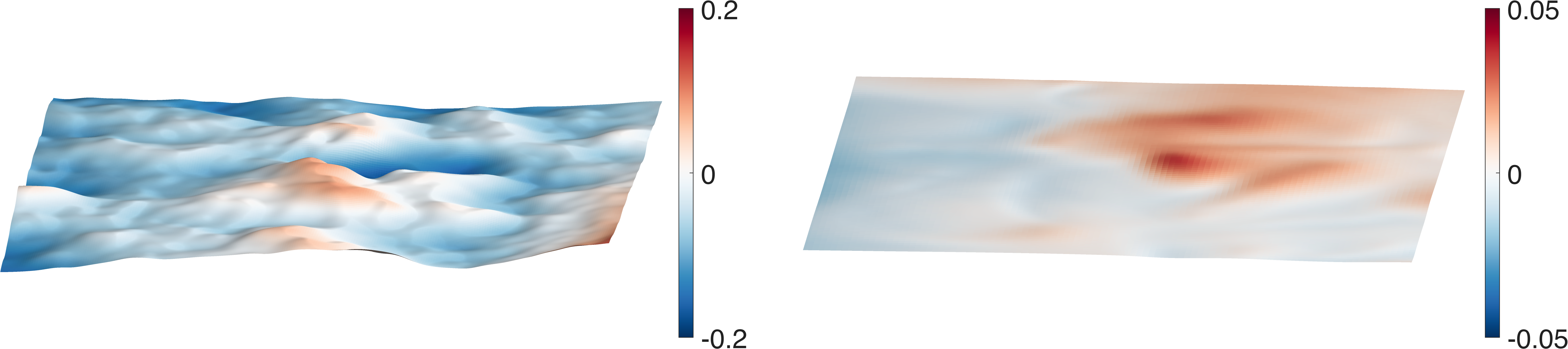}
        \put(0, 23) {a)}
        \put(53, 23) {b)}
    \end{overpic}\\ 
    \caption{Snapshots of the surface elevation from (a) experiments and (b) DNS datasets. 
    Simulations are dimensionalized following Appendix~\ref{app:dimensional}. Both show a 100 mm $\times$ 100 mm cross-section of their respective domain.
    The vertical displacements across both figures are 33 times the horizontal distances (exaggerated for visibility), with colorbars showing elevation in millimetres (scaled differently between the cases for visibility).}
    \label{fig:flow_surface}
\end{figure}

Figure \ref{fig:flow_surface} depicts snapshots of the free surface in the experiments and the DNS. There are clear differences in appearance which illustrate the contrast between ``real world" air and water properties in experiments and the idealized numerical case. The increase in Reynolds number in the experiments also manifests a surface with much more complex topology, an indication of the significant change in the separation of scales in the flow. While in Fig.~\ref{fig:flow_surface}b the sub-surface flow features may be apparent from the surface, it is not clear the same would hold for Fig.~\ref{fig:flow_surface}a.  In this study, we present a careful comparison and synthesis of the experimental and numerical results, thus contributing to bridging the gap that often separates the two. 

\begin{figure}
    \centering
    \includegraphics[width=.5\linewidth]{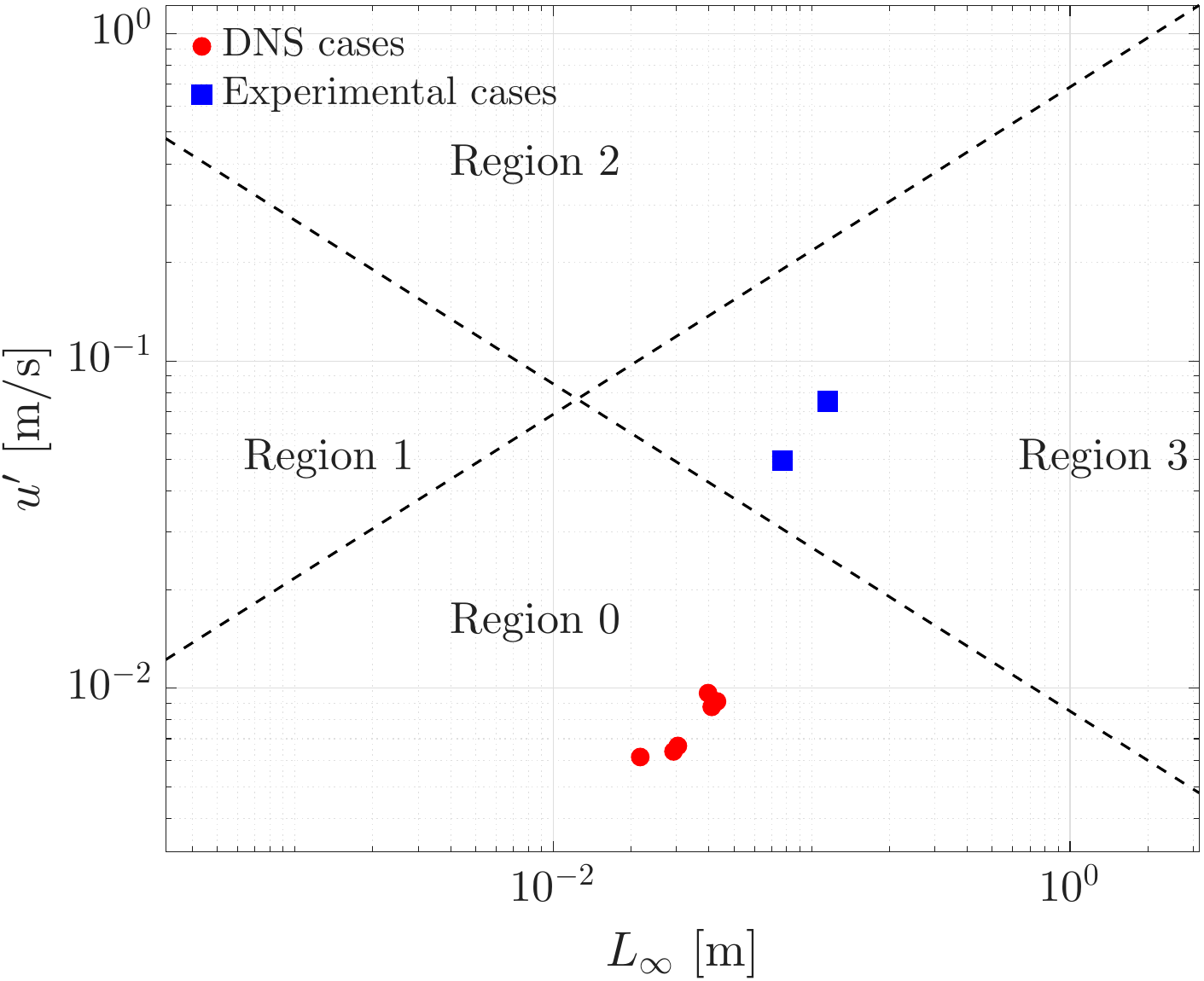}
    \caption{Experimental and numerical cases placed in a Brocchini-Peregrine diagram \cite{brocchini2001a}. See Tab.~\ref{tab:flopProp} for flow parameters. Dashed lines are indicative of boundaries between the four qualitatively different flow regimes. Experimental cases are indicated with square blue markers, DNS cases with round red ones. }
    \label{fig:Brocchini-Peregrine}
\end{figure}

\citet{brocchini2001a} introduced a useful way to characterize free-surface flows, suggesting four qualitatively different flow regimes based on the representative velocity, $u'$, and the integral length scale, $\Lint$ (see also \cite{muraroFreesurfaceBehaviourShallow2021}). Using the values of $u'$ and $\Lint$ obtained as detailed in Sec.~\ref{sub-sec:flow-parameters}, we have placed the experimental and DNS cases in the Brocchini-Peregrine diagram in Fig.~\ref{fig:Brocchini-Peregrine}. Both of the experimental cases fall within ``Region 3" in the chart, the more energetic of the two clearly so, while the DNS cases are decidedly in ``Region 0". Region 0 characterises `weak turbulence' with only very small surface deformations. 
Region 3, on the other hand, is what Brocchini and Peregrine call `gravity dominated flow' characterised by clearly visible surface indentations. It is the typical state of natural flows \cite{brocchini2001a}. Figure \ref{fig:Brocchini-Peregrine} thus indicates how, by extending the analysis to the laboratory, a major step is taken towards real-world application of our methods and results. 

A multitude of experimental studies of free-surface homogeneous turbulence, where turbulence is generated far beneath the water surface and diffuses towards it, can be found in the literature. In the majority of studies, turbulence is generated by an oscillating grid at the bottom of a water tank and allowed to diffuse towards the surface \citep[][]{Thompson1975, hopfinger1976, brumley1987, McKenna2004,herlina2008,Chiapponi2012, Lacassagne2017}, or by jets with zero net flow \citep[][]{variano2008,asher2012,variano2013,carter2016,Jamin2024, ruth2024}. Several studies have measured sub-surface velocity fields affected by a free surface \cite{brumley1987, variano2008, ruth2024}, sometimes along with dissolved gas concentration fields \cite{herlina2008, Lacassagne2017}. 
Some work has been done on measuring the tangential velocity fields at the surface itself \cite{McKenna2004, qi2025, li2024}, sometimes along with the free surface deformations \cite{chatellier2013parametric}.
Simultaneous measurements of the free surface topology and the velocity field below were reported using a laser line-scanning technique \cite{savelsberg2006, 
savelsberg2009}
or by measuring the two-dimensional surface field directly \cite{dabiri2001, dabiri2003, fouras2008, ng2011experimental, gomit2013free, steinmann2021} (see \cite{gomit2022free} for comparisons between the different methods). However, no experimental studies that we are aware of have attempted to identify surface topology features from time- and space-resolved surface height measurements and correlate them directly to the sub-surface velocity field, as we do in the present study. 
In our experiments, we generate turbulence in a jet-stirred tank and take measurements of the surface elevation in a 2-D plane and the velocity field in a plane directly beneath, simultaneously. 
We use a method recently developed by \Semati, where a horizontal slice of the flow close to the surface is measured using Particle Image Velocimetry (PIV), while the surface itself is measured using two-dimensional profilometry. This allows us to directly couple the observed features on the surface with the flow underneath at high Reynolds numbers.

It is necessary to supplement direct numerical simulation studies with experiments because it is not a foregone conclusion that the correlations between surface features and velocity field observed in the DNS considered in BA remain strong for much higher Reynolds numbers. Similarly, the collapse of the surface-to-bulk statistics for different DNS datasets to one curve with the right scaling demonstrated in AA cannot be extrapolated to far higher turbulence levels without careful enquiry. It is well known that low-Reynolds number phenomena can differ from those at high-Reynolds numbers \citep[see, e.g.,][]{antonia2017}. In large, this is believed to be related to a reduction in the dependence of the turbulence on the boundary conditions as $\ReynoldsNumber$ increases. For instance, there is a $\ReynoldsNumber$ dependence of the dissipation scaling of the cascade \citep{sreenivasan1998} and the decay exponent of turbulence also appears to have a $\ReynoldsNumber$ dependence \citep{sinhuber2015}. In boundary layers, the well-known hairpin structures \citep{adrian2000} are also known to be less coherent as the wall Reynolds number increases \citep{LozanoDuran_Jimenez_2014}. For free-surface turbulent flows specifically, a recent study of mass transfer across the free surface \cite{herlina2019} found different trends for turbulence with low and high Reynolds numbers (corresponding to $\Reint \leq 500$ and $1440 \leq \Reint \leq 1856$, respectively, in their simulations, where $\Reint$ is the tubulent Reynolds number defined in Section \ref{sub-sec:flow-parameters} below.). Moreover, existing experiments at higher $\ReynoldsNumber$ \citep{savelsberg2009} found that there is relatively little correlation between the free surface and the sub-surface turbulence, in apparent contrast to the lower-Re DNS findings of BA. (Note, however, that \citep{savelsberg2009} considered point-to-point correlations, not plane-averaged values like BA, making direct comparison challenging.) 

In the present study, we focus on the dimples and scars on the surface and horizontal divergence fields below the surface, as we extend previous analysis of surface patterns and their correlation to subsurface flow to much higher Reynolds numbers. 
We express DNS results in dimensional units in such a way that they may be directly compared to the results from experiments.
Next, we investigate the extent to which the strong correlation between surface features and horizontal divergence extends downwards into the flow, finding considerable covariance even at depths beyond the blocking layer thickness, with excellent agreement between simulation and experiment despite lying in disparate Reynolds number regimes.
The study reaffirms the conclusion that observations of surface features could be a promising avenue for remote sensing of sub-surface turbulence.
It also illustrates that although global point-by-point covariance between surface and bulk quantities may be small because the majority of the surface bears no visible mark of the sub-surface turbulence, surface features reveal flow information deep beneath the surface. 

The paper is structured as follows. Section \ref{sec:data_acquis_&_charac} concerns our flow data; the experiment is presented in Sec.~\ref{sub-sec:Exp-setup}, while the DNS data, which has been published previously, is briefly summarized in section \ref{sub-sec:DNS-data}. The datasets are then characterized in terms of turbulent quantities in section \ref{sub-sec:flow-parameters} so that they can be compared to each other as well as to studies by other authors. In Sec.~\ref{sub-sec:detection-method}, the identification procedures that provide datasets of dimples and scars---previously employed for DNS in BA and AA---are reviewed and adapted for experimental data. Section \ref{sec:results} presents the results of correlation analysis between surface features and horizontal divergence. The velocity field measured in the plane closest to the surface is considered in Sec.~\ref{sub-sec:covarainve}, while the covariances with deeper planes are studied and compared in Sec.~\ref{sub-sec:depth-dependence}. Conclusions are drawn in Section~\ref{sec:conclusions}.


\section{Experimental and numerical flow data: acquisition and characterization}
\label{sec:data_acquis_&_charac}


\subsection{Experimental data} 
\label{sub-sec:Exp-setup}

We conducted the measurements in a turbulence tank at the Norwegian University of Science and Technology. As noted in the introduction, similar tank set-ups have been widely used in the past \citep[e.g.][]{variano2008,asher2012,carter2016, ruth2024, Jamin2024}, using submerged jets to create approximately isotropic and homogeneous turbulence without mean flow. 
High and adjustable turbulence levels can be readily created. We choose pump power such that distinct surface deformations and visible surface imprints are present while ensuring turbulent scales remain small compared to tank dimensions. This is a deliberate choice as we intend for the surface to deform. This approach contrasts with that of some earlier studies, e.g., \cite{ruth2024}, where the intention was to study surface turbulence on an approximately flat interface. To achieve the goal of deforming the surface, we had to relax our requirements for homogeneity below the surface. Note, this facility has a second set of nozzles designed to produce nearly homogeneous and isotropic turbulence, but they were not used in the present study.

\begin{figure}%
    \centering
    \begin{overpic}[width=0.4\linewidth]{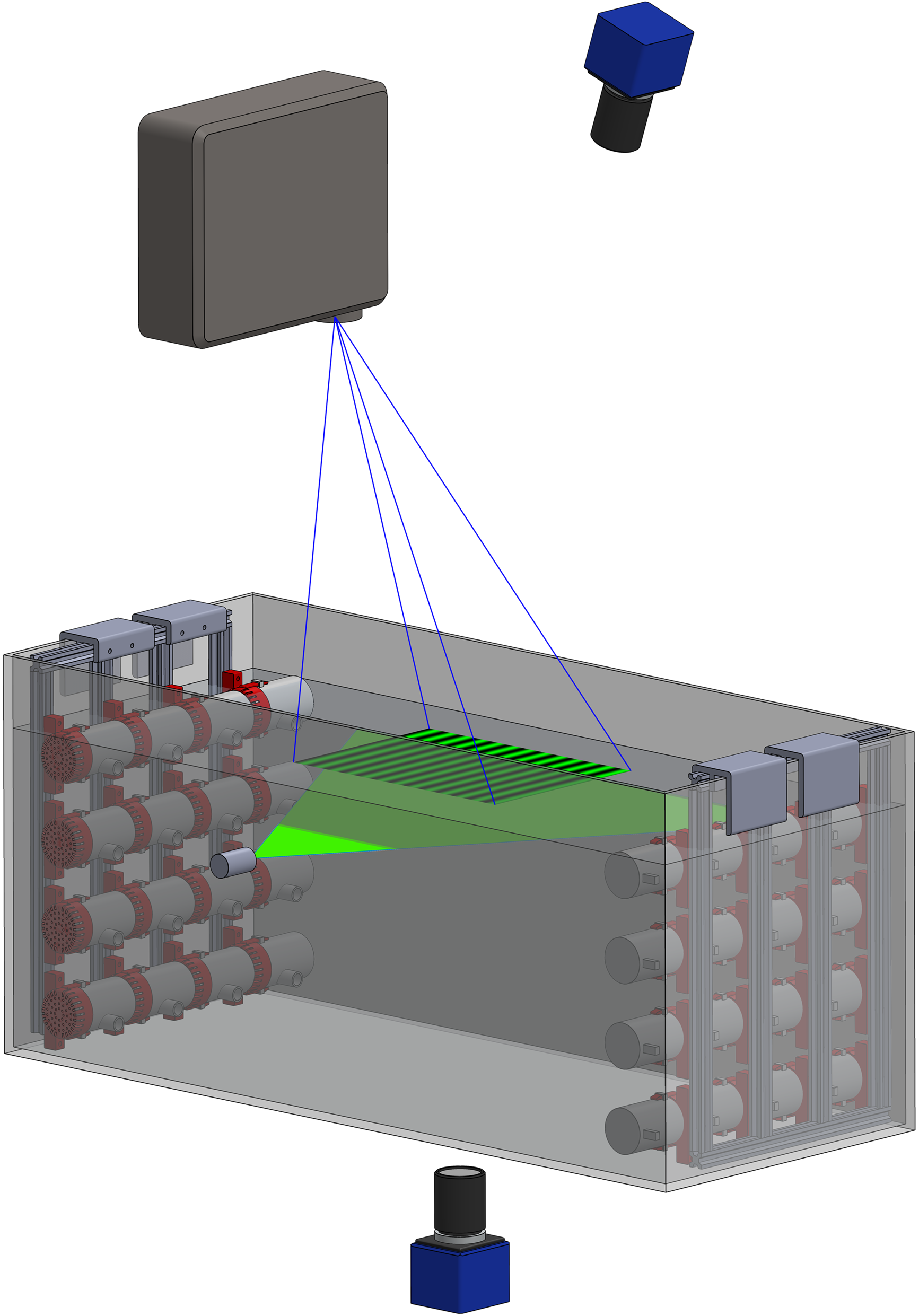}
        \put (2, 100) {a)}
        \put(3.5, 88.5){
        \begin{tikzpicture}
          \filldraw[white, opacity=0.8] (0,0) rectangle (1.5, 0.45);
        \end{tikzpicture}
        }
        \put (5, 90) {Projector}
        \put (40, 85) {Profilometry}
        \put (42, 80) {camera}

        \put(15, 27){
        \begin{tikzpicture}
          \filldraw[white, opacity=0.7] (0,0) rectangle (2.6, 0.45);
        \end{tikzpicture}
        }
        \put (17, 28) {PIV Laser sheet}
        \put (9, 4) {PIV Camera}
    \end{overpic}
    \begin{overpic}[width=0.55\linewidth]{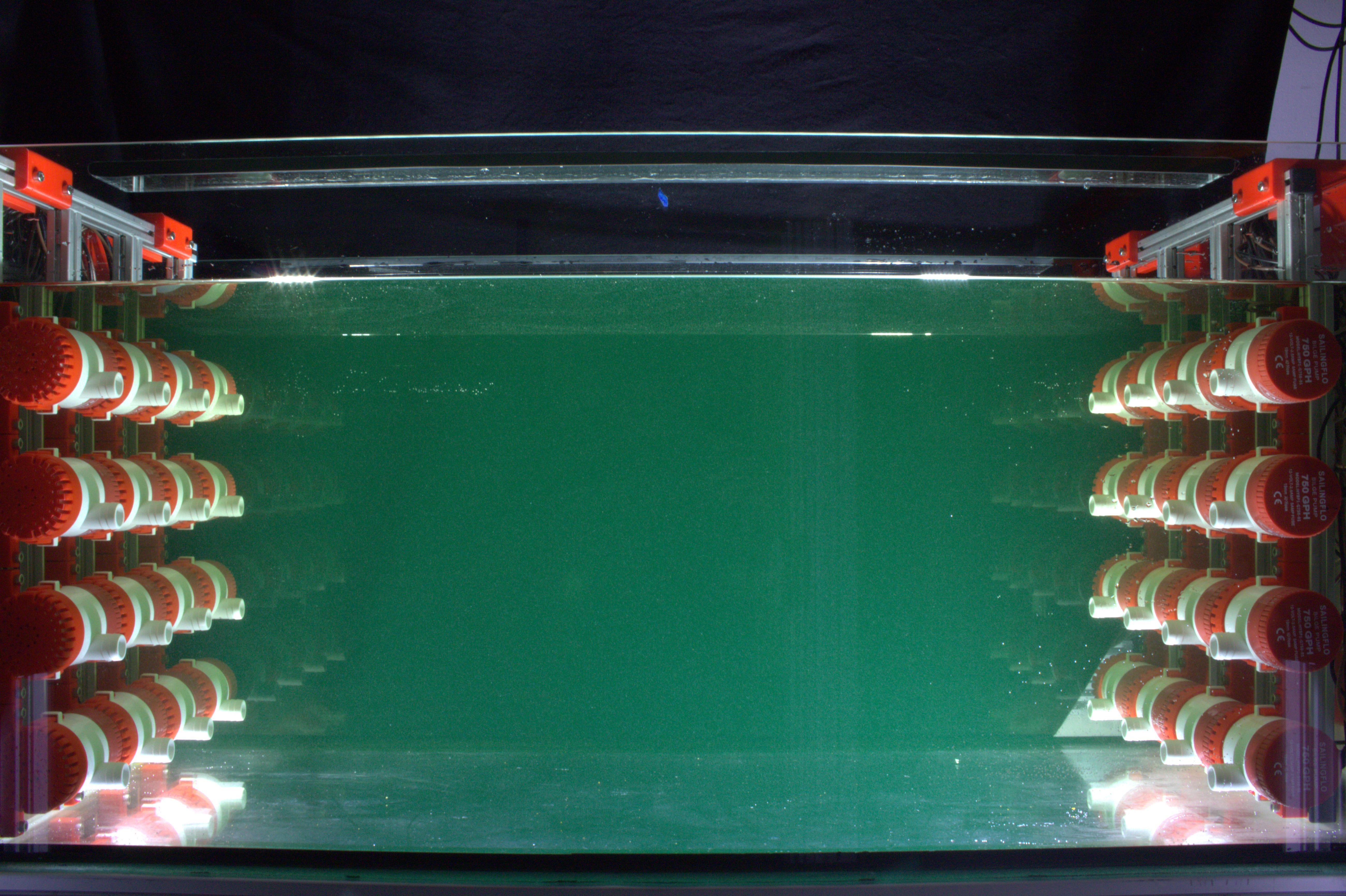}
        \put (2, 60) {\color{white}b)}
    \end{overpic}
    
    \caption{(a) Schematic of the experiment, showing the projector, profilometry camera above the tank, projected fringe pattern on the surface, PIV camera, and horizontal laser sheet. (b) Picture of the turbulence tank with the 32 submersible bilge pumps. Only the two lowest rows of pumps were used in the present experiments.}%
    \label{fig:T-tank}
\end{figure}

The tank is made entirely of glass with dimensions of 1100~mm $\times$ 448~mm $\times$ 500~mm. The tank is based on the designs of \cite{bellani_turbulence_2013} and \cite{esteban_laboratory_2019}, where turbulence is generated by two planes of randomly actuated jets placed on both sides of the tank. Each plane of jets consists of 16 submersible bilge pumps (Sparelys Norway, 12V) with a maximum pump capacity of 47~litres/min. The pumps are arranged in a $4\times4$ array configuration as shown in Fig.~\ref{fig:T-tank}, with a spacing of 107.5~mm from each other. The nozzles of the bottom-row pumps are centred $38.5$~mm above the bottom glass wall. Each pump draws water radially from one side and discharges it axially through a cylindrical nozzle with an inner diameter of 15~mm. In other words, the pump draws fluid from and returns it to the same control volume, maintaining mass conservation. This zero-net-mass-flux behaviour allows each pump to function as a synthetic jet, injecting momentum into the flow to induce turbulence (see also the review by \cite{nezami2023laboratory} on other approaches in generating zero-mean-flow homogeneous isotropic turbulence).

Each bilge pump is connected to an individual solid-state relay, enabling independent on/off control. These relays are triggered by digital signals from an NI sbRIO9636 controller board programmed in LabVIEW. The controller executes a pre-programmed sequence of time-stamped commands based on the ``sunbathing" algorithm introduced by \cite{variano2008}. This algorithm controls the firing pattern of the synthetic jets by independently turning each pump on or off for durations randomly sampled from Gaussian distributions with defined means and variances for the on- and off-times. This decoupled actuation prevents synchronized firing across the array and maintains a steady statistical fraction of active jets, promoting approximately homogeneous, isotropic turbulence (HIT) with minimal mean flow in the central region of the tank, although this is not explicitly the objective here. In order to produce a flow roughly similar to the simulation of \citet{guoInteractionDeformableFree2010} (the same data as used by BA and AA), we only actuate the two bottom rows of pumps on both sides (sixteen in total) to generate turbulence in the deepest part of the tank and allow that flow to advect upwards and impinge on and deform the surface. 
This means that the topmost actuated pumps are located at $265$~mm from the surface.

The produced flows are turbulent, but are not HIT, nor are they meant to be, as HIT in this facility does not have sufficient energy to significantly deform the surface. The mean on-duration of the actuation was $3$\,s, with a standard deviation of $1$~s. This corresponds to a source fraction of $\phi = 0.25$, representing the average portion of active jets across the full array. While a pump is active, its power level is controlled by adjusting the amplitude of the signal, which regulates the effective voltage supplied to the pump, and thus controls the jet velocity. In this study, we investigate two pump power settings, where all active pumps operated at either $50\%$ power (Case A), or $25\%$ power (Case B) to examine the effect of jet strength on the resulting turbulence structures.

We measured the sub-surface flow field via Particle Image Velocimetry (PIV), using a 25-megapixel LaVision Imager CX2-25MP camera. The flow is illuminated with a double-pulsed Nd:YAG laser (Litron Nano L 200-15 PIV) at a wavelength of $532$\,nm and seeded with $40\,\rm{\mu m}$ spherical polystyrene particles, using LaVision DaVis software version 11.0 to record the images.
First, we considered a vertical PIV plane aligned with the jet stream direction to characterize the vertical turbulence evolution in the tank. This imaging plane extends from the air--water interface down to $z=-293$~mm, covering a total area of approximately $200\times300$~mm.
For the simultaneous measurements of surface topology and flow velocity, we investigated several horizontal planes parallel to the undisturbed water surface at depths $z = [-10, -25, -50, -100]~\mathrm{mm}$.
Each horizontal plane had a field of view spanning approximately $250\times 250$\,mm.
We computed the velocity fields using the LaVision multi-pass cross-correlation algorithm. For the vertical plane, the initial interrogation window size was $64\times 64$\,pixels, and the final pass used 32 $\times$ 32 pixels, resulting in a spatial resolution of $1.34\times 1.34$\,mm. For the surface-parallel planes, the interrogation window size was fixed at $32\times32$ pixels, resulting in a spatial resolution of $2.24\times 2.24$\,mm. In all cases, a 50\% overlap was applied between interrogation windows. 

PIV measurements of the flow in a turbulence tank present challenges due to both its highly three-dimensional nature and the intermittency of the pumps. This unsteady forcing creates a flow with a high dynamic range. Consequently, the pulse separation time ($0.5$--$2$\,ms in our experiments) had to be selected to balance competing constraints: it must be short enough to minimize particle loss due to out-of-plane motion, yet large enough to ensure sufficient pixel displacement despite the varying velocities. Given these challenges, we conducted a rigorous error analysis of the results. Vector validation by universal outlier detection \citep{Westerweel2005} resulted in the rejection of less than $0.4\%$ of vectors across all cases. Uncertainty fields were computed from the correlation statistics \citep{Wieneke2015}; spatio-temporal averaging of the uncertainty magnitude for the different cases yielded a maximum uncertainty of $4.6$\,mm/s. In relative terms, the peak uncertainty among the different cases was $5.7$\%, calculated by dividing the temporal average of the uncertainty magnitude by the temporal average of the velocity magnitude, and subsequently averaging in space.

We employ Fringe Projection Profilometry (FPP) to measure the moving free surface, as introduced for this purpose by \cite{cobelli2009}. The FPP technique consists of projecting a fringe pattern onto the water surface and capturing the pattern deformations by means of a downward-pointing camera installed above the tank (see experimental setup in Fig.~\ref{fig:T-tank}a). A change in water height induces a phase shift in the projected pattern, allowing us to obtain the 3D surface topology. So that FPP can be used simultaneously with PIV, the method had to be modified to solve the following challenge: the water surface must be opaque to the projected light while translucent to the PIV laser sheet. We achieve colour-dependent opacity by adding a fluorescent dye to the water, fluorescein disodium salt hydrate (uranine) at $12$\,ppm. The incoming light from the projector is tuned near fluorescein's excitation peak by using a $20$\,nm band-pass filter centered at $490$\,nm.
To capture the fluorescein emission peak while avoiding specular reflections, we equipped the filometry camera, another LaVision Imager CX2-25MP, with a 510~nm long-pass filter.
Fluorescein is also transparent to the 532~nm laser light, allowing sub-surface flow visualisation.
However, since its emission peak, at 520~nm, lies close to the laser wavelength, we installed an ultra-narrow band-pass filter ($4$\,nm) on the PIV camera to prevent contamination from dye emission in the particle images. 
Methods based on refraction or reflection have been used successfully in the past for simultaneous velocity and flow measurements in channel flow \citep{dabiri2001,dabiri2003,fouras2008,ng2011experimental}. The profilometry technique has the distinct advantage over these methods that it is not limited to low surface slopes; indeed with our highest turbulence levels surface deformations were too abrupt for Schlieren-based techniques to be applicable, at least without significant modification. We estimate the spatial resolution of the profilometry to be $4.5$\,mm (corresponding to twice the fringe spacing of $2.25$\,mm), with the maximum surface elevation error not exceeding $50$\,{\textmu}m. The full details on the technique for simultaneous PIV/FPP measurements are presented in a separate paper \citep{semati2025}.

Both surface and velocity fields were measured simultaneously at a sampling frequency of $15$\,Hz, with 20 one-minute cycles per case, leading to 18~000 realizations per test case. We introduced a short time delay of approximately $3$\,ms between the acquisition of the PIV and the surface topology to prevent laser light saturation in the profilometry images.


\subsection{DNS data}
\label{sub-sec:DNS-data}

We use the same DNS datasets as AA for comparison with the experimental data, generously provided by Prof.\ Lian Shen and his group at the University of Minnesota; we refer to AA and references therein for the full technical details of the simulation and only iterate the essentials here. The simulations are based on box turbulence interacting with a free surface using the numerical code and set-up as detailed in Refs.\ \citep{guoInteractionDeformableFree2010, xuanConservativeSchemeSimulation2019}. The flow is artificially agitated in the centre of the domain by random linear forcing \citep[see][]{rosalesLinearForcingNumerical2005}. This forcing is damped as the upper and lower boundaries of the box are approached, where the regions closest to the surface and to the bottom boundary experience no forcing. The turbulence then naturally spreads due to diffusion \citep{guoInteractionDeformableFree2010, xuanConservativeSchemeSimulation2019}. 
The simulation dimensions are \(L_x, L_y, L_z = 2\pi L, 2\pi L, 5\pi L\), where $L$ is a characteristic length scale.
We limit our analysis to data from the upper region, that is, for $z \in [-\pi L / 2,\eta]$.
The simulations were run for two nominal Reynolds numbers, \(\ReL = (2500, 1000)\), where \(\ReL = \mathit{U} \mathit{L}/\nu\), with characteristic velocity scale \(\mathit{U}\) and kinematic viscosity \(\nu\), three Weber numbers \(\WeL = (\infty, 20, 10)\), where \(\WeL = \rho \mathit{U}^2 \mathit{L} /\sigma\), with density \(\rho\) and surface tension \(\sigma\) (\(\WeL = \infty\) representing no surface tension), all with the same Froude number (\(\FrL = \mathit{U}/ \sqrt{g\mathit{L}} = 0.1\), with gravity \(g\)). 
For \(Re = 2500\), the mesh size is \(256 \times 256 \times 660\), while for \(Re = 1000\) it is \(128 \times 128 \times 348\), where the mesh in the vertical direction is refined as the surface is approached and undulates to always adhere to the surface.
These parameters are only used to set the simulations in motion. Of more interest are the turbulent parameters presented in the next section. To avoid confusion when comparing simulated cases to the experimental ones we henceforth do not use these nominal dimensionless numbers, but rather the dimensional groups based on turbulent length and velocity scales as explained in Sec.~\ref{sub-sec:flow-parameters} and App.~\ref{app:dimensional}.


\subsection{Characterization of the turbulent flows}
\label{sub-sec:flow-parameters}

To compare the experimental flow with the simulations and other studies, we need to analyse the turbulent properties of the flows in terms of appropriate length scales and nondimensional groups. 
Numerical figures are listed in Tab.~\ref{tab:flopProp} for later reference. 
There, \(\Reint = 2u'\Lint/\nu\) is the turbulent Reynolds number, where $\Lint$ is the integral scale of the turbulence and \(u'\) is the root-mean square (rms) of the fluctuating velocity which we take as our `representative velocity' \citep[][Sec.~3.2]{tennekes1972}, \(\Rel = u' \Taylor / \nu\) is the Reynolds number based on the Taylor microscale, \(\Taylor = u' \sqrt{15 \nu/ \epsilon}\), where \(\epsilon\) is the dissipation. $\Frt = u'/\sqrt{2g\Lint} $ and $\Wet = 2\rho u^{\prime 2} \Lint/\sigma $ are the turbulent Froude and Weber numbers, respectively. The dissipation is also used to compute the Kolmogorov length scale, \(L_\mathrm{K} = \left(\nu^3/\epsilon\right)^{1/4}\). Finally, $\Tint = l/u'$ is integral time scale of the turbulence.

The integral length scale, $\Lint$ is found from autocorrelations of the spatial velocity fluctuations (details below). This method of integral scale estimations is widely used to characterize the turbulence, and we apply it here to both the experimental and DNS data. 

All turbulent properties are computed outside the surface-influenced region, unless explicitly stated otherwise (e.g., when the depth-dependence of $\Lint$ is considered in Fig.\ \ref{fig:exp_dns_comp}b). The surface-influenced region of a turbulent free-surface flow is the region with a thickness $\mathcal{O}\left(\Lint\right)$ near the surface wherein the the flow becomes significantly anisotropic. It can be classified as consisting of a blockage layer and a viscous (sub)layer \citep[see ][]{hunt1978, hunt1984,shenSurfaceLayerFreesurface1999,calmet2003, magnaudet2003}. The former results from the kinematic boundary condition blocking fluid from penetrating the surface and forcing all motion to be horizontal there, while the latter is a very thin region close to the surface wherein tangential stresses tend rapidly to zero due to the dynamic free-surface boundary condition. The turbulence is characterized outside this region, where the flow is approximately isotropic.

In the following we detail how the various flow quantities were calculated for the experimental flows and the DNS data. 

\begin{table}
    \centering
     \begin{tabular}{p{2.5em} p{3em} p{3em} p{3em} p{3em} p{3em} p{3em} p{3em} p{3em} p{2em}}
    \hline
        Case & $\Reint$ & $\Rel$ & $\Frt$ & $\Wet$ & $u'$ & $\Lint$ & $\Taylor$ & $L_\mathrm{K}$ & $\Tint$\\
        &&&&&[mm/s]&[mm]&[mm]&[mm]&[s]\\
        \hline
        A & 17369 & 397 & 0.050 & 18.1 & 75.3 & 115.4 & 5.3 & 0.13 & 3.0 \\
        B & 7636 & 275 & 0.040 & 5.2 & 49.5 & 77.1 & 5.5 & 0.17 & 3.1  \\
        \hline        1 & 782 & 84 & 0.010 & $\infty$ & 9.10 & 43.0 & 9.18 & 0.51 & 9.4 \\
        2 & 719 & 77 & 0.010 & 0.80 & 8.76 & 41.0 & 8.83 & 0.51 & 9.4        \\
        3 & 766 & 85 & 0.011 & 0.47 & 9.65 & 39.7 & 8.80 & 0.49 & 8.2        \\
        4 & 267 & 47 & 0.009 & $\infty$ & 6.15 & 21.7 & 7.59 & 0.56 & 7.0    \\
        5 & 403 & 51 & 0.009 & 1.16 & 6.65 & 30.3 & 7.71 & 0.54 & 9.1        \\
        6 & 373 & 50 & 0.009 & 0.52 & 6.40 & 29.2 & 7.79 & 0.56 & 9.1        \\
        \hline
    \end{tabular}
    \caption{Summary of turbulent properties for experimental cases (A and B) and simulated  cases (1--6). From left: case name, turbulent Reynolds number, Taylor Reynolds number, turbulent Froude number, turbulent Weber number, representative velocity, integral length scale, Taylor microscale, Kolmogorov length scale and integral time scale. All length scales in are given in dimensional units as indicated; for details on dimensionalizetion of cases 1--6, see App.~\ref{app:dimensional}. Note the $\Lint$ in cases 1--6 differ from those reported in AA due to a different estimation procedure.}
    \label{tab:flopProp}
\end{table}

\subsubsection{Experimental flows}
\label{sub-sub-sec:experimental_flows}

Pertinent quantities for the experimental cases A and B are listed in Table~\ref{tab:flopProp}. This section details how these quantities were obtained from our measurement data.

The turbulent flow in the two experimental cases was characterized by taking planar PIV measurements in a vertical $xz$-plane, and averaged quantities were evaluated at a depth where the influence of the surface is minimal. This was chosen to be around $z=\zref=-$120\,mm from the surface in both experimental flow cases. 

We use a Reynolds decomposition of the velocity such that \(U_i = \meanT{u_i} + u_i\), where \(U_i\) is the measured velocity,  $\langle\cdots\rangle$ denotes average in time, and \(u_i\) is the fluctuating component of the velocity. Here and henceforth subscript $i$ and $j$ denote vector components, $i,j\in(1,2,3)\leftrightarrow (x,y,z)$. We let \(u_i'\) denote the root-mean-square (RMS) over a single depth and in time: $u_x^{\prime}(z) = \mean{u_x^2(x,y,z,t)}^{1/2}$ and so on, where an overbar $\overline{\cdots}$ denotes average over the horizontal plane (the $xy$-plane for horizontal-plane PIV, the $x$ direction for vertical-planes). We also denote the representative velocity in the experiments as \(u' =  {\textstyle \frac12} \left(u_x'+u_z'\right)\), all measured at the reference depth $\zref$. 

\begin{figure}
    \centering
    \begin{overpic}[width=0.49\linewidth]{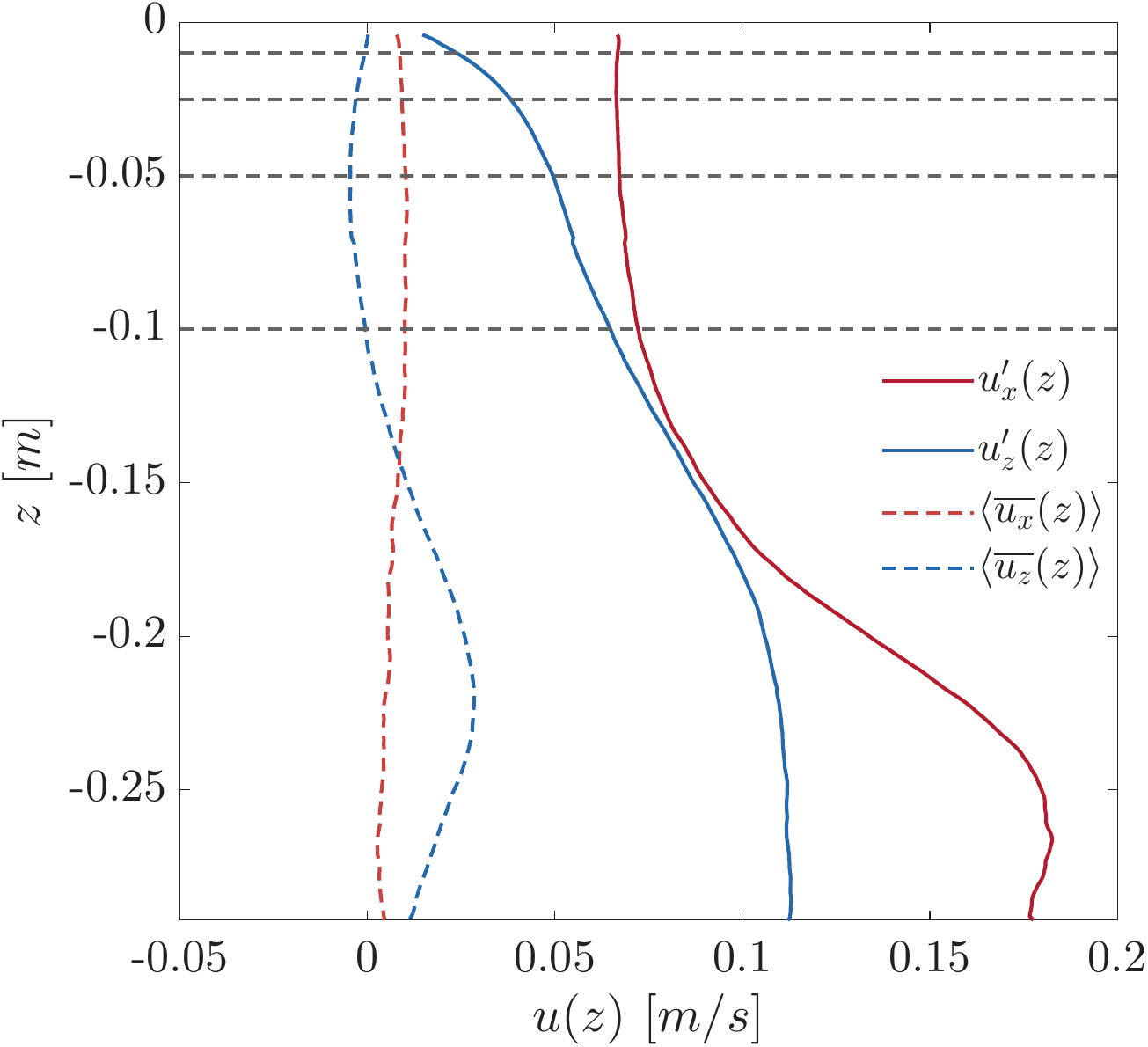}
        \put (4, 92) {a)}
    \end{overpic}
    \begin{overpic}[width=0.49\linewidth]{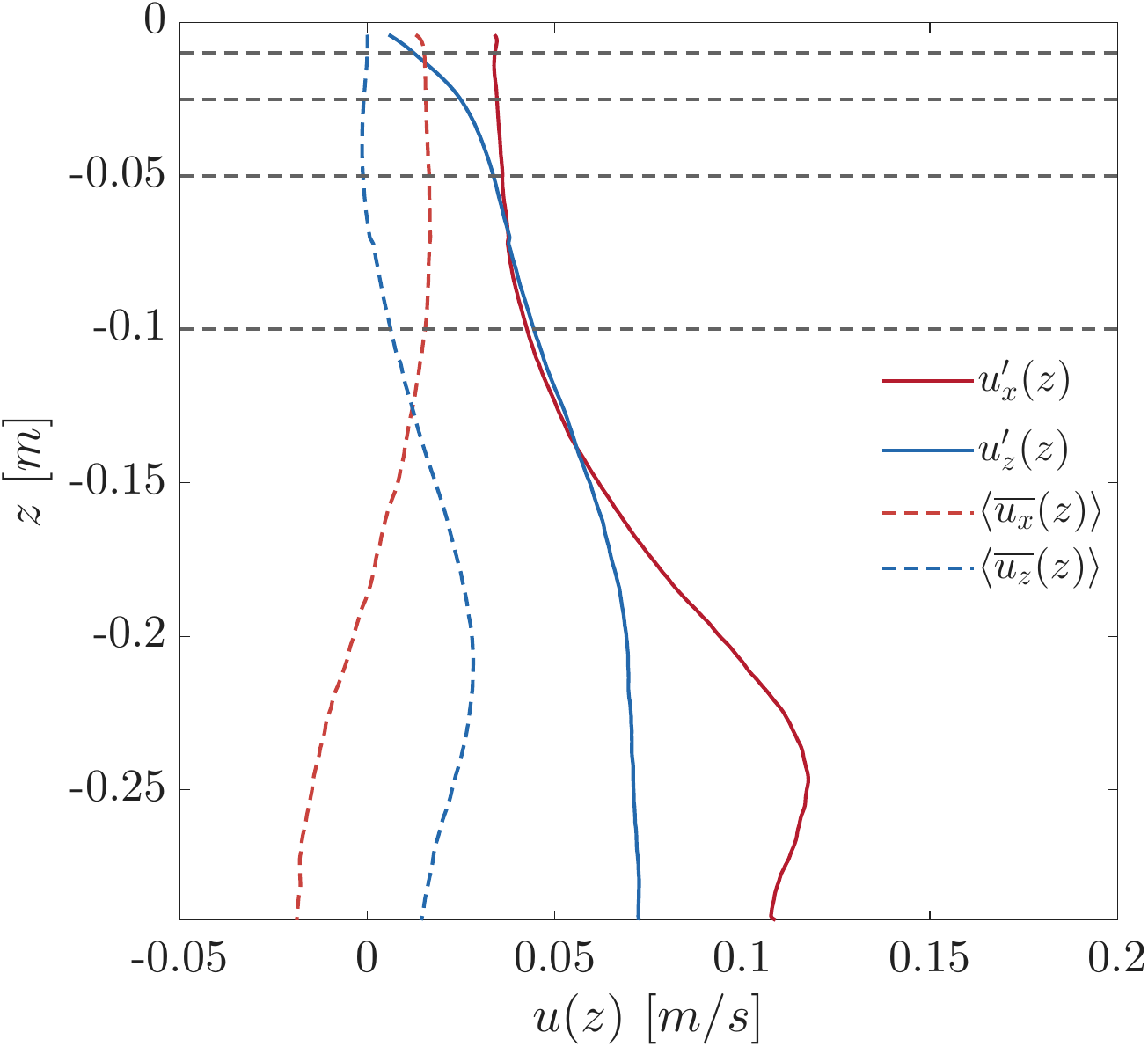}
        \put (4, 92) {b)}
    \end{overpic} \\ 
    \vspace{10pt}
    \caption{Mean (dashed line) and root-mean-square (solid lines) of turbulent velocity components in the vertical (blue) and horizontal (red) directions, as a function of depth for the experimental cases A (a) and B (b). Mean velocities are calculated along the vertical centreline of the field of view. 
    Horizontal dashed lines represent the depths of the horizontal PIV sheets.The uppermost activated pumps were located at $z = -0.265$~m.}
    \label{fig:exp-vel-profile}
\end{figure}

Figure \ref{fig:exp-vel-profile} shows profiles of mean and RMS turbulent velocities as functions of depth. Some anisotropy is seen below around $z=-200$\,mm, caused by the proximity of the flow pumps at that depth, while approximately isotropic flow is observed between $z=-100$\,mm and $z=-200$\,mm. In the figure, this is illustrated by the relative proximity of the RMS of the orthogonal velocity components. Above $z=-100$\,mm, surface blocking starts to affect the flow, with a marked decrease in vertical fluctuations, while the horizontal fluctuations remain relatively constant.
The depth set to $z=-$120\,mm is outside of the blockage layer, and thus the flow characterization is representative of bulk flow, where the flow is approximately isotropic. As the turbulence decays away from the pumps, it is important to set the depth for the characterization close to, but still outside the surface-influence region. We opted to use the same characterization depth for cases A and B consistency, even though the surface-influenced region is smaller for Case B.

\begin{figure}
    \centering
    \begin{overpic}[height=0.5\linewidth]{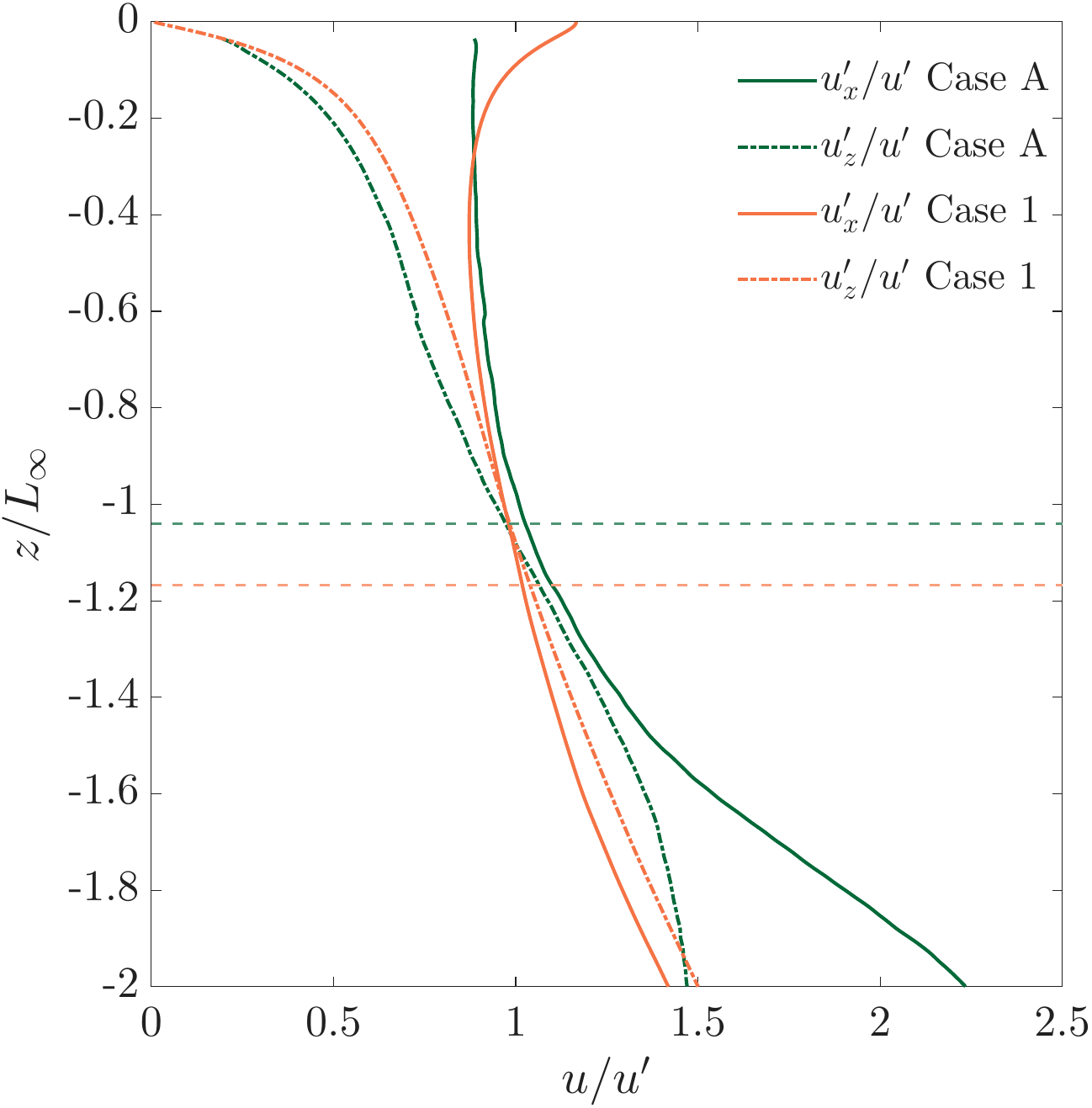}
        \put (2, 95) {a)}
    \end{overpic}
    \begin{overpic}[height=0.5\linewidth]{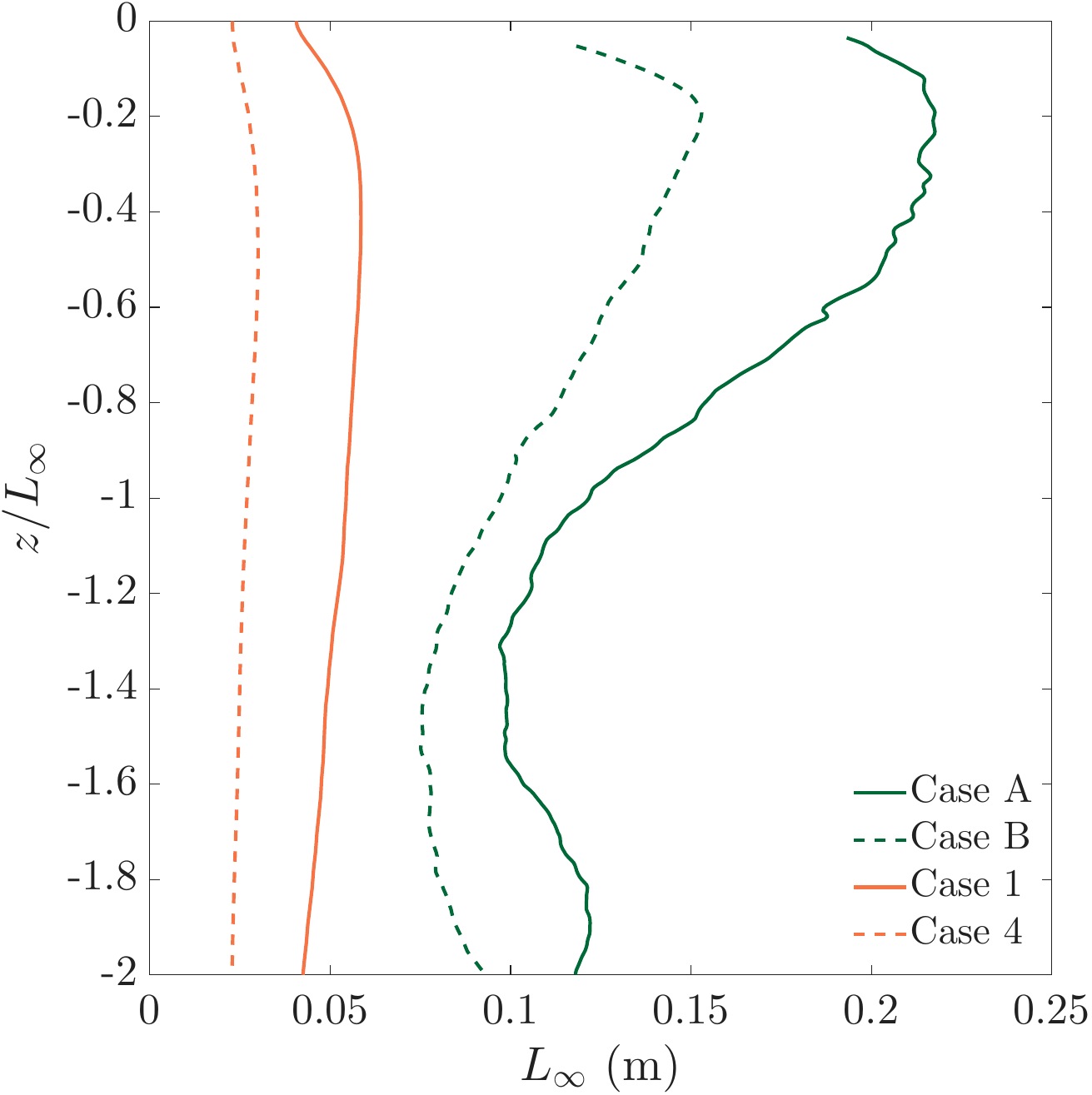}
        \put (2, 95) {b)}
    \end{overpic}
    \caption{(a) Root-mean-square of turbulent velocity fluctuations in the vertical (dash-dot) and horizontal (solid) lines for experiments (Case A, green) and a representative DNS (Case 1, orange) as a function of depth; velocities are normalised by each case's representative velocity $u'$. Horizontal dashed lines represent the depths used for flow characterization. (b) Integral length scale as a function of depth for Case A (solid green), Case B (dashed green), Case 1 (solid orange) and Case 4 (dashed orange). Depths are normalised by the integral scale computed at the reference depth.}
    \label{fig:exp_dns_comp}
\end{figure}

For the following characterization of the experimental flows our requirements for isotropy and homogeneity are considerably lower than studies where the nature of the turbulence itself is the focus. We wish to study how a subsurface structure manifests on the surface and as such, having structures with sufficient strength to deform the surface is of greater import than producing idealised HIT.
Minor deviation from homogeneity and isotropy should not affect the flow-to-surface coupling mechanism of structures, 
and a small mean flow will merely advect persistent structures and features slowly, which affects our results negligibly. For reference, some metrics of homogeneity and isotropy are found in App.~\ref{app:HITmetrics}. 

It is worth bearing in mind that our region of interest is primarily the near-surface layer in which the flow is significantly influenced by the surface. In this region, the surface physics dominate, and the RMS horizontal and vertical turbulent velocities, $u'_x(z)$ and $u'_z(z)$, have very similar behaviour between experiment and DNS, as is clear from Fig.\ \ref{fig:exp_dns_comp}a where $u'_x/u'$ and $u'_z/u'$ are plotted as functions of $z/\Lint$ for cases A and 1, which are representative examples. (We slightly foreshadow section \ref{sub-sub-sec:DNS_turb_param} where details for the latter case are given. Compare with Fig.\ 3 from AA \cite{aarnes2025} for a fuller picture). This establishes confidence that a comparison is reasonable.


We calculate the integral scale, $\Lint$, based on the autocorrelation of the velocity fluctuations in space,
\begin{equation}
    \rho_{ij} (r) = \frac{\mean{u_i(\mathbf{x}) u_i(\mathbf{x} + r \mathbf{e}_{j})}}{u_i'^2},
    \label{eq:vel_autocorr}
\end{equation}
where \(\mathbf{e}_{j}\) is the unit vector in the \(j\)-th direction.
We calculate the autocorrelation \(\rho_{xx}\), where \(\mean{u_x(x,\zref) u_x(x + r, \zref)}\) is averaged over all $x$ and in time, and is then divided by \(u_x'^2\). An exponential of the form \(a e^{-bx}\) is then fitted to the average autocorrelation, whereupon the integral scale is estimated as \(\Lint=a/b\) (following, e.g., \cite{fuchs2022}). Figure \ref{fig:exp_dns_comp}b depicts the integral scale as a function of depth, for the two experimental flows and two DNS cases. The strong similarity across all cases further strengthens our confidence in comparisons across the experiments and DNS.

\begin{figure}
    \centering
    \includegraphics[width=0.7\linewidth]{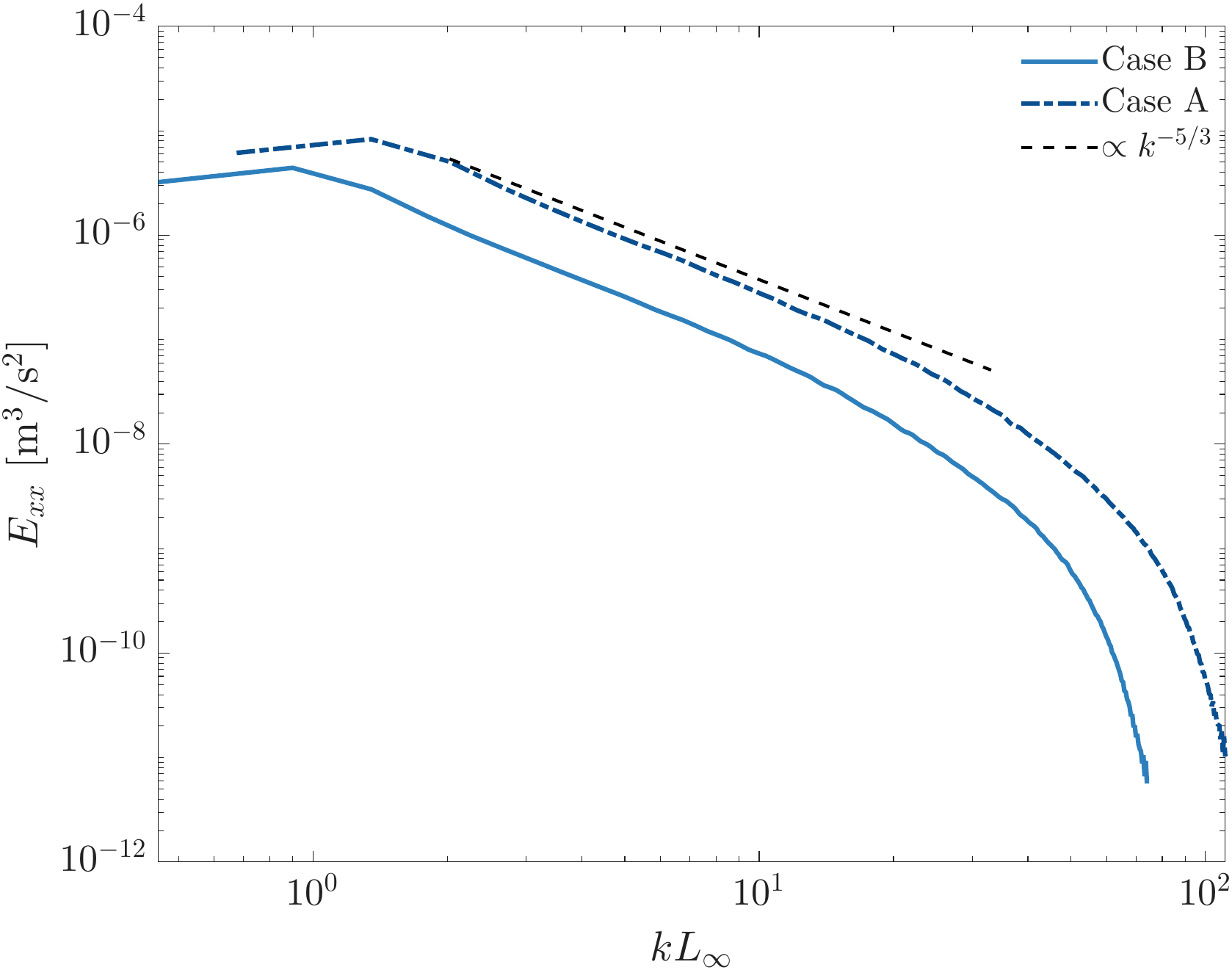}
    \caption{Energy spectrum for experimental cases A and B calculated along a line of the vertical PIV fields at \(z=\zref\). The Kolmogorov scaling $E_{xx}\sim k^{-5/3}$ is shown for reference.}
    \label{fig:ek}
\end{figure}

The other turbulent properties can be calculated given an estimate of the viscous dissipation $\epsilon$, a somewhat subtle point since our flow is not strictly HIT. After comparing several approaches detailed in Appendix~\ref{app:viscous_dissipation}, we employ a fit to the energy spectrum of the turbulent fluctuations $E_{xx}$, such that the inertial range of the turbulence follows the power law $E_{xx} (k) = C_K \epsilon^{2/3} k^{-5/3}$ where $k$ is the wave number and $C_K \approx 0.5$. Figure \ref{fig:ek} shows the energy spectra for both forcing cases, with a dashed line representing the $-5/3$ power-law decay.

\subsubsection{DNS flows}
\label{sub-sub-sec:DNS_turb_param}

Pertinent quantities for the DNS cases, denoted cases 1--6, are listed in Tab.~\ref{tab:flopProp}. For the DNS data, the representative velocity is computed as
\(u'^2 = \frac13\mean{u_i u_i}\) and the dissipation
\(\epsilon\) is computed directly using \(\epsilon = 2 \nu \mean{s_{ij}s_{ij}}\), where \(s_{ij} = (\partial_i u_j +\partial_j u_i)/2\). An overbar denotes average taken in the horizontal direction(s) (the $xy$-plane for horizontal-plane PIV, the $x$ direction for vertical-planes), and $\langle\cdot\rangle$ denotes average in time. The integral length scale is computed based on autocorrelations of the velocity, as detailed above. This way of computing $\Lint$ is selected to allow the most direct comparison possible between the experimental and DNS flows. 
In BA and AA, $\Lint$ was computed directly from the Taylor scales, by $\Lint = \Taylor \Rel/30$ \cite[][\S 3.2]{tennekes1972}; the different procedures result in minor discrepancies in the turbulent Reynolds numbers reported here as compared to those in AA for the same DNS datasets. This has no significant bearing on our results.
An overbar denotes average taken in the horizontal direction(s) (the $xy$-plane for horizontal-plane PIV, the $x$ direction for vertical-planes), and $\langle\cdot\rangle$ denotes average in time.

In order to represent the bulk flow, turbulent properties and non-dimensional groups are calculated at a depth of \(z=\zref=-\pi/2.5\) from the surface (approximately \(1.2\, \Lint\) below the surface for Case 1), which is just below the edge of the blockage layer for all DNS cases, minimising the effect of the surface while still being well away from the numerical forcing region. When referring to the DNS datasets, the depth \(z\) is denoted from the surface directly above the point of interest at that time, such that \(z = 0\) is at the surface, instead of at the mean surface level (see AA for more details).



\subsection{Detection and classification of relevant surface structures}
\label{sub-sec:detection-method}

To detect and classify dimples and scars on the free surface from surface elevation data, we apply the computer vision method developed and described in detail in BA.
Previous application of this detection algorithm has been limited to DNS datasets with low Reynolds numbers (Cases 1--6 here, $207 \leq \Reint \leq 570$; 
see AA for details). As is apparent from the snapshots of the free surface in Fig.~\ref{fig:flow_surface}, there is a striking difference between the surface in a low--Reynolds number DNS and a high--Reynolds number experimental case. The much more violently disturbed air--water interface and larger scale separation in the experiments comprises a greater challenge for feature tracking and identification from surface elevation measurements. To add to this challenge, the measurements done at the surface have a substantially coarser time resolution than the DNS data, making the surface appear to flutter since the most rapid timescales, those that characterise capillary waves, are underresolved.

Drawing on the observation that dimples and scars are surface indentation with relatively small variations in shape, the method makes use of a 2-D wavelet transform to highlight areas of the surface with indentations of the right size and curvature which are thus candidates for dimples and scars. Furthermore, the detection algorithm sorts the detected surface areas by into groups with low and high eccentricity, identifying low-eccentricity (high circularity) features as potential dimples and high-eccentricity ones as potential scars, whereupon it uses thresholds to remove candidate detections that have short lifetime and/or very small area. The result is a filtering out of noise, waves (if present) and large surface structures without sharp edges (which the wavelets do not highlight), eliminating the vast majority of potential false positives (which are small and short-lived) and allows us to identify each remaining pattern as either a dimple or a scar. 

Although tracking individual structures in the experimental data might be possible with a more sophisticated method than ours (e.g, through data-driven interpolation), we find that it is sufficient for our purposes to remove the lifetime criterion altogether for detection in the experiments. Setting a stricter wavelet threshold and enforcing the minimum area threshold for detected features filters out most fast fluctuations since their imprints are comparatively weak. A side effect is that many scars appear as two or more disjointed areas and that weakly defined features may flicker in and out of view. Both of these effects have a limited effect on the relative surface area covered by dimples and scars, which we correlate to sub-surface measurements (as discussed below). 
In the inevitable compromise between false positives and false negatives, we found it preferable for the experimental data with a larger separation of scales that some of the weaker structures were not detected. The results presented in Section~\ref{sec:results} confirm that the accuracy is sufficiently high to obtain a strong correlation between surface features and the sub-surface flow.

For the DNS data, where the time resolution is very fine, we enforce a minimum-lifetime threshold for dimples and scars of $10$ and $5$ recorded frames in the simulation (corresponding to approximately \(0.04\, T_\infty\) and \(0.02\,T_\infty\), respectively, for Case 1). 
The minimum area covered by a detected feature for it to be identified by a dimple or scar was set to $15$ data points in the experiments (corresponding to a circle of radius $\approx 0.16\Taylor$ -- $0.17 \Taylor$) and $10$ data points in simulation data (corresponding to radius $\approx 0.12\Taylor$ -- $0.20\Taylor$).

Following BA, we use the `Mexican hat' wavelet with a width of \(\approx 0.7\Taylor\) for the experiments, and $\approx 0.6\Taylor$ -- $1.0 \Taylor$ for the DNS. These are near the smallest wavelet scales available for our analysis, and as explained in BA, at these small scales, detection results are not sensitive to the exact scale used. (The same may not hold for significantly higher Reynolds numbers than ours where there is a greater disparity of lengthscales and using more than one wavelet size might then be necessary).

\begin{figure}
    \centering
    \begin{overpic}[width=0.8\linewidth]{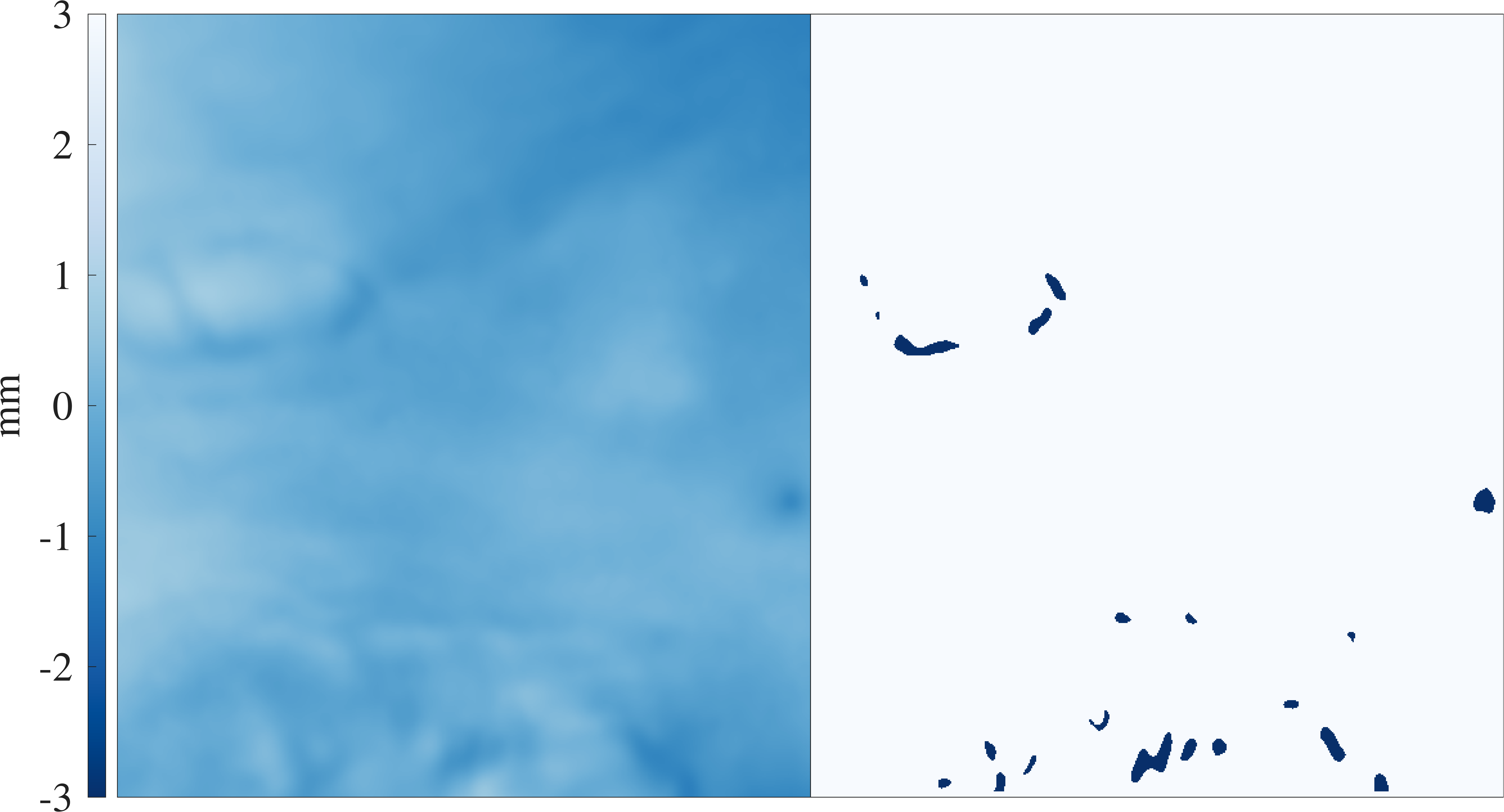}
        \put (8, 50) {\color{white}a)}
        \put (55, 50) {b)}
    \end{overpic}
    
    \vspace{3mm}
    \begin{overpic}[width=0.8\linewidth]{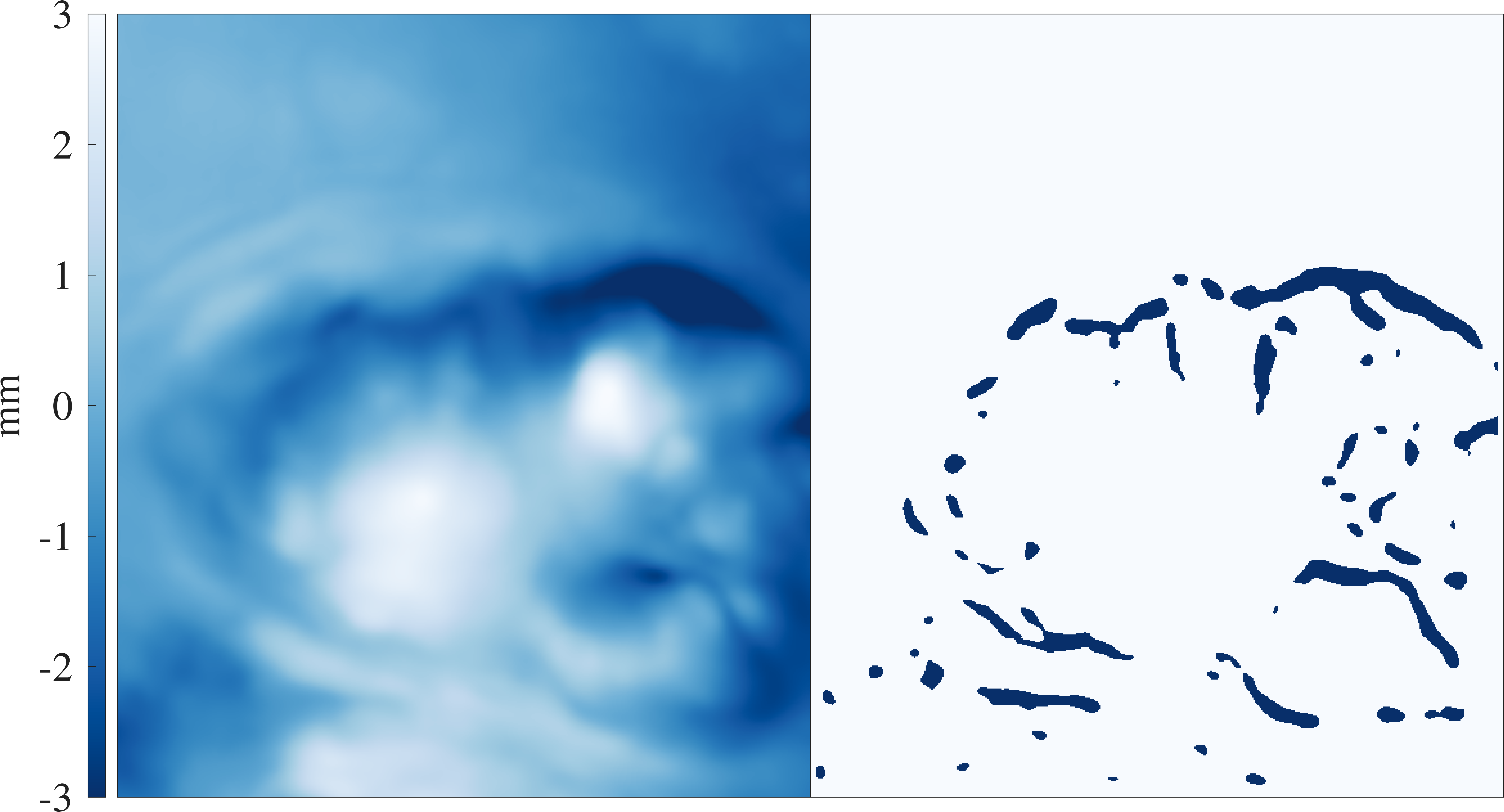}
        \put (8, 50) {\color{white}c)}
        \put (55, 50) {d)}
    \end{overpic}
    \caption{Example of surface elevation from the profilometry measurements (a,c) and the identified structures on the surface (b,d) of two time steps from Case A.}
    \label{fig:exp_struct_example}
\end{figure}

To get a qualitative picture of the detection method, consider the two snapshots showing feature detection applied to experimental data in Fig.~\ref{fig:exp_struct_example}, at representative instances with few and many features present, respectively. The left panels display the measured instantaneous surface elevation and the right panels display the identified structures at the same instance. 

When we analyse the results from the surface structure detection, we make the following two modifications: First, rather than count the relevant surface structures (as done for dimples in BA), we calculate the relative surface area covered by these features, a dimensionless quantity that is more easily compared between simulation and experiment and which is less sensitive to the choice of threshold parameters. Secondly, instead of distinguishing between scars and dimples when correlating them to the sub-surface flow, we aggregate the total area of each surface feature-type in our analysis.  As can be seen in Fig.\ \ref{fig:exp_struct_example}d, our strict wavelet threshold means there is a tendency to split long scars into smaller pieces, some of which may be misattributed as dimples. Counting the number of scars is thus not robust. The \emph{area} of scars and dimples put together, however, is far less sensitive to this and is, in fact, an equally good or even better proxy for surface divergence, as we detail further in Sec.~\ref{sec:results} and App.~\ref{app:num_vs_area_det_strcuct}. 
We emphasize, as discussed in Appendix~\ref{app:window_size}, that due to the different motion of dimples and scars, a sufficiently large interrogation area is necessary for the combined quantity to be successful, which is satisfied in our case.


\section{Results}
\label{sec:results}

In this section, we present the results of identifying surface structures and their relationships with sub-surface flow. 
To wit, we study correlation between time series of the relative area of surface structures, $\Area(t)$, and the horizontal divergence field, $\beta(z)$, at different depths $z$, defined in Eq.\ \eqref{eq:div_hor}, equal to the surface divergence, $\beta_s$, at $z=0$.
Our lab measurements are made a little way below the interface, making it necessary to employ $\beta(z)$, which signifies the local rate of change in vertical velocity through the horizontal plane at depth $z$, indicative of upwelling ($\beta>0$) or downwelling ($\beta<0$) motions, turbulent rushes of fluid towards or away from the surface, respectively.

In \ref{sub-sec:covarainve}, we consider how the horizontal divergence field of the flow is correlated to the dimples and scars at the surface. 
In \ref{sub-sec:depth-dependence} we extend this analysis by investigating how rapidly the correlation declines for horizontal planes increasingly far into the bulk.


\subsection{Covariance of horizontal divergence and surface features}
\label{sub-sec:covarainve}

\begin{figure}
    \centering
    \begin{overpic}[width=\linewidth]{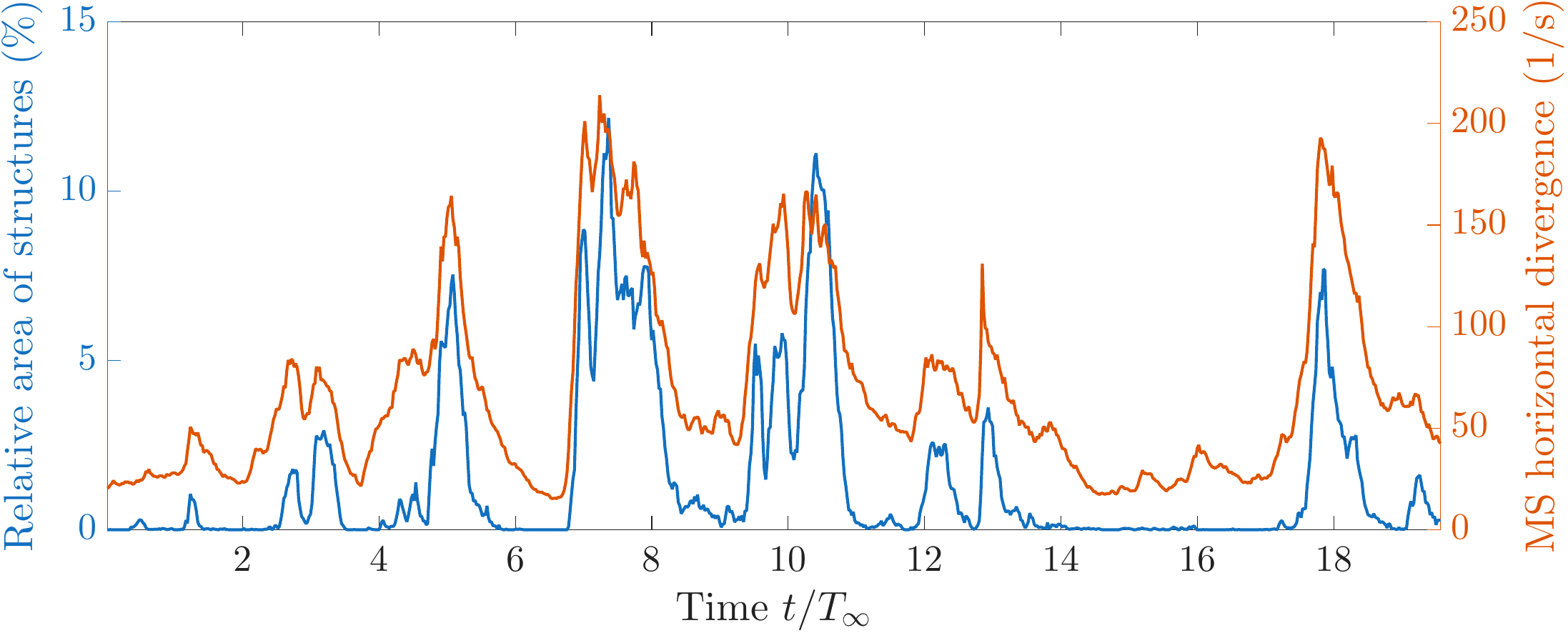}
    \put (8, 36) {a)}
    \end{overpic} 
    
    \vspace{4mm}
    \begin{overpic}[width=0.9\linewidth]{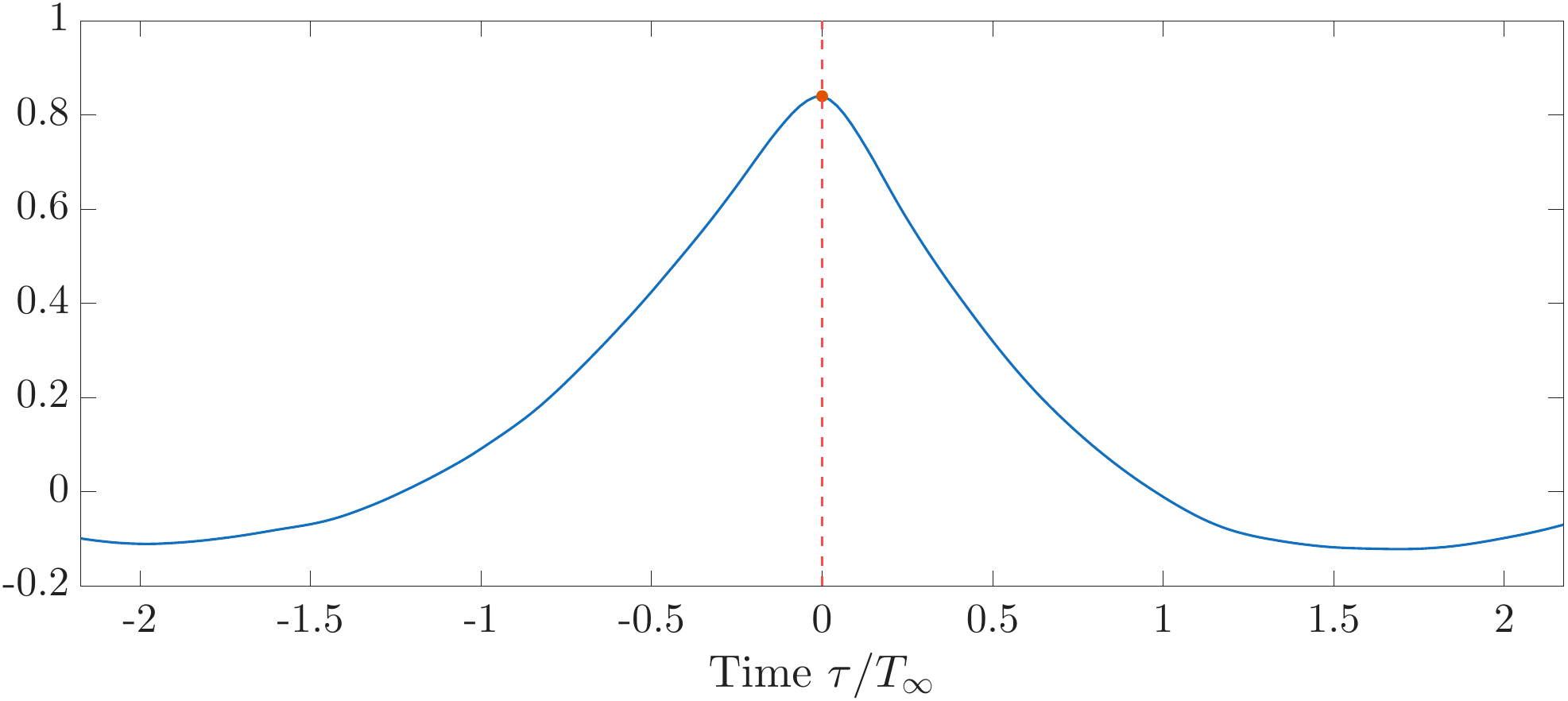}
    \put (7, 38) {b)}
    \end{overpic}
    \hspace{7mm}
    
    \caption{(a) Time series of the relative area of the interrogation window covered by dimples or scars (blue) and the mean-squared horizontal divergence (red) at depth \(z \approx -1.9\, \Taylor\), for one cycle of experimental Case A; (b) average cross-correlation across all cycles at depth \(z \approx -1.9 \,\Taylor\) for Case A, as function of time lag $\tau$ in units of integral time scale $\Tint=2\Lint/u'$. (b) has a correlation peak of $0.84$.}
    \label{fig:structure_divergence_corr_exp}
\end{figure}

Figure \ref{fig:structure_divergence_corr_exp}a depicts two time series for the highest--Reynolds number case (experimental Case~A, see Tab.~\ref{tab:flopProp}), over one measurement cycle. The time series are (1) the relative area of the interrogation window covered by dimples or scars, $\Area(t)$, and (2) the instantaneous mean-squared horizontal divergence in a plane \(z \approx -1.9 \,\Taylor\). The figure resembles Fig.~3a in BA, which compared time series of the mean-square surface divergence and instantaneous number of dimples detected at the surface.
BA found a high level of agreement in a DNS dataset equivalent to Case 1 in the present study, with a correlation coefficient with a peak of 0.9 at 0.3 $\Tint$ (listed as $\Tint =0.8$ in BA due to a different conventions for integral length and time scale). 
In figure \ref{fig:structure_divergence_corr_exp}b we plot the cross-correlation between the relative surface area covered by dimples and scars and surface divergence at height $z$ as 
\begin{equation}
    \label{eq:CC_tau}
     \CAb(z; \tau) = \frac{\Bigl\langle \left( \Area(t+\tau) - \meanT{\Area}\right) \left(\MSbeta(t) - \meanT{\MSbeta} \right) \Bigl\rangle} {\sqrt{\Bigl\langle \left( \Area - \meanT{\Area}\right)^2 \Bigl\rangle} \sqrt{\Bigl\langle \left(\MSbeta - \meanT{\MSbeta} \right)^2 \Bigl\rangle}} \, ,
\end{equation} 
where the parameter $\tau$ is a time lag between $\MSbeta$ and $\Area$, evaluated at \(z \approx -1.9\, \Taylor = 1\, \mathrm{cm}\) for the measurement cycle considered in Fig.~\ref{fig:structure_divergence_corr_exp}(a). We include also the ensemble averaged cross-correlation over $20$ measurement cycles in Fig.~\ref{fig:structure_divergence_corr_exp}(e).

The correlation considered in BA was $\CNb$ rather than $\CAb$, with the instantaneous number of dimples $N(t)$ replacing $\Area(t)$ and surface divergence $\beta_s(t)$ replacing horizontal divergence $\beta(z,t)$ in Eq.~\eqref{eq:CC_tau}.
The cross-correlation in the present figure data [depicted in Fig.~\ref{fig:structure_divergence_corr_exp}(b)] does not differ substantially in magnitude from that in BA except that in the latter case the maximum of $\CNb$ was with a significant positive time lag $\tau$, which now is essentially gone.

Although the results resemble those of BA, they in fact differ substantially on three points, making the correlation between the time series all the more striking. 

First, consider the blue curve in Fig.\ \ref{fig:structure_divergence_corr_exp}a. As mentioned in Sec.~\ref{sub-sec:detection-method}, we do not count the number of dimples or scars on the surface, but rather use the relative area of the interrogation window covered by dimples and scars at each time step. Including the scars in the analysis has a large effect on the time-lag (or lack thereof) in the cross-correlation. Scars occur nearly simultaneously as upwellings, the main drivers of surface renewal, whereas the `daisy chain' of dimples at the outskirts of upwellings appears only gradually, with the most persistent dimples far outliving the upwelling that spawned them.
Following the arguments in AA, dimples and scars are also surface imprints of turbulent vortices lying in the close vicinity of the surface, meaning that the relative area of dimples and scars relates to the relative amount of vortices close to the surface. Using the relative area also simplifies the analysis, as it eschews the need to distinguish strictly between dimples and scars which can be challenging when one of each appear close together (see AA) or are, in fact, parts of the selfsame detected region. See Appendix~\ref{app:num_vs_area_det_strcuct} for details. 

Second, regarding now the orange curve in Fig.\ \ref{fig:structure_divergence_corr_exp}a, it shows not surface divergence as in BA, but horizontal divergence $\beta(z)$ a considerable way beneath the surface, \(z \approx -1.9 \,\Taylor\) ($\approx \Lint/12$ for the present case). For comparison this is more than seven times the thickness of the viscous surface layer wherein the flow changes character as the viscous surface-tangential stresses vanish \cite{shenSurfaceLayerFreesurface1999,aarnes2025}. 
The correlation presented here is a strong indication of good agreement between the trends of the surface divergence and horizontal divergence a considerable way downward into the bulk, which we study further in section \ref{sub-sec:depth-dependence}.

\begin{figure}[ht!]
    \centering
    
    \begin{overpic}[width=0.95\linewidth]{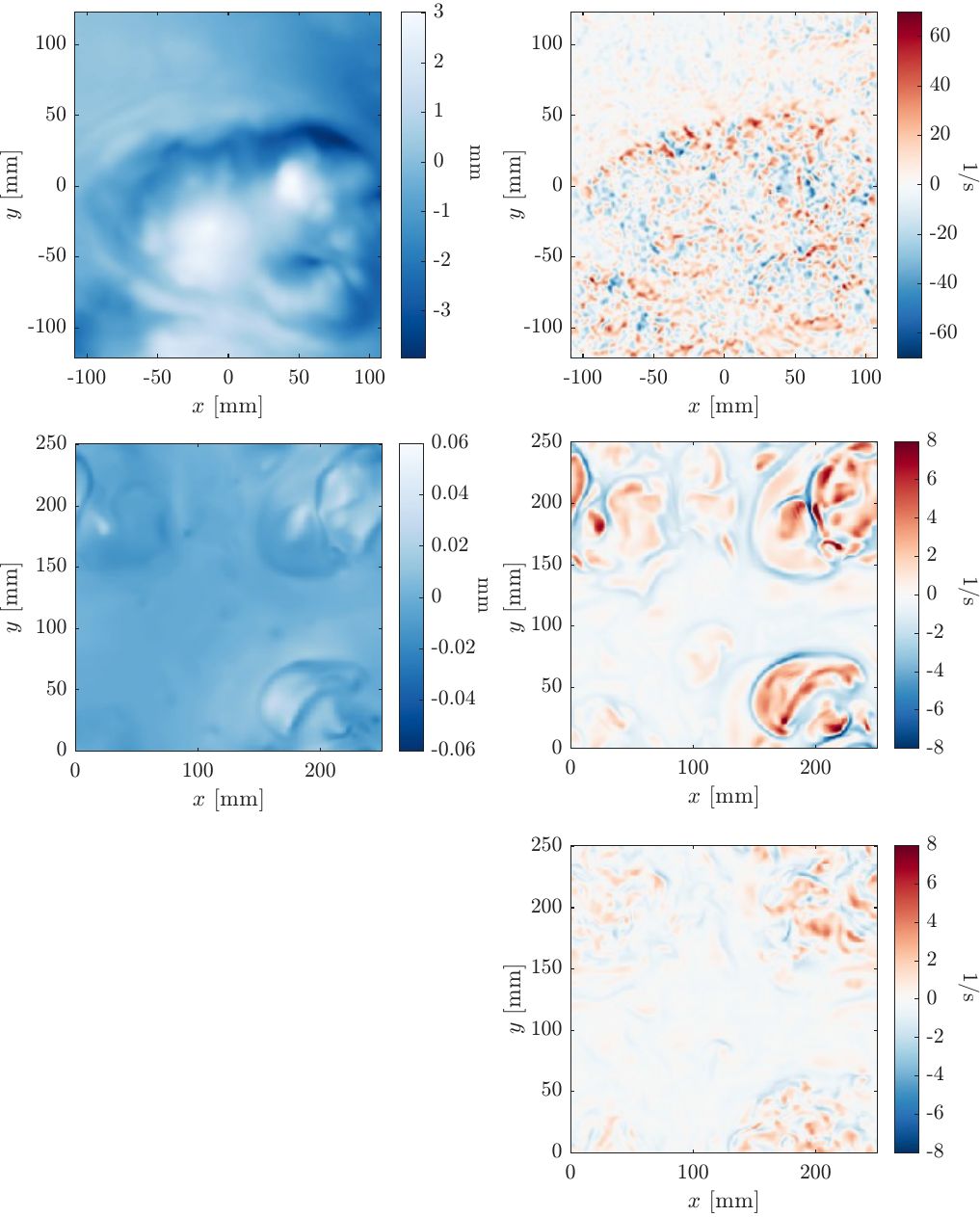}
    \put (4,  100) {a)}
    \put (43, 100) {b)}
    \put (4,  66) {c)}
    \put (43, 66) {d)}
    \put (43, 33) {e)}
    \end{overpic}
      
    \caption{Simultaneous snapshots from Case A [(a) and (b)] and Case 2 (c, d, e) of the surface elevation [(a) and (c)], the horizontal divergence field at \(z \approx -1.9\, \Taylor\) below the surface (b, e) and the surface divergence (d). Simulations are dimensionalized following Appendix~\ref{app:dimensional}.}
    \label{fig:surf_hordiv_sidebyside}
\end{figure}

Such a strong correlation between surface features and the horizontal divergence field at this depth is not obvious. For an impression of the contrast between surface features, surface divergence, $\beta_s(x,y)$, and the horizontal divergence, $\beta(x,y,z)$ at $z \approx -1.9\,\Taylor$, consider Fig.\ \ref{fig:surf_hordiv_sidebyside}. Figures \ref{fig:surf_hordiv_sidebyside}(a) and (c) depict the surface elevation during upwelling events for experimental case A and DNS case 2, respectively. For both cases, we include the horizontal divergence at $z \approx -1.9\,\Taylor$ measured simultaneously with the surface elevation, in panels (b) (Case A) and (e) (Case 2); for the simulation (Case 2), we also include the surface divergence in panel (d). The surface divergence is known to take high positive values near the centre of upwelling boils (e.g.\ \citet{guoInteractionDeformableFree2010}). In the snapshot of the surface divergence displayed in Fig.~\ref{fig:surf_hordiv_sidebyside}d, we note that $\beta_s$ is increased in upwelling regions and decreased at the edges, where downwelling areas are found, corroborating the results from the literature. The spatial structure of the horizontal divergence at $z\approx-1.9\,\Taylor$ is, however, qualitatively quite different. For the high Reynolds number case, the structure of the horizontal divergence (panel b) is much smaller than the imprint of the surface boil, with strong positive values found around the periphery of the boil, roughly underneath where scars appear. For the DNS case, which has a much lower Reynolds number than the experimental case, the difference in scales of the divergence regions is evident when comparing the surface divergence and the subsurface horizontal divergence in panels d and e. We also note that the match between surface imprints and the surface divergence (comparing panels c and d) is clearer than the match between the surface and the horizontal divergence at $z \approx -1.9\,\Taylor$ (panels c and e). There is, however, a general increase in $|\beta|$, and sharp changes in $\beta$, within the entire welling regions in Figs.~\ref{fig:surf_hordiv_sidebyside}b and e, compared to the calmer surface around them. This suggests that $\MSbeta$ is indeed enhanced under upwelling events even at $z \approx -1.9\,\Taylor$.

Third, our experimental flow in Case A has turbulent Reynolds number, $\Reint$, twenty times higher than the simulation case for which a similar correlation was found (for $\MSbetas(t)$, not $\MSbeta(t)$) in BA. That such a strong correlation between similar quantities exists at a much higher Reynolds number is a non-trivial result, in particular, in light of earlier research \cite{herlina2019} which found different scaling for mass transfer over free surfaces for flows with $\Reint \leq 500$ and $1440 \leq \Reint \leq 1856$. We remark in passing that, since we do not specifically target the mass transfer rate here, we do not contest the results in \cite{herlina2019}; we limit our tentative conclusion to confirming that the correlation we observe carries over from low to high Reynolds number flow, from surface divergence to (subsurface) horizontal divergence, and from dimples only to dimples and scars combined. The persistence of surface-to-bulk correlations for much higher $\ReynoldsNumber$ underlines a fundamental difference between a freely moving surface and a solid-wall no-slip boundary layer, where vortical structures can depend strongly on Reynolds number (e.g.\ \cite{LozanoDuran_Jimenez_2014}).

To further investigate the covariance between horizontal divergence and surface features, we make use of the DNS datasets where velocity data for all relevant $z$ is available.
Fair comparison across different datasets requires that their respective domain sizes be similar in all cases, measured in units of the most relevant length scale,
which based on AA we find is the integral scale $\Lint$.
Since the experiments have the largest integral scale-to-domain ratio, we use a cropped section of the DNS domain to 
limit our field of view (FOV) to approximately \(2.9 \,\Lint\ \times 2.9 \,\Lint\), about half of the horizontal plane in all simulation cases. The surface size in the experimental datasets is around \(1.9 \,\Lint\ \times 2.1 \,\Lint\) in Case A and \(2.8 \,\Lint\ \times 3.2 \,\Lint\) in Case B. Details on the effects of limiting the FOV for the DNS data can be found in App.~\ref{app:window_size}. We find that our field of view is sufficiently large and that the results in this section are not sensitive to its exact size.

\begin{figure}
    \centering
    \begin{overpic}[width=.8\linewidth]{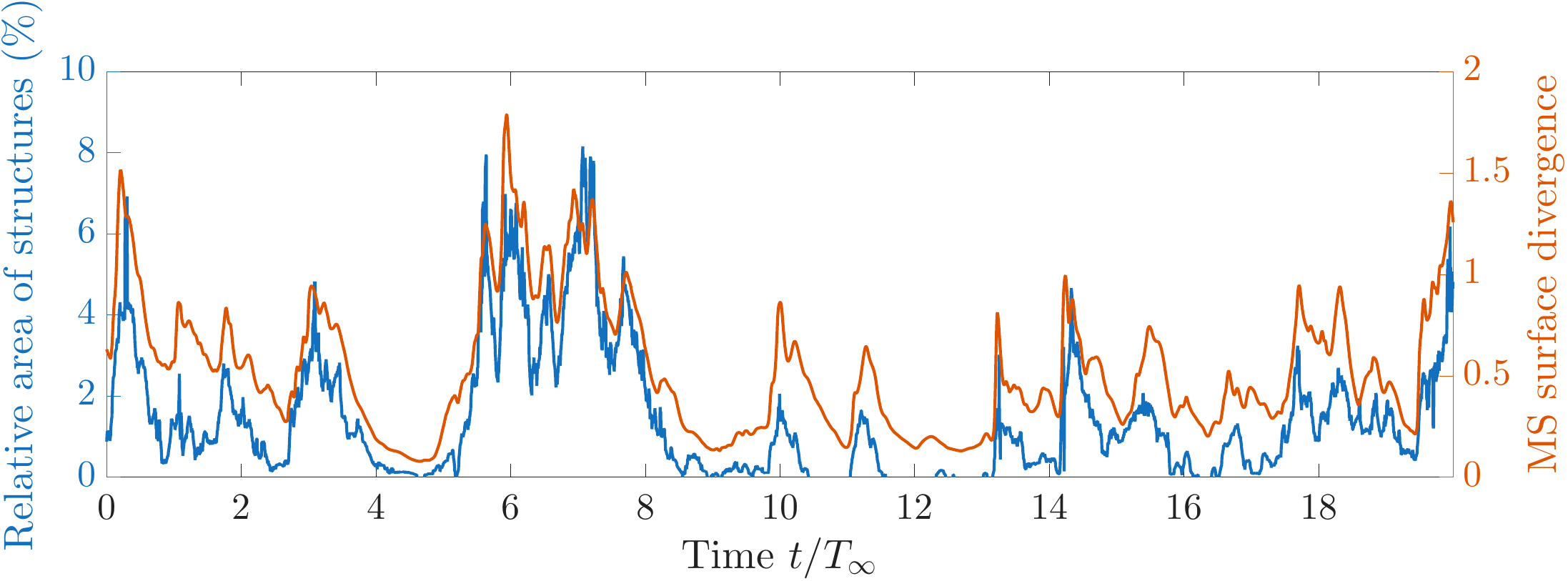}
        \put (8, 30) {a)}
    \end{overpic}\vspace{0.3em}
    \begin{overpic}[width=.8\linewidth]{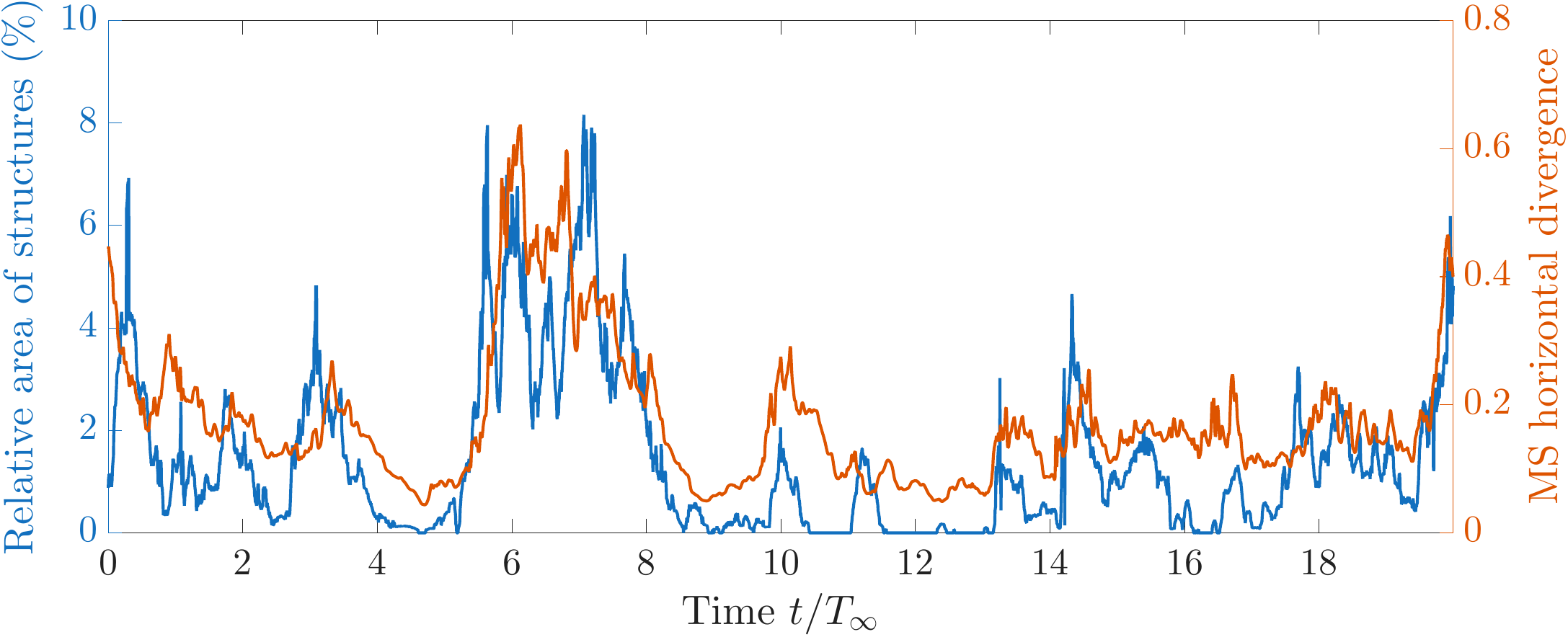}
        \put (8, 30) {b)}
    \end{overpic}\vspace{0.3em}
    \caption{Time series of the relative area of the field of view covered by dimples or scars (blue) and the mean-squared horizontal divergence (red) for the DNS Case 1, (a) at the surface (i.e., $\beta=\beta_s)$ and (b) at depth \(z \approx -1.9 \,\Taylor\).}
    \label{fig:timeseries-struct-hor-div}
\end{figure}

Figure~\ref{fig:timeseries-struct-hor-div} shows a segment of the time series of the relative area covered by dimples and scars alongside the time series of the surface divergence in panel (a) and the horizontal divergence at depth \(z \approx -1.9 \,\Taylor\) in panel (b), both for DNS Case 1. The length of the segment is selected to be comparable to one measurement cycle in the experiments.

In agreement with the experimental results discussed above, we find very strong cross correlation for the surface features and the surface/horizontal divergence. We observe that the correlation is stronger for the surface divergence, with a normalized cross correlation peak of approximately $0.9$ when the entire time series is included, and drops to approximately $0.7$ at the lower height. 
That the covariance between the time series lessens must be expected since deeper fluid layers are also penetrated by structures from beneath with no surface imprint, while far from all surface-attached vortices reach this deep (as discussed in AA). Nevertheless, peaks in the prevalence of surface structures clearly correspond to peaks in horizontal divergence also at this level and the cross-correlation remains significant. We return to this point in the next subsection.

Comparing horizontal divergence at \(z \approx -1.9 \,\Taylor\) between experiment (figure \ref{fig:structure_divergence_corr_exp}a) and DNS (figure \ref{fig:timeseries-struct-hor-div}b) the similarity in behaviour is striking.  
For the far higher Reynolds number ($\Reint= 17369$ in Case A, compared to $\Reint= 782$
in DNS Case 1),
the relative influence of large intermittent events and smaller-scale structures which affect horizontal divergence might have been expected to be more dissimilar. For example, as discussed in the introductory section, structures like the hairpin vortices from no-slip boundary layers are found to be less coherent at large Reynolds number \cite{LozanoDuran_Jimenez_2014}. This suggests a fundamental difference in the structures found in free-surface flows and no-slip boundary layers. The high peaks in the time series are associated with large upwelling events that trigger the formation of scars and dimples. These have no equivalent at a rigid interface. 

Figures~\ref{fig:structure_divergence_corr_exp}--\ref{fig:timeseries-struct-hor-div} provide further, and more general, evidence that the area of turbulent surface imprints can be used as a \emph{proxy} for sub- and cross-surface processes governed by turbulent motions also at Reynolds numbers approaching those found in natural flows (typically `gravity-driven flows' found in Region 3 of the Brocchini-Peregrine diagram of Fig.~\ref{fig:Brocchini-Peregrine} \cite{brocchini2001a}), holding promise for the use of surface imaging for practical remote-sensing purposes.


\subsection{Depth-dependence of surface-to-bulk correlations}
\label{sub-sec:depth-dependence}

To answer a main question of this study, we consider how the cross correlation between surface-feature area $\Area(t)$ and mean-square (MS) horizontal divergence $\MSbeta(z,t)$, falls off when moving away from the surface. 

\subsubsection{Surface divergence vs horizontal divergence}

We begin by shedding light on the relation between the surface divergence $\beta_s(x,y,t)$ and the horizontal divergence $\beta(x,y,z,t)$ below the surface. The two differ greatly when compared point-to-point, as discussed in connection with figure \ref{fig:surf_hordiv_sidebyside}, yet their spatial averages $\MSbetas(t)$ and $\MSbeta(z,t)$ have similar behaviour as figure \ref{fig:timeseries-struct-hor-div} shows. 
To quantify this observation, we calculate the normalized correlation between \(\MSbetas(t)\) and \(\MSbeta(z,t)\) at various depths, defined as
\begin{equation}
    \label{eq:CC_betaS_beta}
     \Cbb(z) = \frac{\Bigl\langle \left( \MSbetas - \meanT{\MSbetas}\right) \left(\MSbeta - \meanT{\MSbeta} \right) \Bigl\rangle} {\sqrt{\Bigl\langle \left( \MSbetas - \meanT{\MSbetas}\right)^2 \Bigl\rangle} \sqrt{\Bigl\langle \left(\MSbeta - \meanT{\MSbeta} \right)^2 \Bigl\rangle}} \, ,
\end{equation}
(compared to Eq.\ \eqref{eq:CC_tau} no time lag is used since the instantaneous correlation is essentially always highest in this case) and contrast it with the point-to-point correlation
\begin{equation}
    \label{eq:CCp2p}
    \Cbbpp (z) = \frac{\Bigl\langle\overline{ 
    \beta_s^2(x,y,t)\beta^2(x,y,z,t)} \Bigl\rangle} {\sqrt{\mean{\beta_s^4} \sqrt{\mean{\beta^4(z)}}}}\, .
\end{equation}

The normalized correlation, $\Cbb(z = -1.9\,\Taylor)$, for Case 1 corresponds to correlating the time series of the mean-squared surface divergence in Fig.\,\ref{fig:timeseries-struct-hor-div}a to the mean-squared horizontal divergence at \(z \approx -1.9 \,\Taylor\) in Fig.\,\ref{fig:timeseries-struct-hor-div}b with no lag between the two. For different depths $z$, the horizontal divergence at that depth is used to compute $\Cbb(z)$, while the time series of the mean-squared surface divergence remains the same. The point-to-point correlation, $\Cbbpp$, on the other hand, compares the whole surface divergence field point for point to the horizontal divergence field at a certain depth. For example, $\Cbbpp(z = -1.9\,\Taylor)$ for Case 2 corresponds to squaring the values in Fig.\,\ref{fig:surf_hordiv_sidebyside}d and e and correlating them to each other for all snapshots in Case 2.

The statistical investigation in AA indicated that even the strongest vertical `bathtub' vortices attached to dimples in the DNS reached considerably less than an integral scale into the deep when measured straight downwards, and that the horizontal vortices giving rise to scars were always found to lie near the bottom of the viscous layer $z\approx -0.26\Taylor$, far from the bottom of the blockage layer. Indeed, the blockage layer is normally understood to comprise the volume within which the turbulence is significantly influenced by the surface.
One might then expect the temporal connection between surface features and sub-surface velocity properties like $\beta(z)$ would be insignificant outside of the blockage layer. Such, however, is not the case.

\begin{figure}
    \centering
    \begin{overpic}[width=0.485\linewidth]{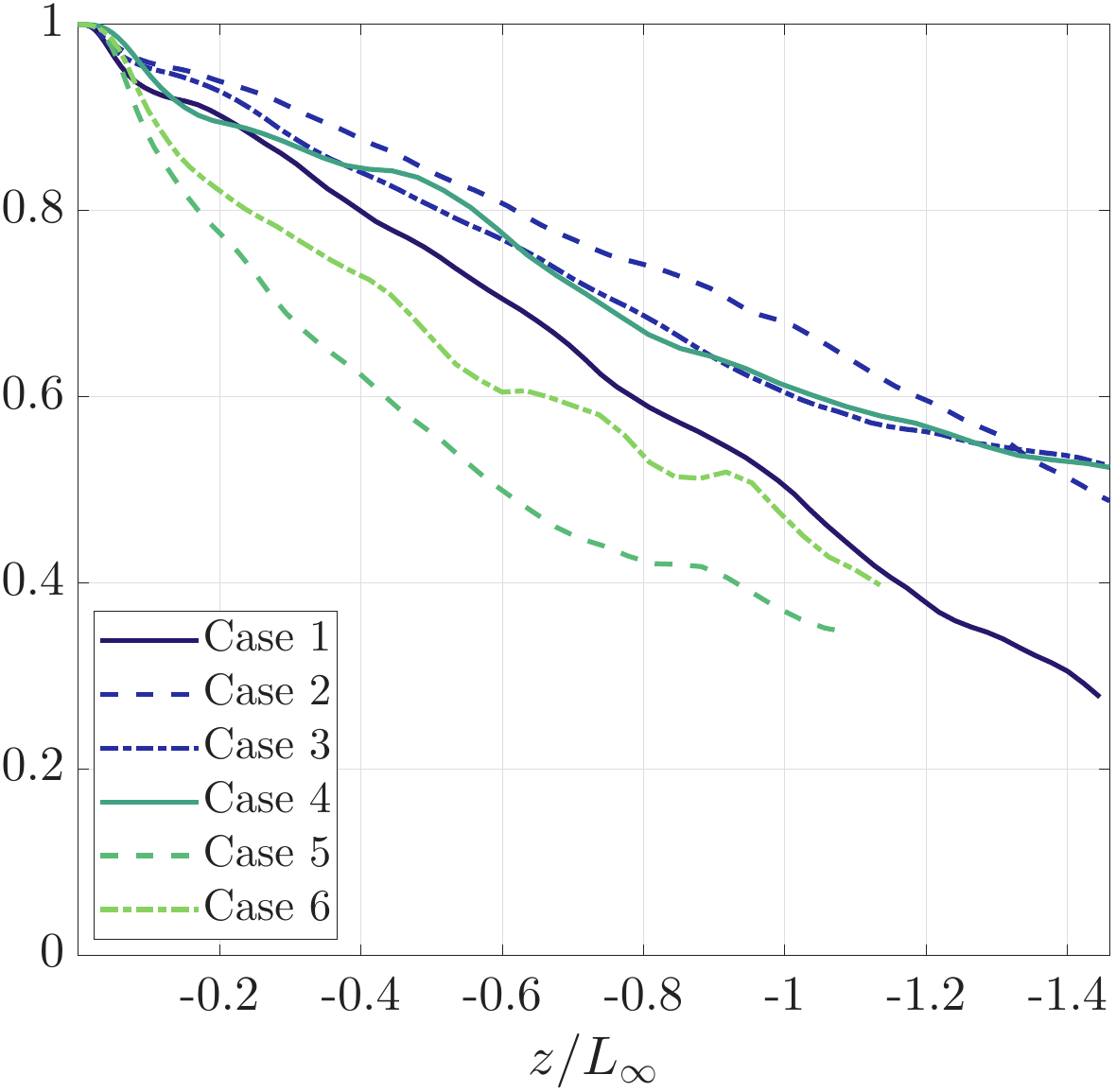}
        \put (90, 87) {a)}
    \end{overpic}
    \begin{overpic}[width=0.49\linewidth]{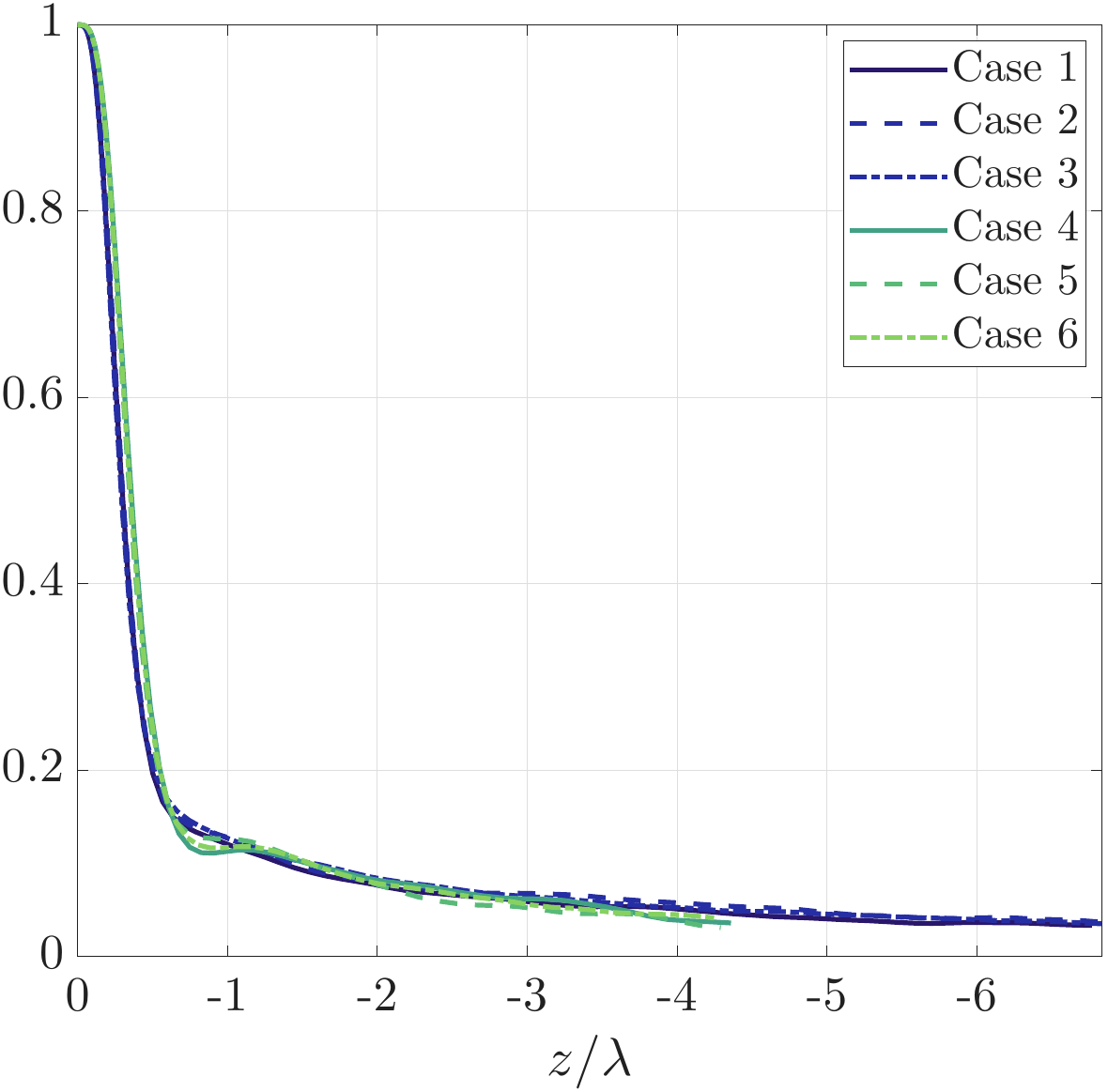}
        \put (70, 87) {b)}
    \end{overpic}
       
    \caption{Autocorrelation of the time series of (a) horizontal divergence, $\Cbb$, and (b) the point-to-point correlation, $\Cbbpp$, of the horizontal divergence for DNS cases 1-6 as a function of depth in units of the integral length scale (a), and Taylor microscale (b).}
    \label{fig:horDiv-autocorr}
\end{figure}

We plot $\Cbb$ from the surface down to $z=-1.42\,\Lint$ in Fig.~\ref{fig:horDiv-autocorr}a, while Fig.~\ref{fig:horDiv-autocorr}b shows the point-to-point correlation of $\beta_s^"$ to $\beta^2$, defined in Eqs.\ \eqref{eq:CC_betaS_beta} and \eqref{eq:CCp2p}, respectively. Not only does $\Cbb$ decline more slowly than $\Cbbpp$ (as might to some extent be expected), but the functional dependence on $z$ is radically different.
The cross correlations in Fig.\ \ref{fig:horDiv-autocorr}a for all simulation cases show an approximately linear decrease, with cross correlations remaining well over $0.3$ throughout the blockage layer, $z\gtrsim-\Lint$. Different cases show slightly different slopes. A plausible explanation is that, due to the strong intermittency of large upwelling events, statistics are not fully converged and conjecture that, given a much longer time series, the slopes would become more similar. This is supported by redoing the analysis for parts of the time series individually, which changes the slope while the linear trend is robust. 
The point-to-point correlations in Fig.~\ref{fig:horDiv-autocorr}b fall off much more rapidly, collapsing when $z$ is scaled with $\Taylor$ (noting from Tab.~\ref{tab:flopProp} that $\Lint\sim3\Taylor$--$5\,\Taylor$ in cases 1--6).
$\beta = -\partial_z u_z$ scales with $\Taylor$ near the surface \citep[se e.g.][]{aarnes2025, calmet2003, ruth2024}.

\subsubsection{Surface features vs horizontal divergence}

We consider the correlation between $\Area(t)$ and \(\MSbeta(z,t)\) at various depths,
\begin{equation}
    \label{eq:CC_z}
    \CAb     (z) = \frac{\Bigl\langle \left( \Area(t) - \meanT{\Area}\right) \left(\MSbeta(z,t) - \meanT{\MSbeta} \right) \Bigl\rangle} {\sqrt{\Bigl\langle \left( \Area(t) - \meanT{\Area}\right)^2 \Bigl\rangle} \sqrt{\Bigl\langle \left(\MSbeta(z,t) - \meanT{\MSbeta} \right)^2 \Bigl\rangle}}.
\end{equation}
This is a formal measure of the similarity between the blue and orange graphs shown for singular particular depths in Figures \ref{fig:structure_divergence_corr_exp}--\ref{fig:timeseries-struct-hor-div}, computed for all available depths down to $1.4$ integral scales.

\begin{figure}
    \centering
    \includegraphics[width=0.6\linewidth]{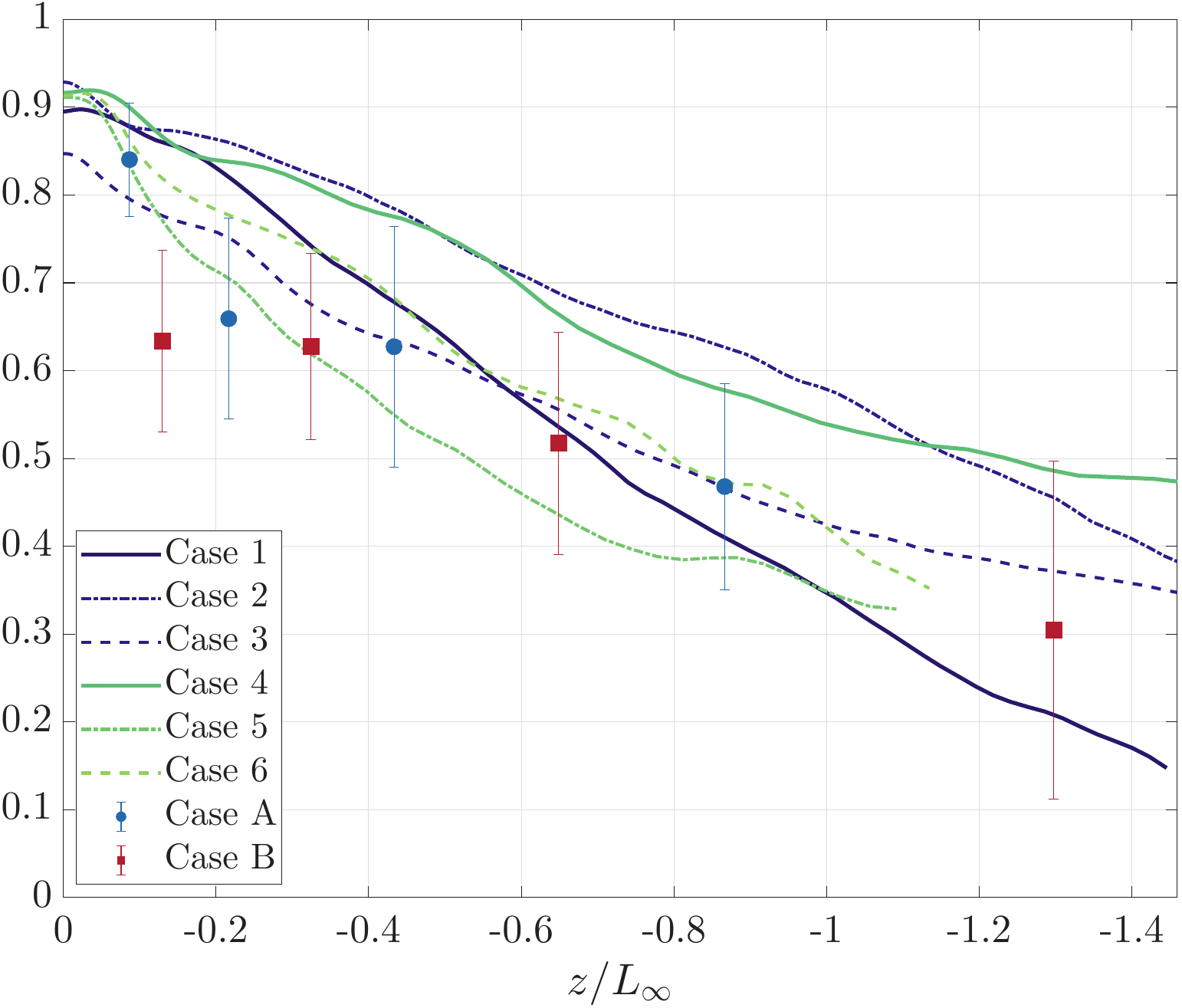}
    \caption{Cross-correlation values between the time series of mean-square horizontal divergence and the relative area of structures. The depth is normalized by the integral length scales of the various datasets. Error bars in the experiments are the standard deviations of the correlation values across the 15-20 cycles for each case and depth.}
    \label{fig:xCorr-peaks-lInt}
\end{figure}

Figure~\ref{fig:xCorr-peaks-lInt} contains some of the key conclusions in the present study. Here, $\CAb$ is plotted as a function of depth for all DNS and experimental cases.
Lines represent the DNS cases, and blue circles and orange squares represent the experimental cases A and B, respectively.

As expected, similar to Fig.~\ref{fig:horDiv-autocorr}, the correlation values in Fig.~\ref{fig:xCorr-peaks-lInt} are highest close to the surface and thence steadily decrease. The blockage layer extends to $z\sim-\Lint$, and beyond this depth, at $z\approx -1.4\,\Lint$, $\CAb$ is in the range $0.1$ -- $0.5$ for all cases, a significant degree of correlation. 
Equally interesting is the observation that the correlation appears to fall off linearly within this range across all cases, even though the statistical properties investigated by AA decay much more rapidly. There, the authors studied how the flow field behaved directly beneath surface features; what is clear from Fig.~\ref{fig:horDiv-autocorr} is that surges in horizontal divergence and increased abundance of surface features coincide in time, but not in horizontal space. This is likely at least part of the explanation for the low correlation values found by \citet{savelsberg2009} who also considered point-to-point correlations between points with the same $(x,y)$ coordinates.

Therefore, if the aggregated horizontal divergence is of interest, observation of surface features can provide a reasonable indication of it as deep as $ z\sim - \Lint$ (approximately the extent of the blockage layer)
with normalised cross-correlation of $0.35 $--$ 0.60$. That the correlation values remain significant at such depths can be ascribed to the key role of intermittency: Fig.~\ref{fig:xCorr-peaks-lInt} clearly suggests that the abundance of surface features inside a horizontal area of the size we consider is dominated by large intermittent upwelling events causing both $\Area(t)$ and $\MSbeta(z,t)$ to surge. These events, which are the main source of renewal of the surface and driver of exchanges between air and water \cite{turney2013}, can be large enough to cause coherent motion throughout the blockage layer and beyond. In contrast, surface imprints of eddies which are significantly smaller than an integral scale are frequently too weak to be detected, and live for a shorter time before diffusing.

Another important takeaway from Fig.~\ref{fig:xCorr-peaks-lInt} is the agreement between high Reynolds number experimental data and low to moderate Reynolds number DNS data. The observation is significant not least because it is evidence that a number of previous observations of interdependence of surface features and sub-surface flow made in low-$\ReynoldsNumber$ DNS in BA and AA hold for far more intense turbulence also. Although there are differences between the flows in DNS and experiments as discussed in section \ref{sec:data_acquis_&_charac}, the close agreement on display in Fig.~\ref{fig:xCorr-peaks-lInt} provides a level of validation of the simulations and their ability to provide insight into near-surface turbulence well beyond their limited parameter regimes.

One will notice that the cross correlations in the experimental cases generally lie somewhat lower than those from DNS, and that the points pertaining to Case B lie slightly below those of Case A. These observations are as can be expected due to the greater difficulty in detecting surface structures in the experiments than in DNS as discussed in Sec.~\ref{sub-sec:detection-method}, particularly in Case B where the lower turbulence level (compared to Case A) makes surface imprints much less pronounced. Although the turbulent Reynolds number of Case B is an order of magnitude larger than the highest intensity DNS case, smaller surface tension combined with fine grid size allows much shallower imprints to be detectable in the DNS data. 
The shallowest depth measurement in Case B in Fig.~\ref{fig:xCorr-peaks-lInt} appears as something of an outlier.
We attribute this to the difficulty of detecting the smallest surface features in this case
due to higher measurement noise levels caused by higher concentrations of tracer particles necessary for the shallowest measurement. Case A, having more easily detectable features due to stronger turbulence, was not affected by this challenge.

%
%

\section{Conclusions}
\label{sec:conclusions}

We have performed experiments with simultaneous measurements of the surface elevation and the sub-surface velocity field in a jet-stirred turbulence tank. The surface was measured using a profilometry method, while the flow was measured using PIV. We have implemented a previously developed, simple computer vision method that utilizes 2-D wavelet transformations to detect structures on the surface \cite{babiker2023}, namely dimples and scars, imprints associated with gravity-dominated free-surface turbulence flows. Dimples are small circular imprints of vertical vortices that have attached themselves to the surface, whereas scars are sharp, curved, elongated indentations left by horizontal vortices close beneath the surface \cite{aarnes2025}.
We demonstrate the method successfully for the subtantially more challenging case of experimental data, allowing us to correlate surface features to sub-surface velocity fields for the first time to our knowledge.

Throughout the study, we compare the experimental measurements with a set of Direct Numerical Simulation (DNS) data (see \cite{aarnes2025, guoInteractionDeformableFree2010,xuanConservativeSchemeSimulation2019}) of approximately homogeneous, isotropic turbulence impinging on a free surface. Although experimental Reynolds numbers are up to two orders of magnitude higher than those in DNS, all the physical insights we report concerning the coupling between surface imprints and bulk flow hold for both and behave qualitatively very similarly. 

It was found by \Babiker\ from DNS data that there was a high degree of correlation between the development in time of the mean-square surface divergence and the number of dimples observed on the surface. The observation was a promising indication of the possibility of observing sub-surface flow from observations of the surface only, but being based on DNS, was limited to Reynolds numbers far removed from those encountered in natural flow. Our experiments allow access to far higher Reynolds numbers, allowing us to test the DNS observations under conditions which would be prohibitively expensive for DNS.

Rather than counting surface features, we find here that the fraction of the free surface area covered by dimples and scars combined, $\Area(t)$, is a more practical and robust measure and shows equally strong surface-to-bulk correlations. Since experimental measurements were by necessity done a small distance underneath the surface itself, consider the horizontal divergence $\beta(z,t)=\partial_x u + \partial_y v$ in planes at different depths (of which surface divergence is a special case), allowing us to directly compare DNS and experiments, and to study how the close correlation between surface features and sub-surface horizontal divergence depends on depth.

A strong correlation---normalized cross-correlation (NCC) of approximately $0.84$---is found between $\Area(t)$ and $\MSbeta(z,t)$ at the shallowest measurement level, approximately \(1.9 \,\Taylor\) below the surface, $\Taylor$ being the Taylor length scale. 
This demonstrates that the relative area of dimples and scars, which can be estimated solely based on the surface elevation, can be used as a ``proxy" for the changes in the mean-square horizontal divergence, itself shown to be a good indicator of the mean-square surface divergence which is known to be a predictor of gas transfer \cite{Mcready1986,turney2005,turney2013}.

By studying the NCC between \(\Area(t)\) and \(\MSbeta(z,t)\) at increasing depth, we were able to study numerically as well as experimentally how the correlation decayed away from the surface. 
We find in all cases that the NCC falls off approximately linearly as a function of depth, and remains significant, at the level of $0.1$ and above, down to a depth of 1.9 turbulent integral scales, significantly deeper than the thickness of the `blockage layer' in which the turbulent statistics are altered by the presence of the surface \citep[see e.g.][]{nagaosa1999, calmet2003}. The very deep range of correlations can be attributed to the large, intermittent upwelling events which involve structures considerably greater than the bulk integral scale, and which cause surges in both horizontal divergence and free-surface imprints. Dimples and scars themselves are much smaller, on the order of the viscous layer thickness, \(\approx 0.26 \,\Taylor\); however, many of these imprints can be seen on the surface around a single large upwelling.

Despite very disparate Reynolds numbers, NCC as a function of depth in experiments and simulations show excellent agreement as a function of depth measured in units of the integral scale. NCC falls off linearly away from the surface with the same slope, indicating that conclusions regarding surface-to-bulk correlations based on surface feature detection which were drawn from DNS simulations extend to far higher Reynolds numbers.
This demonstrates fundamental differences between free-surface turbulence and other types of boundary layers. 
In contrast, correlations between surface divergence $\beta_s(x,y,t)$ and horizontal divergence directly beneath, $\beta(x,y,z,t)$, decays rapidly with depth beyond the thin viscous sub-layer; the close correlation between flow fields at the surface and in horizontal planes below it is highly non-local in space. 
Thus, the relatively weak correlations between surface elevation and sub-surface velocity field quantities directly beneath found in previous studies \citep{savelsberg2009} should not be interpreted as a weak coupling between the two in general. Singling out the areas of the surface where the turbulence leaves clear imprints and accounting for the fact that the interplay is non-local in space and (to a lesser degree) time, surface shape can provide significant information about the large, energy-carrying turbulent eddies surprisingly far beneath.
The results confirm experimentally, at Reynolds numbers close to natural flows, that optical observation of the free surface holds promise for remote sensing purposes for a wide range of Reynolds numbers.
%
%

\subsubsection*{Acknowledgements}
DNS data was generously shared with us by Prof.\ Lian Shen and Dr.\ Anqing Xuan at the University of Minnesota. We benefited from discussions and input from Daniel Kjellevold and Prof.\ Ingelin Steinsland.
The research was co-funded by the Research Council of Norway (\emph{iMOD}, grant 325114) and the European Union (ERC CoG, \emph{WaTurSheD}, grant 101045299; ERC StG, \emph{GLITR}, grant 101041000; MSCA-PF, \emph{InMyWaves}, grant 101107440).
Views and opinions expressed are however those of the authors only and do not necessarily reflect those of the European Union or the European Research Council. Neither the European Union nor the granting authority can be held responsible for them.

\subsubsection*{Author contributions}
A.S., A.F., and Y.H.T.\ developed the experimental methods, set up and performed the experiments. R.J.H.\ supervised the experiments. O.M.B.\ and J.R.A.\ analysed the data and performed the tracking and statistical calculations. The experimental facility was developed by Y.H.T.\ and R.J.H. S.{\AA}.E.\ led and supervised this project and contributed discussions throughout. O.M.B., J.R.A., A.F., and Y.H.T.\ wrote the manuscript with edits and contributions from all authors.

\subsection*{Datasets}
Simulation data available at \cite{aarnes25data} and experimental data available at \cite{babiker26data}.

%
%

\appendix


%
%
\section{Dimensionalization of DNS results and preservation of turbulent dimensionless groups}
\label{app:dimensional}

The DNS data are non-dimensional by construction, characterized by nominal dimensionless groups $\ReL = U^*L^*/\nu^*$, $\WeL = \rho^* {U^*}^2 L^* / \sigma^*$ and $\FrL = U^*/ \sqrt{g^*L^*}$, detailed in section \ref{sub-sec:DNS-data}, all made up of reference scales (we mark non-dimensional parameters with an asterisk in this Appendix). 
For convenience, $L^*$, $U^*$ and $\rho^*$ are set to 1. The fluid properties $\nu^*$ and $\sigma^*$ are used to vary the Reynolds and Weber numbers from case to case, giving $\ReL=1/\nu^*$, $\WeL=1/\sigma^*$, while $g^* = 1/\FrL^2$ 
is kept the same for all cases. 
Hence, $\nu^* = 4\times 10^{-4}$ and $\nu^* = 1\times 10^{-3}$ for cases 1-3 and 4-6, respectively, $\sigma^*$-values in the different cases are $0, 0.05, 0.1$, and $g^*=100$ for all cases (in practical terms, these are all inputs in the simulation).

To transform the results to dimensional quantities in a lab-relevant framework, we introduce reference physical scales $ L\rf,\, U\rf,\, \nu\rf,\, g\rf,\, \sigma\rf,\, \rho\rf $. Dimensional and non-dimensional variables satisfy $ x = L\rf\,x^*, \, u = U\rf\, u^*, \, t = \left(L\rf/ U\rf\right)\,t^*$. The remaining scales are coupled through the dimensionless flow groups. 
We require that the turbulent 
Reynolds, Froude and Weber numbers remain invariant under the transformation. In short, $\Reint =  2 u'^* \Lint^*/\nu^* = 2u' \Lint / \nu$, where $u' = U\rf u'^*$, $\Lint = L\rf \Lint^*$, $\nu = \nu\rf$, and similarly for $\Frt$ and $\Wet$. Solving for $U\rf\, L\rf$ and $\sigma\rf$ yields
\[ U\rf = \left(\frac{g\rf \nu\rf}{g^* \nu^*}\right)^{1/3}  \, , \quad
L\rf = \frac{\nu\rf}{\nu^* U\rf} \, , \quad \sigma\rf= \sigma^*\rho\rf U\rf^2 L\rf \, ,\] 
where we note that reference scales for velocity and length are independent of surface tension and density (i.e., independent of Weber number).

Inserting values for viscosity, gravity and density corresponding to water---$\nu\rf = 10^{-6}\ {\rm{m^2/s}}$, $g\rf = 9.81\ {\rm {m/s^2}}$ and $\rho\rf = 1000\ {\rm {kg/m^3}}$---yields the numerical values for reference scales listed in Tab.~\ref{tab:ref_scales}. Using the reference length scale the simulation domain has side lengths of approximately $25$\,cm for cases 1--3 and $14$\,cm for cases 4--6.

\begin{table}[h!]
\centering
\begin{tabular}{c c c c c c}
\hline
Case & $U\rf$ [mm/s] & $L\rf$ [mm] & $\sigma\rf$ [N/m] \\
\hline
1 & 62.6 & 39.9 & 0 \\
2 & 62.6 & 39.9 & $7.82 \times 10^{-3}$ \\
3 & 62.6 & 39.9 & $1.56 \times 10^{-2}$ \\
\hline
4 & 46.1 & 21.7 & 0 \\
5 & 46.1 & 21.7 & $2.31 \times 10^{-3}$ \\
6 & 46.1 & 21.7 & $4.62 \times 10^{-3}$ \\
\hline
\end{tabular}
\caption{Dimensional reference scales for the six simulation cases using
$\nu\rf=10^{-6}\,\mathrm{m^2/s}$, $g\rf=9.81\,\mathrm{m/s^2}$,
$\rho\rf=1000\,\mathrm{kg/m^3}$, and preserving $\Reint,\Frt,\Wet$.}
\label{tab:ref_scales}
\end{table}

%
%

\section{Metrics for homogeneity and isotropy}
\label{app:HITmetrics}

Following \citet{ruth2024} We quantify deviations from zero-mean-flow HIT using three different metrics defined as

\begin{gather}
    \text{mean-flow factor (MFF) } = \frac{\sqrt{\meanT{u_x}^2 + \meanT{u_z}^2}}{\sqrt{u_x'^2 + u_z'^2}}, \\
    \text{normalized Reynolds stress (NRS) } = \frac{\meanT{|u_x u_z|}}{ \meanT{u_x^2} + \meanT{u_z^2}}, \\
    \text{mean strain-rate factor (MSRF) } = \frac{\sqrt{ (\partial_x \meanT{u_x})^2 + (\partial_z \meanT{u_z})^2 }}{\sqrt{ \meanT{(\partial_x u_x)^2} + \meanT{(\partial_z u_z)^2}}}.
\end{gather}

\citet{ruth2024} include a factor 2 in front of the z-direction quantities by using a horizontal PIV plane to show that the statistics qualitatively match in the $y$- and $z$-direction, while the $x$-direction differed. They do this by comparing the root-mean-square velocity fluctuations, and showing that $u_x'> u_z'\approx u_y'$. Since our topmost pumps are below the characterisation depth, the anisotropy is less pronounced. For example, for Case A, we have $u_x' = 77.3$~mm/s, $u_z'=73.2$~mm/s at $z=\zref$. We have therefore omitted the factor 2.

The values are calculated locally, then averaged horizontally at \(z=\zref\), and are tabulated in Tab.~\ref{tab:tank_flow_param}. In perfectly homogeneous isotropic zero-mean turbulence, all these quantities would be zero, and values well below unity are desired.

\begin{table}
    \centering
    \begin{tabular}{p{4em} p{4.5em} p{4.5em} p{4.5em}}
        \hline
         Case & \(MFF\) & \(NRS\) & \(MSRF\) \\
         A    & 0.12   &  0.32   & 0.0017   \\
         B    & 0.29   &  0.33   & 0.0030   \\
         \hline
    \end{tabular}
    \caption{Parameters to characterize homogeneity and isotropy of the bulk turbulent flow in the experiments.}
    \label{tab:tank_flow_param}
\end{table}

Both MFF and MSRF have similar values as reported in \cite{ruth2024} at comparable Reynolds numbers \(\Reint\), whereas NRS is an order of magnitude higher, indicating some production of turbulent kinetic energy (TKE) in the bulk flow. 
The production of TKE, $\Pi = -\meanT{u_iu_j} \partial_j \meanT{u_i}$ in the region below the surface was evaluated and is significant throughout the sub-surface region with the ratio of production to dissipation being \(\Pi / \epsilon \approx 0.7-0.9\) (depending on the case) at \(z=\zref\).

\section{Estimating the viscous dissipation}
\label{app:viscous_dissipation}

An estimate of viscous dissipation \(\epsilon\) of turbulent kinetic energy (TKE) was required for experimental flow characterization. 
The PIV measurements do not resolve the smallest scales (the Kolmogorov scales) of the turbulence \citep[as is typical of such experiments; see e.g.][]{poelma2006, lavoie07}, so direct calculations of the measured shear stress will not give reliable results. 

With homogeneous isotropic turbulence (HIT), we can use the scaling argument
\begin{equation}
    \epsilon_s = C_\epsilon \frac{u'^3}{L_\infty},
    \label{eq:scaling_argument}
\end{equation}
where \(C_\epsilon\) is a constant commonly set to \(0.5\) \citep{sreenivasan1998, hearst2014, vassillicos15}.

As discussed in Sec.~\ref{sub-sub-sec:experimental_flows} our experiments do not fully satisfy the criteria for HIT, so we also consider other physics-based approaches. We draw on the longitudinal and transverse structure functions, respectively,
\begin{subequations}\label{eq:structure_function}
    \begin{align}
        D^2_\parallel(r,z)=&\meanT{\left(u_x(\mathbf{x} + r \mathbf{e}_x) - u_x(\mathbf{x}) \right)^2} \, , \\
        D^2_\perp(r,z)=&\meanT{\left(u_x(\mathbf{x} + r \mathbf{e}_z) - u_x(\mathbf{x}) \right)^2}  \, ,
    \end{align}
\end{subequations}
which we calculate in the vertical PIV plane at $z=\zref$.

Following \citet{kolmogorov41a}, at sufficiently high Reynolds numbers the turbulence is locally homogeneous and isotropic leading to the scaling laws in the inertial range 
\(D_\parallel^2 =C_2 (\epsilon r)^{2/3}\) and the transverse \(D_\perp^2 = \frac43 C_2 (\epsilon r)^{2/3}\) directions, where \(C_2 \approx 2\) is a constant. The longitudinal structure function \(D_\parallel^2\) is related to the energy spectrum such that the latter also exhibits a power law of the form \(E_{xx} (k) = C_K \epsilon^{2/3} k^{-5/3}\), with \(C_K = \frac{\Gamma(2/3)}{\pi \sqrt{3}} C_2 \approx 0.5\) where $\Gamma$ is the gamma function.
\begin{figure}
    \centering
    \includegraphics[width=0.7\linewidth]{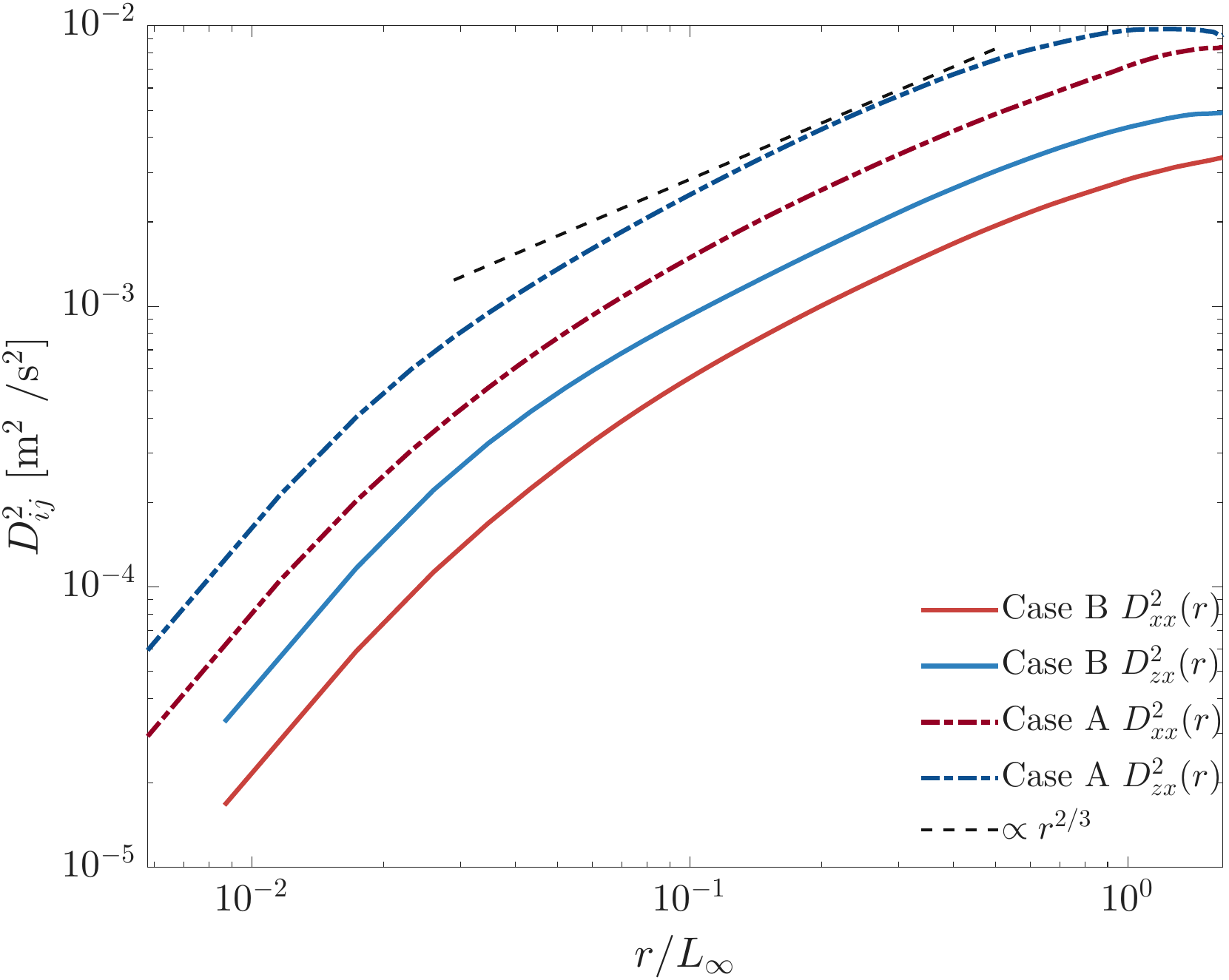}
    \caption{Longitudinal ($D^2_\parallel$) and transverse ($D^2_\perp$) structure functions as defined in Eqs.\ \eqref{eq:structure_function} for both experimental cases.}
    \label{fig:sf}
\end{figure}
The dissipation \(\epsilon\) can then be estimated for each experimental case by fitting a line to the inertial range of either the energy spectrum shown in Fig.~\ref{fig:ek} or the longitudinal and transverse structure functions, shown in Fig.~\ref{fig:sf}. The values are presented in Tab.~\ref{tab:vis_diss}, where \(\epsilon_{sf,\parallel}\) is the dissipation estimate based on a fit to the longitudinal structure function, \(\epsilon_{sf,\perp}\) is based on a fit to the transverse structure function and $\epsilon_{E_k}$ is based on a fit to the energy spectrum.

From the energy spectrum, it is also possible to evaluate the dissipation spectrum, where a new estimate for the dissipation can be calculated as \citep{Mora2019, fuchs2022}
\begin{equation}
    \epsilon_D = \int_0^{\infty} 2 \nu \kappa^2 E(\kappa) d\kappa \, ,
    \label{eq:dissipation_spectrum}
\end{equation}
\(\epsilon_D\) is then the estimate based on the integral of the dissipation spectrum, also tabulated in Tab.~\ref{tab:vis_diss}.
Since the highest wave-numbers in the energy and dissipation spectra depend on the PIV resolution, we use a fit in the dissipation region to integrate up to the wave-number corresponding to the Kolmogorov scale (the precise choice of integration limit had very little effect in our case). Finally, we also include the scaling argument \(\epsilon_s\) for completeness in Tab.~\ref{tab:vis_diss}.

\begin{table}
    \centering
    \begin{tabular}{p{3em} p{4em} p{4em} p{4em} p{4em} p{4em}}
        \hline
        Case & $\epsilon_{sf,\parallel}$ & $\epsilon_{sf,\perp}$ & $\epsilon_{E_k}$      & $\epsilon_{D}$       & $\epsilon_{s}$ \\
        A    &  0.0021   & 0.0028 & 0.0030 & 0.0039 & 0.0018 \\
        B    &  0.00078  & 0.0010 & 0.0012 & 0.0022 & 0.00079\\
        \hline    
    \end{tabular}    
    \caption{Estimates of the viscous dissipation in \(\rm{m}^2 \rm{s}^{-3}\). From left to right, estimate based on: longitudinal structure function, transverse structure function, energy spectrum, dissipation spectrum, and scaling argument.}
    \label{tab:vis_diss}
\end{table}

The various methods give a range of estimates of dissipation due to the different assumptions they are based on, as well as differences in the calculations and averaging procedures. We see, for example, that \(\epsilon_{D}\) seems to be much larger than the other values, indicating overestimation, while \(\epsilon_{sf,\parallel}\) is lower than the others in both cases. We use the estimate based on \(\epsilon_{E_K}\), this value falls in the middle, mitigating over- and underestimation errors.

%
%

\section{Number and cumulative area of detected structures}
\label{app:num_vs_area_det_strcuct}

\begin{figure}
    \centering
    \begin{overpic}[width=\linewidth]{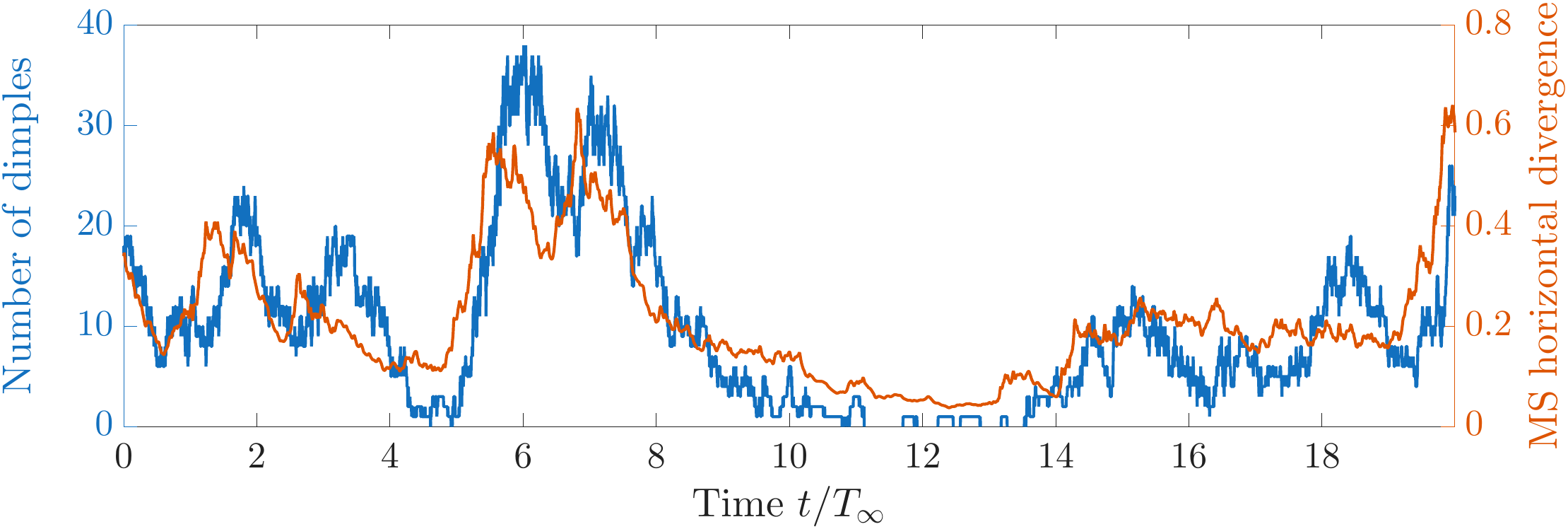}
        \put (9, 30) {a)}
    \end{overpic}\vspace{0.3em}
    \begin{overpic}[width=\linewidth]{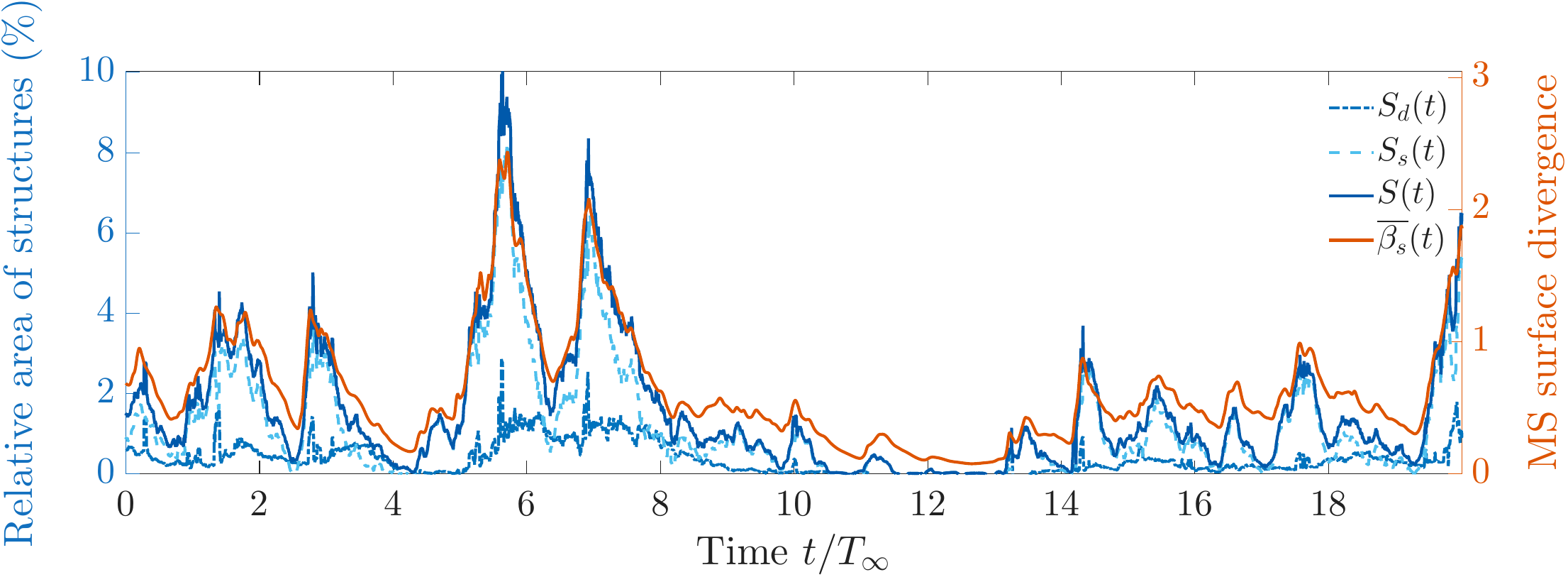}
        \put (9, 30) {b)}
    \end{overpic}
    \caption{Time series of the mean-squared surface divergence (red) plotted against (a) the number of detected dimples in the field of view and (b) the relative area of dimples and/or scars, for DNS Case 1. In the bottom panel, \(\Area_d(t)\) is the contribution from dimples, \(\Area_s(t)\) is from the scars, and \(\Area(t)\) is the aggregate of these.}
    \label{fig:DNS-hor-div-area-combo}
\end{figure}

In Sections~\ref{sec:results}, we use the relative area covered by detected structures when correlating surface structures to the horizontal divergence. In contrast, BA used the number of detected dimples for the same purpose. Here, we investigate the effect of using relative area rather than the number of identified structures. We use the DNS dataset denoted Case 1 in Tab.~\ref{tab:flopProp}, the same dataset used in BA. Unlike the results in Sec.~\ref{sec:results}, we do not limit the results to match the experimental field of view here, but use rather the whole horizontal domain to make the results directly comparable to the work in BA.

Figure \ref{fig:DNS-hor-div-area-combo}a is a reproduction of Fig.~3(a) in BA, depicting the time series of the number of detected dimples at the surface, \(N_d(t)\) (blue curve), alongside the mean-square (MS) surface divergence $\MSbetas(z)$ (red curve). The two curves are closely correlated, yet the dimple count lags behind the surface divergence: dimples arrive a little after the upwelling event (and therefore the peak in surface divergence) and then persist for some time after. 
The same time series for surface divergence is plotted in Fig.~\ref{fig:DNS-hor-div-area-combo}b alongside curves for time series of the relative area covered by dimples and scars, $\Area(t)$, (blue, solid curve) and time series of the relative area of dimples or scars only, $\Area_d(t)$ (blue, dot-dashed curve) and $\Area_s(t)$ (light blue, dashed curve), respectively. Considering first $\Area(t)$ and $\MSbetas(z)$, the correlation is even more pronounced than that in panel (a), particularly since there is no longer any time lag between the two curves. Considering the relative contributions from $\Area_d(t)$ and $\Area_s(t)$ separately, instead of their aggregate $\Area(t)$, sheds light on what causes this change. 
The time evolution of the dimple area $S_d(t)$ \emph{does} display the same lag as in the panel above, Figure\, \ref{fig:DNS-hor-div-area-combo}a---indeed, the two graphs are very similar but for an overall scaling---yet this is not the case for the area covered by scars, $S_s(t)$. The latter overlaps temporally with the surface divergence. This is consistent with our general observations of the turbulent free-surface: scars appear at the same time as an upwelling boil, yet as the upwelling boil they encircle expands, they become too shallow to detect before the upwelling event is over. Dimples, in contrast, are delayed and contribute only a small fraction of the area most of the time, yet their ability to linger after the upwelling event is over counteracts the early disappearance of scars. 

In conclusion, Fig.~\ref{fig:DNS-hor-div-area-combo} shows that, at least in this example case, using the combined area of dimples and scars is not only an easier, but also a better proxy of surface divergence than the dimple count. The same observation holds for all our DNS datasets, not only at the surface but also in correlation with $\MSbeta(z)$ at depths underneath the surface. 

The sharp spikes seen in the dimple-area graph, which appear directly beneath surges in total area, are near-circular scars misattributed as dimples of large area. This false-positive error could be removed fairly easily with a more thoughtful sorting criterion than simply the eccentricity, yet we leave them in to show the advantage of not needing to distinguish between the types.

\section{Field of view effects on correlations}
\label{app:window_size}

\begin{figure}
    \centering
    \includegraphics[width=0.6\linewidth]{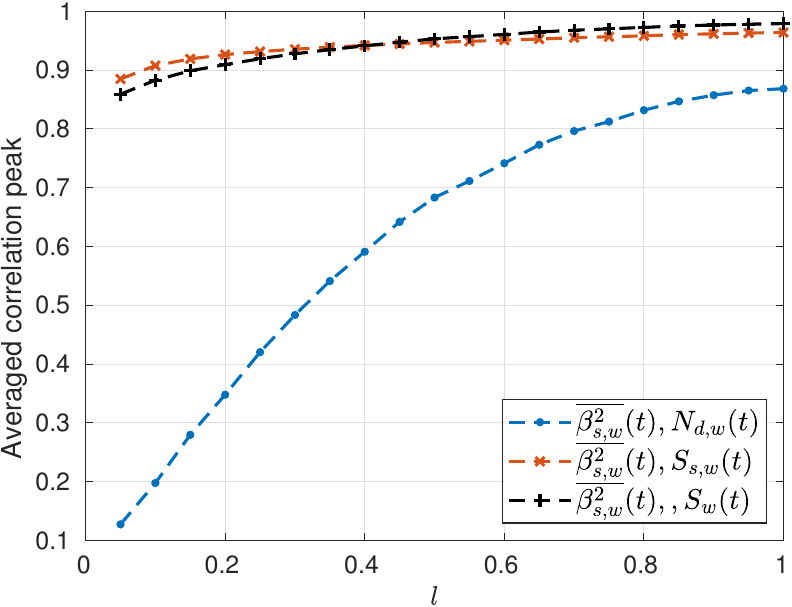}
    \caption{Averaged correlation peak between time series of the mean square surface divergence \(\meanH{\beta_{s,w}^2}(t)\) and number of dimples \(N_{d,w}(t)\) (blue), area of scars \(S_{s,w}(t)\) (red), or area of dimples and scars \(S_{w}(t)\) for FOV with normalized side length $l = L_{FOV}/L_{DNS}$.}
    \label{fig:windowed_correlations}
\end{figure}

In our study, the detection and counting of surface features as well as averaging of horizontal divergence was performed in a field of view (FOV) adjusted so that data from DNS and experiments could be fairly compared. Clearly, a FOV of appropriate size is necessary: A window that is too small will capture only parts of large upwelling events and potentially miss spatially nonlocal correlations. A window much bigger than ours, large enough to contain several independent upwelling boils at the same time, would lose the ability to ascribe a local surge in surface features with its concomitant upwelling event. We therefore briefly demonstrate that the size of our FOV is appropriate for our purposes for the correlation between the time series of (1) the surface divergence and detected surface structures and (2) the surface divergence and the subsurface horizontal divergence.

Figure~\ref{fig:windowed_correlations} depicts the effect of changing the size of the field of view by showing the highest values of the normalized cross-correlation between the time series of mean-square surface divergence \(\meanH{\beta_{s,w}^2}(t)\) and time series of detected structures on the surface for different $l = L_{FOV}/L_{DNS}$ for Case 1, where $L_{FOV}$ and $L_{DNS}$ denote the horizontal length of the (quadratic) FOV and $xy$-plane in the DNS, respectively, and the subscript $w$ denotes that the results are for a specific window size. The time series for surface structures are either time series of the number of detected dimples, \(N_{d,w}(t)\), the area covered by scars, \(S_{s,w}(t)\), or the total area of structures (dimples and scars), \(S_{w}(t)\), within the FOV. While the number of dimples is straightforward to identify, area is a better measure for scars, since small parameter changes in the scar detection may result in splitting up scars into multiple or combining several scars into one.
Every data point is computed using sliding window averaging, with a step size of $0.05$ times the horizontal domain length. The peak value of the correlation between the time series is used in the averaging.

The results show two clear trends: The correlation between \(\meanH{\beta_{s,w}^2}(t)\) and \(N_{d,w}(t)\) is strongly dependent on the FOV size; the correlation between \(\meanH{\beta_{s,w}^2}(t)\) and the area fraction covered by scars \(\Area_{s,w}(t)\) is all but independent of the FOV size. The latter has a high correlation ($> 0.88$) even for the smallest FOV tested, which spans only 5\% of the horizontal domain. 

The graphs accord well with the `cartoon' narrative presented in BA: surges in surface divergence are associated with upwelling events which has a core---signified by a 'boil'---with high positive values of $\beta_s$ and an area with corresponding downwelling surrounding the boil. Scars, as detailed in AA, are positioned in between the boil and the downwelling region outside. Dimples appear around the edge of the upwelling boil a little after it appears, and then persist for some time while being advected with the surrounding flow. According to this picture, for a strong time correlation to be observed between mean-square surface divergence and a type of surface feature, the dimple or scar and (at least parts of) the up- or downwelling region they are associated with must \emph{both} appear inside the FOV (although not necessarily simultaneously, as for dimples). For scars, this is nearly always the case, even for small windows, because they lie in between the areas where $\MSbetas$ is enhanced. Dimples, on the other hand, can drift some distance away from where they originated, and a high level of correlation requires a window significantly larger than the size of large upwelling boils. 

\begin{figure}
    \centering
    \begin{overpic}[width=0.47\linewidth]{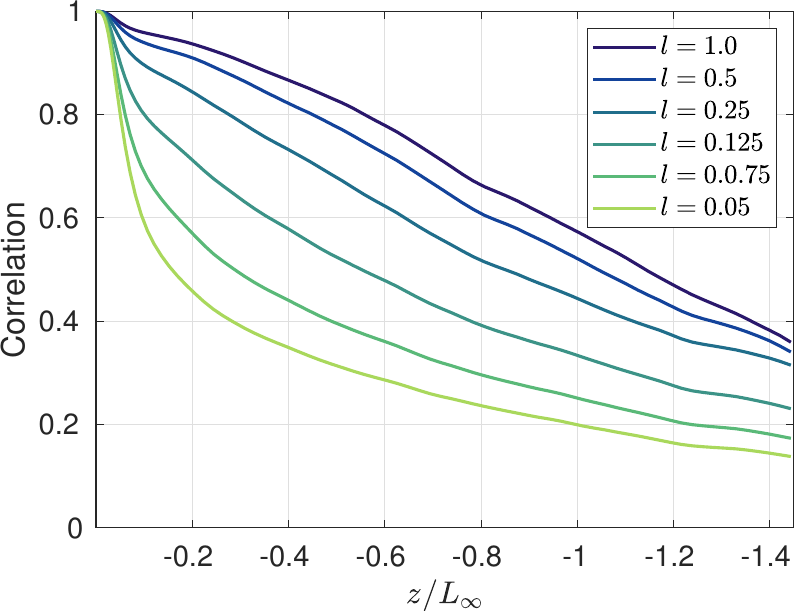}
        \put (9, 82) {a)}
    \end{overpic} ~{} 
    \begin{overpic}[width=0.49\linewidth]{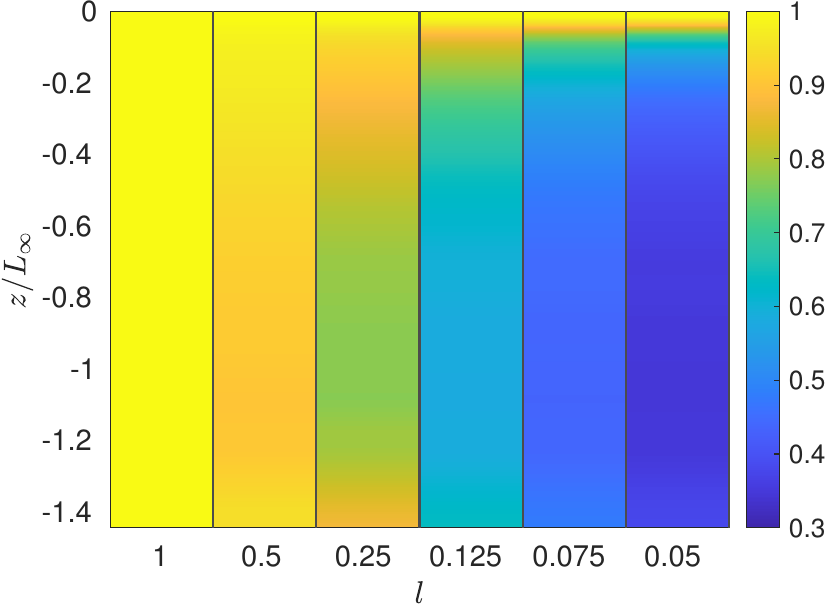}
        \put (9, 82) {b)}
    \end{overpic}
    \caption{Correlation of the time series of the mean-squared surface divergence \(\meanH{\beta_{s,w}^2}(t)\) and the horizontal divergence  \(\meanH{\beta_{w}^2}(z,t)\), for FOV with normalized side length $l = L_{FOV}/L_{DNS}$: (a) Correlation curves as a function of normalized depth, $z/\Lint$; (b) correlation relative relative to the correlation for the full window FOV.}
    \label{fig:horDiv-corr_windowed}
\end{figure}

Moving on to how the FOV affects the correlation of surface to subsurface (horizontal) divergence, we direct the attention to Fig.~\ref{fig:horDiv-corr_windowed}. The figure shows the correlation between the time series of the mean-squared surface divergence  \(\meanH{\beta_{s,w}^2}(t)\) and the horizontal divergence  \(\meanH{\beta_{w}^2}(z,t)\) at increasing depth $z$ (normalized with the integral length scale, $\Lint$), for Case 1. The side lengths of the quadratic FOV is  $l = L_{FOV}/L_{DNS} = [1, 0.5, 0.25, 0.125, 0.075, 0.05]$, i.e., a FOV with area $100\%, 25\%, 6.25\%, 1.56\%, 0.56\%, 0.25\%$ of the full window, corresponding to $L_{FOV}/L_\infty = 5.83, 2.91, 1.46, 0.73, 0.44, 0.29$, respectively. Figure \ref{fig:horDiv-corr_windowed}a shows the correlation curves while panel b highlights the difference between the full window results and the smaller FOV results by displaying the correlation curves normalized with the full window result, that is, $C_l(z)/C_{l=1}(z)$, where $C_l(z)$ is the depth-dependent correlation curve for a FOV with side length $l$.

The correlation curves in Fig.~\ref{fig:horDiv-corr_windowed}a indicate that a reduction of the FOV from the full window to one with side lengths $l \geq 0.25$ results in only a minor attenuation of the correlation, but yield no qualitative changes in the results.
For a smaller FOV, the drop in correlation is much steeper near the free surface, an effect that is particularly visible in panel b, where the three rightmost columns ($l = 0.125, 0.075, 0.05$) deviate significantly from the full window results everywhere but in the immediate vicinity of the free surface. The results suggest that large horizontal structures dominate near the surface, with strong vertical coupling in the immediate vicinity of the surface, captured across all window sizes. However, the coherence is lost as early as $z/\Lint \lesssim -0.1$, for fields of view which have $l < 0.25$, indicating horizontally localized, vertically intermitted structures through the blockage layer. We find such structures in the visualization of the surface divergence and subsurface divergence in Fig.~\ref{fig:surf_hordiv_sidebyside}c--e. The upwelling boils visible at the surface (panel c) are coupled to large structures of strong, positive surface divergence localized where the boils occur (panel d). Below the surface, we see that the structures of horizontal divergence are much smaller (panel e), which will lead to less correlation between the mean-squared values in these snapshots for FOV that do not cover the main body of an upwelling boil (in said snapshots the largest upwelling boil has an area fraction on the order of 0.1 of the full window).

The the correlation results for surface to subsurface divergence for different FOV sizes suggest that an appropriate FOV for our datasets has $l\geq 0.25$. We may add that a slightly larger FOV may be desirable, such that it is large enough to enclose the major part of the surface manifestations of large upwelling boils. Moreover, the results point to near-surface structures with a larger horizontal decorrelation length than the integral length scale measured at the reference depth ($\Lint/L_{DNS} \approx 0.17$ for Case 1). This is consistent with the increase in the integral length scale which occurs as the surface is approached from below, evident in Fig.~\ref{fig:exp_dns_comp} (see also the depth dependence of the ``local integral length scale" in Fig.~7a in \cite{herlina2014}).

The overall conclusions of this Appendix and Appendix \ref{app:num_vs_area_det_strcuct} is that the best correlation between surface divergence and surface observation/subsurface horizontal divergence is found when the FOV is sufficiently large, and the sum of the surface areas of dimples and scars together is used as the proxy. We also show that our window size is appropriate, as indicated by the close correlations we have observed. A detailed investigation into spatially local dependences as well as temporal will be considered in a future publication.


\begin{thebibliography}{76}%
\makeatletter
\providecommand \@ifxundefined [1]{%
 \@ifx{#1\undefined}
}%
\providecommand \@ifnum [1]{%
 \ifnum #1\expandafter \@firstoftwo
 \else \expandafter \@secondoftwo
 \fi
}%
\providecommand \@ifx [1]{%
 \ifx #1\expandafter \@firstoftwo
 \else \expandafter \@secondoftwo
 \fi
}%
\providecommand \natexlab [1]{#1}%
\providecommand \enquote  [1]{``#1''}%
\providecommand \bibnamefont  [1]{#1}%
\providecommand \bibfnamefont [1]{#1}%
\providecommand \citenamefont [1]{#1}%
\providecommand \@href[1]{\@@startlink{#1}\@@href}%
\providecommand \@@href[1]{\endgroup#1\@@endlink}%
\providecommand \@sanitize@url [0]{\catcode `\\12\catcode `\$12\catcode `\&12\catcode `\#12\catcode `\^12\catcode `\_12\catcode `\%12\relax}%
\providecommand \@@startlink[1]{}%
\providecommand \@@endlink[0]{}%
\providecommand \@url [1]{\endgroup\@href {#1}{\urlprefix }}%
\providecommand \urlprefix  [0]{URL }%
\providecommand \Eprint [0]{\url }%
\providecommand \doibase [0]{https://doi.org/}%
\providecommand \selectlanguage [0]{\@gobble}%
\providecommand \bibinfo  [0]{\@secondoftwo}%
\providecommand \bibfield  [0]{\@secondoftwo}%
\providecommand \translation [1]{[#1]}%
\providecommand \BibitemOpen [0]{}%
\providecommand \bibitemStop [0]{}%
\providecommand \bibitemNoStop [0]{.\EOS\space}%
\providecommand \EOS [0]{\spacefactor3000\relax}%
\providecommand \BibitemShut  [1]{\csname bibitem#1\endcsname}%
\let\auto@bib@innerbib\@empty
\bibitem [{\citenamefont {Banerjee}(1994)}]{banerjee1994}%
  \BibitemOpen
  \bibfield  {author} {\bibinfo {author} {\bibfnamefont {S.}~\bibnamefont {Banerjee}},\ }\bibfield  {title} {\bibinfo {title} {Upwellings, downdrafts, and whirlpools: Dominant structures in free surface turbulence},\ }\href {https://doi.org/10.1115/1.3124398} {\bibfield  {journal} {\bibinfo  {journal} {Appl. Mech. Rev.}\ }\textbf {\bibinfo {volume} {47}},\ \bibinfo {pages} {S166} (\bibinfo {year} {1994})}\BibitemShut {NoStop}%
\bibitem [{\citenamefont {Brocchini}\ and\ \citenamefont {Peregrine}(2001)}]{brocchini2001a}%
  \BibitemOpen
  \bibfield  {author} {\bibinfo {author} {\bibfnamefont {M.}~\bibnamefont {Brocchini}}\ and\ \bibinfo {author} {\bibfnamefont {D.~H.}\ \bibnamefont {Peregrine}},\ }\bibfield  {title} {\bibinfo {title} {The dynamics of strong turbulence at free surfaces. {Part} 1. {Description}},\ }\href {https://doi.org/10.1017/S0022112001006012} {\bibfield  {journal} {\bibinfo  {journal} {J. Fluid Mech.}\ }\textbf {\bibinfo {volume} {449}},\ \bibinfo {pages} {225} (\bibinfo {year} {2001})}\BibitemShut {NoStop}%
\bibitem [{\citenamefont {Babiker}\ \emph {et~al.}(2023)\citenamefont {Babiker}, \citenamefont {Bjerkeb{\ae}k}, \citenamefont {Xuan}, \citenamefont {Shen},\ and\ \citenamefont {Ellingsen}}]{babiker2023}%
  \BibitemOpen
  \bibfield  {author} {\bibinfo {author} {\bibfnamefont {O.~M.}\ \bibnamefont {Babiker}}, \bibinfo {author} {\bibfnamefont {I.}~\bibnamefont {Bjerkeb{\ae}k}}, \bibinfo {author} {\bibfnamefont {A.}~\bibnamefont {Xuan}}, \bibinfo {author} {\bibfnamefont {L.}~\bibnamefont {Shen}},\ and\ \bibinfo {author} {\bibfnamefont {S.~{\AA}.}\ \bibnamefont {Ellingsen}},\ }\bibfield  {title} {\bibinfo {title} {Vortex imprints on a free surface as proxy for surface divergence},\ }\href {https://doi.org/10.1017/jfm.2023.370} {\bibfield  {journal} {\bibinfo  {journal} {J. Fluid Mech.}\ }\textbf {\bibinfo {volume} {964}},\ \bibinfo {pages} {R2} (\bibinfo {year} {2023})}\BibitemShut {NoStop}%
\bibitem [{\citenamefont {Aarnes}\ \emph {et~al.}(2025{\natexlab{a}})\citenamefont {Aarnes}, \citenamefont {Babiker}, \citenamefont {Xuan}, \citenamefont {Shen},\ and\ \citenamefont {Ellingsen}}]{aarnes2025}%
  \BibitemOpen
  \bibfield  {author} {\bibinfo {author} {\bibfnamefont {J.~R.}\ \bibnamefont {Aarnes}}, \bibinfo {author} {\bibfnamefont {O.~M.}\ \bibnamefont {Babiker}}, \bibinfo {author} {\bibfnamefont {A.}~\bibnamefont {Xuan}}, \bibinfo {author} {\bibfnamefont {L.}~\bibnamefont {Shen}},\ and\ \bibinfo {author} {\bibfnamefont {S.~{\AA}.}\ \bibnamefont {Ellingsen}},\ }\bibfield  {title} {\bibinfo {title} {Vortex structures under dimples and scars in turbulent free-surface flows},\ }\href {https://doi.org/10.1017/jfm.2025.72} {\bibfield  {journal} {\bibinfo  {journal} {J. Fluid Mech.}\ }\textbf {\bibinfo {volume} {1007}},\ \bibinfo {pages} {A38} (\bibinfo {year} {2025}{\natexlab{a}})}\BibitemShut {NoStop}%
\bibitem [{\citenamefont {Muraro}\ \emph {et~al.}(2021)\citenamefont {Muraro}, \citenamefont {Dolcetti}, \citenamefont {Nichols}, \citenamefont {Tait},\ and\ \citenamefont {Horoshenkov}}]{muraroFreesurfaceBehaviourShallow2021}%
  \BibitemOpen
  \bibfield  {author} {\bibinfo {author} {\bibfnamefont {F.}~\bibnamefont {Muraro}}, \bibinfo {author} {\bibfnamefont {G.}~\bibnamefont {Dolcetti}}, \bibinfo {author} {\bibfnamefont {A.}~\bibnamefont {Nichols}}, \bibinfo {author} {\bibfnamefont {S.~J.}\ \bibnamefont {Tait}},\ and\ \bibinfo {author} {\bibfnamefont {K.~V.}\ \bibnamefont {Horoshenkov}},\ }\bibfield  {title} {\bibinfo {title} {Free-surface behaviour of shallow turbulent flows},\ }\href {https://doi.org/10.1080/00221686.2020.1870007} {\bibfield  {journal} {\bibinfo  {journal} {J. Hydraul. Res.}\ }\textbf {\bibinfo {volume} {59}},\ \bibinfo {pages} {1} (\bibinfo {year} {2021})}\BibitemShut {NoStop}%
\bibitem [{\citenamefont {Longuet-Higgins}(1996)}]{longuet-higgins96}%
  \BibitemOpen
  \bibfield  {author} {\bibinfo {author} {\bibfnamefont {M.~S.}\ \bibnamefont {Longuet-Higgins}},\ }\bibfield  {title} {\bibinfo {title} {Surface manifestations of turbulent flow},\ }\href {https://doi.org/10.1017/S0022112096001371} {\bibfield  {journal} {\bibinfo  {journal} {J. Fluid Mech.}\ }\textbf {\bibinfo {volume} {308}},\ \bibinfo {pages} {15} (\bibinfo {year} {1996})}\BibitemShut {NoStop}%
\bibitem [{\citenamefont {Rashidi}(1997)}]{rashidi1997}%
  \BibitemOpen
  \bibfield  {author} {\bibinfo {author} {\bibfnamefont {M.}~\bibnamefont {Rashidi}},\ }\bibfield  {title} {\bibinfo {title} {Burst--interface interactions in free surface turbulent flows},\ }\href {https://doi.org/10.1063/1.869457} {\bibfield  {journal} {\bibinfo  {journal} {Phys. Fluids}\ }\textbf {\bibinfo {volume} {9}},\ \bibinfo {pages} {3485} (\bibinfo {year} {1997})}\BibitemShut {NoStop}%
\bibitem [{\citenamefont {Kumar}\ \emph {et~al.}(1998)\citenamefont {Kumar}, \citenamefont {Gupta},\ and\ \citenamefont {Banerjee}}]{kumar1998}%
  \BibitemOpen
  \bibfield  {author} {\bibinfo {author} {\bibfnamefont {S.}~\bibnamefont {Kumar}}, \bibinfo {author} {\bibfnamefont {R.}~\bibnamefont {Gupta}},\ and\ \bibinfo {author} {\bibfnamefont {S.}~\bibnamefont {Banerjee}},\ }\bibfield  {title} {\bibinfo {title} {An experimental investigation of the characteristics of free-surface turbulence in channel flow},\ }\href {https://doi.org/10.1063/1.869573} {\bibfield  {journal} {\bibinfo  {journal} {Phys. Fluids}\ }\textbf {\bibinfo {volume} {10}},\ \bibinfo {pages} {437} (\bibinfo {year} {1998})}\BibitemShut {NoStop}%
\bibitem [{\citenamefont {Dolcetti}\ \emph {et~al.}(2016)\citenamefont {Dolcetti}, \citenamefont {Horoshenkov}, \citenamefont {Krynkin},\ and\ \citenamefont {Tait}}]{dolcetti2016}%
  \BibitemOpen
  \bibfield  {author} {\bibinfo {author} {\bibfnamefont {G.}~\bibnamefont {Dolcetti}}, \bibinfo {author} {\bibfnamefont {K.~V.}\ \bibnamefont {Horoshenkov}}, \bibinfo {author} {\bibfnamefont {A.}~\bibnamefont {Krynkin}},\ and\ \bibinfo {author} {\bibfnamefont {S.~J.}\ \bibnamefont {Tait}},\ }\bibfield  {title} {\bibinfo {title} {Frequency--wavenumber spectrum of the free surface of shallow turbulent flows over a rough boundary},\ }\href {https://doi.org/10.1063/1.4964926} {\bibfield  {journal} {\bibinfo  {journal} {Phys. Fluids}\ }\textbf {\bibinfo {volume} {28}},\ \bibinfo {pages} {105105} (\bibinfo {year} {2016})}\BibitemShut {NoStop}%
\bibitem [{\citenamefont {Dolcetti}\ and\ \citenamefont {Garc\'{i}a~Nava}(2019)}]{dolcetti2019}%
  \BibitemOpen
  \bibfield  {author} {\bibinfo {author} {\bibfnamefont {G.}~\bibnamefont {Dolcetti}}\ and\ \bibinfo {author} {\bibfnamefont {H.}~\bibnamefont {Garc\'{i}a~Nava}},\ }\bibfield  {title} {\bibinfo {title} {Wavelet spectral analysis of the free surface of turbulent flows},\ }\href {https://doi.org/10.1080/00221686.2018.1478896} {\bibfield  {journal} {\bibinfo  {journal} {J.~Hydraul. Res.}\ }\textbf {\bibinfo {volume} {57}},\ \bibinfo {pages} {211} (\bibinfo {year} {2019})}\BibitemShut {NoStop}%
\bibitem [{\citenamefont {Mandel}\ \emph {et~al.}(2019)\citenamefont {Mandel}, \citenamefont {Gakhar}, \citenamefont {Chung}, \citenamefont {Rosenzweig},\ and\ \citenamefont {Koseff}}]{mandel2019}%
  \BibitemOpen
  \bibfield  {author} {\bibinfo {author} {\bibfnamefont {T.~L.}\ \bibnamefont {Mandel}}, \bibinfo {author} {\bibfnamefont {S.}~\bibnamefont {Gakhar}}, \bibinfo {author} {\bibfnamefont {H.}~\bibnamefont {Chung}}, \bibinfo {author} {\bibfnamefont {I.}~\bibnamefont {Rosenzweig}},\ and\ \bibinfo {author} {\bibfnamefont {J.~R.}\ \bibnamefont {Koseff}},\ }\bibfield  {title} {\bibinfo {title} {On the surface expression of a canopy-generated shear instability},\ }\href {https://doi.org/10.1017/jfm.2019.170} {\bibfield  {journal} {\bibinfo  {journal} {J. Fluid Mech.}\ }\textbf {\bibinfo {volume} {867}},\ \bibinfo {pages} {633} (\bibinfo {year} {2019})}\BibitemShut {NoStop}%
\bibitem [{\citenamefont {Gakhar}\ \emph {et~al.}(2020)\citenamefont {Gakhar}, \citenamefont {Koseff},\ and\ \citenamefont {Ouellette}}]{gakhar2020}%
  \BibitemOpen
  \bibfield  {author} {\bibinfo {author} {\bibfnamefont {S.}~\bibnamefont {Gakhar}}, \bibinfo {author} {\bibfnamefont {J.~R.}\ \bibnamefont {Koseff}},\ and\ \bibinfo {author} {\bibfnamefont {N.~T.}\ \bibnamefont {Ouellette}},\ }\bibfield  {title} {\bibinfo {title} {On the surface expression of bottom features in free-surface flow},\ }\href {https://doi.org/10.1017/jfm.2020.548} {\bibfield  {journal} {\bibinfo  {journal} {J. Fluid Mech.}\ }\textbf {\bibinfo {volume} {900}},\ \bibinfo {pages} {A41} (\bibinfo {year} {2020})}\BibitemShut {NoStop}%
\bibitem [{\citenamefont {Gakhar}\ \emph {et~al.}(2022)\citenamefont {Gakhar}, \citenamefont {Koseff},\ and\ \citenamefont {Ouellette}}]{gakhar2022}%
  \BibitemOpen
  \bibfield  {author} {\bibinfo {author} {\bibfnamefont {S.}~\bibnamefont {Gakhar}}, \bibinfo {author} {\bibfnamefont {J.~R.}\ \bibnamefont {Koseff}},\ and\ \bibinfo {author} {\bibfnamefont {N.~T.}\ \bibnamefont {Ouellette}},\ }\bibfield  {title} {\bibinfo {title} {Extracting free-surface expressions of underwater features},\ }\href {https://doi.org/10.1007/s00348-022-03491-w} {\bibfield  {journal} {\bibinfo  {journal} {Exp. Fluids}\ }\textbf {\bibinfo {volume} {63}},\ \bibinfo {pages} {138} (\bibinfo {year} {2022})}\BibitemShut {NoStop}%
\bibitem [{\citenamefont {Xuan}\ and\ \citenamefont {Shen}(2023)}]{xuan2023}%
  \BibitemOpen
  \bibfield  {author} {\bibinfo {author} {\bibfnamefont {A.}~\bibnamefont {Xuan}}\ and\ \bibinfo {author} {\bibfnamefont {L.}~\bibnamefont {Shen}},\ }\bibfield  {title} {\bibinfo {title} {Reconstruction of three-dimensional turbulent flow structures using surface measurements for free-surface flows based on a convolutional neural network},\ }\href {https://doi.org/10.1017/jfm.2023.154} {\bibfield  {journal} {\bibinfo  {journal} {J. Fluid Mech.}\ }\textbf {\bibinfo {volume} {959}},\ \bibinfo {pages} {A34} (\bibinfo {year} {2023})}\BibitemShut {NoStop}%
\bibitem [{\citenamefont {Moen}\ \emph {et~al.}(2025)\citenamefont {Moen}, \citenamefont {Aarnes}, \citenamefont {Ellingsen},\ and\ \citenamefont {Kutz}}]{moen2025}%
  \BibitemOpen
  \bibfield  {author} {\bibinfo {author} {\bibfnamefont {K.~S.}\ \bibnamefont {Moen}}, \bibinfo {author} {\bibfnamefont {J.~R.}\ \bibnamefont {Aarnes}}, \bibinfo {author} {\bibfnamefont {S.~{\AA}.}\ \bibnamefont {Ellingsen}},\ and\ \bibinfo {author} {\bibfnamefont {J.~N.}\ \bibnamefont {Kutz}},\ }{\bibinfo {title} {Mapping surface height dynamics to subsurface flow physics in free-surface turbulent flow using a shallow recurrent decoder}} (\bibinfo {year} {2025}),\ \Eprint {https://arxiv.org/abs/2510.06202} {arXiv:2510.06202 [physics.flu-dyn]} \BibitemShut {NoStop}%
\bibitem [{\citenamefont {Ruth}\ and\ \citenamefont {Coletti}(2024)}]{ruth2024}%
  \BibitemOpen
  \bibfield  {author} {\bibinfo {author} {\bibfnamefont {D.~J.}\ \bibnamefont {Ruth}}\ and\ \bibinfo {author} {\bibfnamefont {F.}~\bibnamefont {Coletti}},\ }\bibfield  {title} {\bibinfo {title} {Structure and energy transfer in homogeneous turbulence below a free surface},\ }\href {https://doi.org/10.1017/jfm.2024.1017} {\bibfield  {journal} {\bibinfo  {journal} {J. Fluid Mech}\ }\textbf {\bibinfo {volume} {1001}},\ \bibinfo {pages} {A46} (\bibinfo {year} {2024})}\BibitemShut {NoStop}%
\bibitem [{\citenamefont {Thompson}\ and\ \citenamefont {Turner}(1975)}]{Thompson1975}%
  \BibitemOpen
  \bibfield  {author} {\bibinfo {author} {\bibfnamefont {S.~M.}\ \bibnamefont {Thompson}}\ and\ \bibinfo {author} {\bibfnamefont {J.~S.}\ \bibnamefont {Turner}},\ }\bibfield  {title} {\bibinfo {title} {Mixing across an interface due to turbulence generated by an oscillating grid},\ }\href {https://doi.org/10.1017/s0022112075000341} {\bibfield  {journal} {\bibinfo  {journal} {J. Fluid Mech.}\ }\textbf {\bibinfo {volume} {67}},\ \bibinfo {pages} {349} (\bibinfo {year} {1975})}\BibitemShut {NoStop}%
\bibitem [{\citenamefont {Hopfinger}\ and\ \citenamefont {Toly}(1976)}]{hopfinger1976}%
  \BibitemOpen
  \bibfield  {author} {\bibinfo {author} {\bibfnamefont {E.~J.}\ \bibnamefont {Hopfinger}}\ and\ \bibinfo {author} {\bibfnamefont {J.~A.}\ \bibnamefont {Toly}},\ }\bibfield  {title} {\bibinfo {title} {Spatially decaying turbulence and its relation to mixing across density interfaces},\ }\href {https://doi.org/10.1017/s0022112076002371} {\bibfield  {journal} {\bibinfo  {journal} {J. Fluid Mech.}\ }\textbf {\bibinfo {volume} {78}},\ \bibinfo {pages} {155} (\bibinfo {year} {1976})}\BibitemShut {NoStop}%
\bibitem [{\citenamefont {Brumley}\ and\ \citenamefont {Jirka}(1987)}]{brumley1987}%
  \BibitemOpen
  \bibfield  {author} {\bibinfo {author} {\bibfnamefont {B.~H.}\ \bibnamefont {Brumley}}\ and\ \bibinfo {author} {\bibfnamefont {G.~H.}\ \bibnamefont {Jirka}},\ }\bibfield  {title} {\bibinfo {title} {Near-surface turbulence in a grid-stirred tank},\ }\href {https://doi.org/10.1017/S0022112087002623} {\bibfield  {journal} {\bibinfo  {journal} {J. Fluid Mech.}\ }\textbf {\bibinfo {volume} {183}},\ \bibinfo {pages} {235} (\bibinfo {year} {1987})}\BibitemShut {NoStop}%
\bibitem [{\citenamefont {McKenna}\ and\ \citenamefont {McGillis}(2004)}]{McKenna2004}%
  \BibitemOpen
  \bibfield  {author} {\bibinfo {author} {\bibfnamefont {S.~P.}\ \bibnamefont {McKenna}}\ and\ \bibinfo {author} {\bibfnamefont {W.~R.}\ \bibnamefont {McGillis}},\ }\bibfield  {title} {\bibinfo {title} {The role of free-surface turbulence and surfactants in air--water gas transfer},\ }\href {https://doi.org/10.1016/j.ijheatmasstransfer.2003.06.001} {\bibfield  {journal} {\bibinfo  {journal} {Int. J. Heat Mass Tran.}\ }\textbf {\bibinfo {volume} {47}},\ \bibinfo {pages} {539} (\bibinfo {year} {2004})}\BibitemShut {NoStop}%
\bibitem [{\citenamefont {Herlina}\ and\ \citenamefont {Jirka}(2008)}]{herlina2008}%
  \BibitemOpen
  \bibfield  {author} {\bibinfo {author} {\bibnamefont {Herlina}}\ and\ \bibinfo {author} {\bibfnamefont {G.~H.}\ \bibnamefont {Jirka}},\ }\bibfield  {title} {\bibinfo {title} {Experiments on gas transfer at the air--water interface induced by oscillating grid turbulence},\ }\href {https://doi.org/10.1017/S0022112007008968} {\bibfield  {journal} {\bibinfo  {journal} {J. Fluid Mech.}\ }\textbf {\bibinfo {volume} {594}},\ \bibinfo {pages} {183} (\bibinfo {year} {2008})}\BibitemShut {NoStop}%
\bibitem [{\citenamefont {Chiapponi}\ \emph {et~al.}(2012)\citenamefont {Chiapponi}, \citenamefont {Longo},\ and\ \citenamefont {Tonelli}}]{Chiapponi2012}%
  \BibitemOpen
  \bibfield  {author} {\bibinfo {author} {\bibfnamefont {L.}~\bibnamefont {Chiapponi}}, \bibinfo {author} {\bibfnamefont {S.}~\bibnamefont {Longo}},\ and\ \bibinfo {author} {\bibfnamefont {M.}~\bibnamefont {Tonelli}},\ }\bibfield  {title} {\bibinfo {title} {Experimental study on oscillating grid turbulence and free surface fluctuation},\ }\href {https://doi.org/10.1007/s00348-012-1367-4} {\bibfield  {journal} {\bibinfo  {journal} {Exp. Fluids}\ }\textbf {\bibinfo {volume} {53}},\ \bibinfo {pages} {1515} (\bibinfo {year} {2012})}\BibitemShut {NoStop}%
\bibitem [{\citenamefont {Lacassagne}\ \emph {et~al.}(2017)\citenamefont {Lacassagne}, \citenamefont {El-Hajem}, \citenamefont {Morge}, \citenamefont {Simoens},\ and\ \citenamefont {Champagne}}]{Lacassagne2017}%
  \BibitemOpen
  \bibfield  {author} {\bibinfo {author} {\bibfnamefont {T.}~\bibnamefont {Lacassagne}}, \bibinfo {author} {\bibfnamefont {M.}~\bibnamefont {El-Hajem}}, \bibinfo {author} {\bibfnamefont {F.}~\bibnamefont {Morge}}, \bibinfo {author} {\bibfnamefont {S.}~\bibnamefont {Simoens}},\ and\ \bibinfo {author} {\bibfnamefont {J.-Y.}\ \bibnamefont {Champagne}},\ }\bibfield  {title} {\bibinfo {title} {Study of gas liquid mass transfer in a grid stirred tank},\ }\href {https://doi.org/10.2516/ogst/2017001} {\bibfield  {journal} {\bibinfo  {journal} {Oil \& Gas Sci. Tech.}\ }\textbf {\bibinfo {volume} {72}},\ \bibinfo {pages} {7} (\bibinfo {year} {2017})}\BibitemShut {NoStop}%
\bibitem [{\citenamefont {Variano}\ and\ \citenamefont {Cowen}(2008)}]{variano2008}%
  \BibitemOpen
  \bibfield  {author} {\bibinfo {author} {\bibfnamefont {E.~A.}\ \bibnamefont {Variano}}\ and\ \bibinfo {author} {\bibfnamefont {E.~A.}\ \bibnamefont {Cowen}},\ }\bibfield  {title} {\bibinfo {title} {A random-jet-stirred turbulence tank},\ }\href {https://doi.org/10.1017/S0022112008000645} {\bibfield  {journal} {\bibinfo  {journal} {J. Fluid Mech.}\ }\textbf {\bibinfo {volume} {604}},\ \bibinfo {pages} {1–32} (\bibinfo {year} {2008})}\BibitemShut {NoStop}%
\bibitem [{\citenamefont {Asher}\ \emph {et~al.}(2012)\citenamefont {Asher}, \citenamefont {Liang}, \citenamefont {Zappa}, \citenamefont {Loewen}, \citenamefont {Mukto}, \citenamefont {Litchendorf},\ and\ \citenamefont {Jessup}}]{asher2012}%
  \BibitemOpen
  \bibfield  {author} {\bibinfo {author} {\bibfnamefont {W.~E.}\ \bibnamefont {Asher}}, \bibinfo {author} {\bibfnamefont {H.}~\bibnamefont {Liang}}, \bibinfo {author} {\bibfnamefont {C.~J.}\ \bibnamefont {Zappa}}, \bibinfo {author} {\bibfnamefont {M.~R.}\ \bibnamefont {Loewen}}, \bibinfo {author} {\bibfnamefont {M.~A.}\ \bibnamefont {Mukto}}, \bibinfo {author} {\bibfnamefont {T.~M.}\ \bibnamefont {Litchendorf}},\ and\ \bibinfo {author} {\bibfnamefont {A.~T.}\ \bibnamefont {Jessup}},\ }\bibfield  {title} {\bibinfo {title} {Statistics of surface divergence and their relation to air--water gas transfer velocity},\ }\href {https://doi.org/10.1029/2011JC007390} {\bibfield  {journal} {\bibinfo  {journal} {J. Geophys. Res.: Oceans}\ }\textbf {\bibinfo {volume} {117}},\ \bibinfo {pages} {C05035} (\bibinfo {year} {2012})}\BibitemShut {NoStop}%
\bibitem [{\citenamefont {Variano}\ and\ \citenamefont {Cowen}(2013)}]{variano2013}%
  \BibitemOpen
  \bibfield  {author} {\bibinfo {author} {\bibfnamefont {E.~A.}\ \bibnamefont {Variano}}\ and\ \bibinfo {author} {\bibfnamefont {E.~A.}\ \bibnamefont {Cowen}},\ }\bibfield  {title} {\bibinfo {title} {Turbulent transport of a high-{S}chmidt-number scalar near an air--water interface},\ }\href {https://doi.org/10.1017/jfm.2013.273} {\bibfield  {journal} {\bibinfo  {journal} {J. Fluid Mech.}\ }\textbf {\bibinfo {volume} {731}},\ \bibinfo {pages} {259} (\bibinfo {year} {2013})}\BibitemShut {NoStop}%
\bibitem [{\citenamefont {Carter}\ \emph {et~al.}(2016)\citenamefont {Carter}, \citenamefont {Petersen}, \citenamefont {Amili},\ and\ \citenamefont {Coletti}}]{carter2016}%
  \BibitemOpen
  \bibfield  {author} {\bibinfo {author} {\bibfnamefont {D.}~\bibnamefont {Carter}}, \bibinfo {author} {\bibfnamefont {A.}~\bibnamefont {Petersen}}, \bibinfo {author} {\bibfnamefont {O.}~\bibnamefont {Amili}},\ and\ \bibinfo {author} {\bibfnamefont {F.}~\bibnamefont {Coletti}},\ }\bibfield  {title} {\bibinfo {title} {Generating and controlling homogeneous air turbulence using random jet arrays},\ }\href {https://doi.org/10.1007/s00348-016-2281-y} {\bibfield  {journal} {\bibinfo  {journal} {Exp. Fluids}\ }\textbf {\bibinfo {volume} {57}},\ \bibinfo {pages} {189} (\bibinfo {year} {2016})}\BibitemShut {NoStop}%
\bibitem [{\citenamefont {Jamin}\ \emph {et~al.}(2025)\citenamefont {Jamin}, \citenamefont {Berhanu},\ and\ \citenamefont {Falcon}}]{Jamin2024}%
  \BibitemOpen
  \bibfield  {author} {\bibinfo {author} {\bibfnamefont {T.}~\bibnamefont {Jamin}}, \bibinfo {author} {\bibfnamefont {M.}~\bibnamefont {Berhanu}},\ and\ \bibinfo {author} {\bibfnamefont {E.}~\bibnamefont {Falcon}},\ }\bibfield  {title} {\bibinfo {title} {Experimental study of three-dimensional turbulence under a free surface},\ }\href {https://doi.org/10.1103/PhysRevFluids.10.034608} {\bibfield  {journal} {\bibinfo  {journal} {Phys. Rev. Fluids}\ }\textbf {\bibinfo {volume} {10}},\ \bibinfo {pages} {034608} (\bibinfo {year} {2025})}\BibitemShut {NoStop}%
\bibitem [{\citenamefont {Qi}\ \emph {et~al.}(2025)\citenamefont {Qi}, \citenamefont {Li},\ and\ \citenamefont {Coletti}}]{qi2025}%
  \BibitemOpen
  \bibfield  {author} {\bibinfo {author} {\bibfnamefont {Y.}~\bibnamefont {Qi}}, \bibinfo {author} {\bibfnamefont {Y.}~\bibnamefont {Li}},\ and\ \bibinfo {author} {\bibfnamefont {F.}~\bibnamefont {Coletti}},\ }\bibfield  {title} {\bibinfo {title} {Small-scale dynamics and structure of free-surface turbulence},\ }\href {https://doi.org/10.1017/jfm.2025.139} {\bibfield  {journal} {\bibinfo  {journal} {J. Fluid Mech.}\ }\textbf {\bibinfo {volume} {1007}},\ \bibinfo {pages} {A3} (\bibinfo {year} {2025})}\BibitemShut {NoStop}%
\bibitem [{\citenamefont {Li}\ \emph {et~al.}(2024)\citenamefont {Li}, \citenamefont {Wang}, \citenamefont {Qi},\ and\ \citenamefont {Coletti}}]{li2024}%
  \BibitemOpen
  \bibfield  {author} {\bibinfo {author} {\bibfnamefont {Y.}~\bibnamefont {Li}}, \bibinfo {author} {\bibfnamefont {Y.}~\bibnamefont {Wang}}, \bibinfo {author} {\bibfnamefont {Y.}~\bibnamefont {Qi}},\ and\ \bibinfo {author} {\bibfnamefont {F.}~\bibnamefont {Coletti}},\ }\bibfield  {title} {\bibinfo {title} {Relative dispersion in free-surface turbulence},\ }\href {https://doi.org/10.1017/jfm.2024.637} {\bibfield  {journal} {\bibinfo  {journal} {J. Fluid Mech.}\ }\textbf {\bibinfo {volume} {993}},\ \bibinfo {pages} {R2} (\bibinfo {year} {2024})}\BibitemShut {NoStop}%
\bibitem [{\citenamefont {Chatellier}\ \emph {et~al.}(2013)\citenamefont {Chatellier}, \citenamefont {Jarny}, \citenamefont {Gibouin},\ and\ \citenamefont {David}}]{chatellier2013parametric}%
  \BibitemOpen
  \bibfield  {author} {\bibinfo {author} {\bibfnamefont {L.}~\bibnamefont {Chatellier}}, \bibinfo {author} {\bibfnamefont {S.}~\bibnamefont {Jarny}}, \bibinfo {author} {\bibfnamefont {F.}~\bibnamefont {Gibouin}},\ and\ \bibinfo {author} {\bibfnamefont {L.}~\bibnamefont {David}},\ }\bibfield  {title} {\bibinfo {title} {A parametric {PIV/DIC} method for the measurement of free surface flows},\ }\href {https://doi.org/10.1007/s00348-013-1488-4} {\bibfield  {journal} {\bibinfo  {journal} {Exp. Fluids}\ }\textbf {\bibinfo {volume} {54}},\ \bibinfo {pages} {1488} (\bibinfo {year} {2013})}\BibitemShut {NoStop}%
\bibitem [{\citenamefont {Savelsberg}\ \emph {et~al.}(2006)\citenamefont {Savelsberg}, \citenamefont {Holten},\ and\ \citenamefont {van~de Water}}]{savelsberg2006}%
  \BibitemOpen
  \bibfield  {author} {\bibinfo {author} {\bibfnamefont {R.}~\bibnamefont {Savelsberg}}, \bibinfo {author} {\bibfnamefont {A.}~\bibnamefont {Holten}},\ and\ \bibinfo {author} {\bibfnamefont {W.}~\bibnamefont {van~de Water}},\ }\bibfield  {title} {\bibinfo {title} {Measurement of the gradient field of a turbulent free surface},\ }\href {https://doi.org/10.1007/s00348-006-0186-x} {\bibfield  {journal} {\bibinfo  {journal} {Exp. Fluids}\ }\textbf {\bibinfo {volume} {41}},\ \bibinfo {pages} {629} (\bibinfo {year} {2006})}\BibitemShut {NoStop}%
\bibitem [{\citenamefont {Savelsberg}\ and\ \citenamefont {van~de Water}(2009)}]{savelsberg2009}%
  \BibitemOpen
  \bibfield  {author} {\bibinfo {author} {\bibfnamefont {R.}~\bibnamefont {Savelsberg}}\ and\ \bibinfo {author} {\bibfnamefont {W.}~\bibnamefont {van~de Water}},\ }\bibfield  {title} {\bibinfo {title} {Experiments on free-surface turbulence},\ }\href {https://doi.org/10.1017/S0022112008004369} {\bibfield  {journal} {\bibinfo  {journal} {J. Fluid Mech.}\ }\textbf {\bibinfo {volume} {619}},\ \bibinfo {pages} {95} (\bibinfo {year} {2009})}\BibitemShut {NoStop}%
\bibitem [{\citenamefont {Dabiri}\ and\ \citenamefont {Gharib}(2001)}]{dabiri2001}%
  \BibitemOpen
  \bibfield  {author} {\bibinfo {author} {\bibfnamefont {D.}~\bibnamefont {Dabiri}}\ and\ \bibinfo {author} {\bibfnamefont {M.}~\bibnamefont {Gharib}},\ }\bibfield  {title} {\bibinfo {title} {Simultaneous free-surface deformation and near-surface velocity measurements},\ }\href {https://doi.org/10.1007/s003480000212} {\bibfield  {journal} {\bibinfo  {journal} {Exp. Fluids}\ }\textbf {\bibinfo {volume} {30}},\ \bibinfo {pages} {381} (\bibinfo {year} {2001})}\BibitemShut {NoStop}%
\bibitem [{\citenamefont {Dabiri}(2003)}]{dabiri2003}%
  \BibitemOpen
  \bibfield  {author} {\bibinfo {author} {\bibfnamefont {D.}~\bibnamefont {Dabiri}},\ }\bibfield  {title} {\bibinfo {title} {On the interaction of a vertical shear layer with a free surface},\ }\href {https://doi.org/10.1017/S0022112002003671} {\bibfield  {journal} {\bibinfo  {journal} {J. Fluid Mech.}\ }\textbf {\bibinfo {volume} {480}},\ \bibinfo {pages} {217} (\bibinfo {year} {2003})}\BibitemShut {NoStop}%
\bibitem [{\citenamefont {Fouras}\ \emph {et~al.}(2008)\citenamefont {Fouras}, \citenamefont {Lo~Jacono}, \citenamefont {Sheard},\ and\ \citenamefont {Hourigan}}]{fouras2008}%
  \BibitemOpen
  \bibfield  {author} {\bibinfo {author} {\bibfnamefont {A.}~\bibnamefont {Fouras}}, \bibinfo {author} {\bibfnamefont {D.}~\bibnamefont {Lo~Jacono}}, \bibinfo {author} {\bibfnamefont {G.~J.}\ \bibnamefont {Sheard}},\ and\ \bibinfo {author} {\bibfnamefont {K.}~\bibnamefont {Hourigan}},\ }\bibfield  {title} {\bibinfo {title} {Measurement of instantaneous velocity and surface topography in the wake of a cylinder at low {R}eynolds number},\ }\href {https://doi.org/10.1016/j.jfluidstructs.2008.06.013} {\bibfield  {journal} {\bibinfo  {journal} {J Fluids Struct.}\ }\textbf {\bibinfo {volume} {24}},\ \bibinfo {pages} {1271} (\bibinfo {year} {2008})}\BibitemShut {NoStop}%
\bibitem [{\citenamefont {Ng}\ \emph {et~al.}(2011)\citenamefont {Ng}, \citenamefont {Kumar}, \citenamefont {Sheard}, \citenamefont {Hourigan},\ and\ \citenamefont {Fouras}}]{ng2011experimental}%
  \BibitemOpen
  \bibfield  {author} {\bibinfo {author} {\bibfnamefont {I.}~\bibnamefont {Ng}}, \bibinfo {author} {\bibfnamefont {V.}~\bibnamefont {Kumar}}, \bibinfo {author} {\bibfnamefont {G.~J.}\ \bibnamefont {Sheard}}, \bibinfo {author} {\bibfnamefont {K.}~\bibnamefont {Hourigan}},\ and\ \bibinfo {author} {\bibfnamefont {A.}~\bibnamefont {Fouras}},\ }\bibfield  {title} {\bibinfo {title} {Experimental study of simultaneous measurement of velocity and surface topography: in the wake of a circular cylinder at low {R}eynolds number},\ }\href {https://doi.org/10.1007/s00348-010-0960-7} {\bibfield  {journal} {\bibinfo  {journal} {Exp. Fluids}\ }\textbf {\bibinfo {volume} {50}},\ \bibinfo {pages} {587} (\bibinfo {year} {2011})}\BibitemShut {NoStop}%
\bibitem [{\citenamefont {Gomit}\ \emph {et~al.}(2013)\citenamefont {Gomit}, \citenamefont {Chatellier}, \citenamefont {Calluaud},\ and\ \citenamefont {David}}]{gomit2013free}%
  \BibitemOpen
  \bibfield  {author} {\bibinfo {author} {\bibfnamefont {G.}~\bibnamefont {Gomit}}, \bibinfo {author} {\bibfnamefont {L.}~\bibnamefont {Chatellier}}, \bibinfo {author} {\bibfnamefont {D.}~\bibnamefont {Calluaud}},\ and\ \bibinfo {author} {\bibfnamefont {L.}~\bibnamefont {David}},\ }\bibfield  {title} {\bibinfo {title} {Free surface measurement by stereo-refraction},\ }\href {https://doi.org/10.1007/s00348-013-1540-4} {\bibfield  {journal} {\bibinfo  {journal} {Exp. Fluids}\ }\textbf {\bibinfo {volume} {54}},\ \bibinfo {pages} {1540} (\bibinfo {year} {2013})}\BibitemShut {NoStop}%
\bibitem [{\citenamefont {Steinmann}\ \emph {et~al.}(2021)\citenamefont {Steinmann}, \citenamefont {Casas}, \citenamefont {Braud},\ and\ \citenamefont {David}}]{steinmann2021}%
  \BibitemOpen
  \bibfield  {author} {\bibinfo {author} {\bibfnamefont {T.}~\bibnamefont {Steinmann}}, \bibinfo {author} {\bibfnamefont {J.}~\bibnamefont {Casas}}, \bibinfo {author} {\bibfnamefont {P.}~\bibnamefont {Braud}},\ and\ \bibinfo {author} {\bibfnamefont {L.}~\bibnamefont {David}},\ }\bibfield  {title} {\bibinfo {title} {Coupled measurements of interface topography and three-dimensional velocity field of a free surface flow},\ }\href {https://doi.org/10.1007/s00348-020-03115-1} {\bibfield  {journal} {\bibinfo  {journal} {Exp. Fluids}\ }\textbf {\bibinfo {volume} {62}},\ \bibinfo {pages} {14} (\bibinfo {year} {2021})}\BibitemShut {NoStop}%
\bibitem [{\citenamefont {Gomit}\ \emph {et~al.}(2022)\citenamefont {Gomit}, \citenamefont {Chatellier},\ and\ \citenamefont {David}}]{gomit2022free}%
  \BibitemOpen
  \bibfield  {author} {\bibinfo {author} {\bibfnamefont {G.}~\bibnamefont {Gomit}}, \bibinfo {author} {\bibfnamefont {L.}~\bibnamefont {Chatellier}},\ and\ \bibinfo {author} {\bibfnamefont {L.}~\bibnamefont {David}},\ }\bibfield  {title} {\bibinfo {title} {Free-surface flow measurements by non-intrusive methods: a survey},\ }\href {https://doi.org/10.1007/s00348-022-03450-5} {\bibfield  {journal} {\bibinfo  {journal} {Exp. Fluids}\ }\textbf {\bibinfo {volume} {63}},\ \bibinfo {pages} {94} (\bibinfo {year} {2022})}\BibitemShut {NoStop}%
\bibitem [{\citenamefont {Semati}\ \emph {et~al.}(2026)\citenamefont {Semati}, \citenamefont {Shankaran}, \citenamefont {Smeltzer}, \citenamefont {{\AE}s{\o}y}, \citenamefont {Hearst},\ and\ \citenamefont {Ellingsen}}]{semati2025}%
  \BibitemOpen
  \bibfield  {author} {\bibinfo {author} {\bibfnamefont {A.}~\bibnamefont {Semati}}, \bibinfo {author} {\bibfnamefont {A.}~\bibnamefont {Shankaran}}, \bibinfo {author} {\bibfnamefont {B.~K.}\ \bibnamefont {Smeltzer}}, \bibinfo {author} {\bibfnamefont {E.}~\bibnamefont {{\AE}s{\o}y}}, \bibinfo {author} {\bibfnamefont {R.~J.}\ \bibnamefont {Hearst}},\ and\ \bibinfo {author} {\bibfnamefont {S.~{\AA}.}\ \bibnamefont {Ellingsen}},\ }\bibfield  {title} {\bibinfo {title} {Simultaneous free-surface profilometry and subsurface velocimetry with fringe projection and {PIV}},\ }{\bibfield  {journal} {\bibinfo  {journal} {Exp. Fluids}\ ,\ \bibinfo {pages} {(to appear)}} (\bibinfo {year} {2026})},\ \Eprint {https://arxiv.org/abs/2512.22641} {arXiv:2512.22641 [physics.flu-dyn]} \BibitemShut {NoStop}%
\bibitem [{\citenamefont {Antonia}\ \emph {et~al.}(2017)\citenamefont {Antonia}, \citenamefont {Djenidi}, \citenamefont {Danaila},\ and\ \citenamefont {Tang}}]{antonia2017}%
  \BibitemOpen
  \bibfield  {author} {\bibinfo {author} {\bibfnamefont {R.~A.}\ \bibnamefont {Antonia}}, \bibinfo {author} {\bibfnamefont {L.}~\bibnamefont {Djenidi}}, \bibinfo {author} {\bibfnamefont {L.}~\bibnamefont {Danaila}},\ and\ \bibinfo {author} {\bibfnamefont {S.~L.}\ \bibnamefont {Tang}},\ }\bibfield  {title} {\bibinfo {title} {Small scale turbulence and the finite {R}eynolds number effect},\ }\href {https://doi.org/10.1063/1.4974323} {\bibfield  {journal} {\bibinfo  {journal} {Phys. Fluids}\ }\textbf {\bibinfo {volume} {29}},\ \bibinfo {pages} {020715} (\bibinfo {year} {2017})}\BibitemShut {NoStop}%
\bibitem [{\citenamefont {Sreenivasan}(1998)}]{sreenivasan1998}%
  \BibitemOpen
  \bibfield  {author} {\bibinfo {author} {\bibfnamefont {K.~R.}\ \bibnamefont {Sreenivasan}},\ }\bibfield  {title} {\bibinfo {title} {An update on the energy dissipation rate in isotropic turbulence},\ }\href {https://doi.org/10.1063/1.869575} {\bibfield  {journal} {\bibinfo  {journal} {Phys. Fluids}\ }\textbf {\bibinfo {volume} {10}},\ \bibinfo {pages} {528–529} (\bibinfo {year} {1998})}\BibitemShut {NoStop}%
\bibitem [{\citenamefont {Sinhuber}\ \emph {et~al.}(2015)\citenamefont {Sinhuber}, \citenamefont {Bodenschatz},\ and\ \citenamefont {Bewley}}]{sinhuber2015}%
  \BibitemOpen
  \bibfield  {author} {\bibinfo {author} {\bibfnamefont {M.}~\bibnamefont {Sinhuber}}, \bibinfo {author} {\bibfnamefont {E.}~\bibnamefont {Bodenschatz}},\ and\ \bibinfo {author} {\bibfnamefont {G.~P.}\ \bibnamefont {Bewley}},\ }\bibfield  {title} {\bibinfo {title} {Decay of turbulence at high {R}eynolds numbers},\ }\href {https://doi.org/10.1103/PhysRevLett.114.034501} {\bibfield  {journal} {\bibinfo  {journal} {Phys. Rev. Lett.}\ }\textbf {\bibinfo {volume} {114}},\ \bibinfo {pages} {034501} (\bibinfo {year} {2015})}\BibitemShut {NoStop}%
\bibitem [{\citenamefont {Adrian}\ \emph {et~al.}(2000)\citenamefont {Adrian}, \citenamefont {Meinhart},\ and\ \citenamefont {Tomkins}}]{adrian2000}%
  \BibitemOpen
  \bibfield  {author} {\bibinfo {author} {\bibfnamefont {R.~J.}\ \bibnamefont {Adrian}}, \bibinfo {author} {\bibfnamefont {C.~D.}\ \bibnamefont {Meinhart}},\ and\ \bibinfo {author} {\bibfnamefont {C.~D.}\ \bibnamefont {Tomkins}},\ }\bibfield  {title} {\bibinfo {title} {Vortex organization in the outer region of the turbulent boundary layer},\ }\href {https://doi.org/10.1017/S0022112000001580} {\bibfield  {journal} {\bibinfo  {journal} {J. Fluid Mech}\ }\textbf {\bibinfo {volume} {422}},\ \bibinfo {pages} {1–54} (\bibinfo {year} {2000})}\BibitemShut {NoStop}%
\bibitem [{\citenamefont {{Lozano-Dur\'an}}\ and\ \citenamefont {Jim\'enez}(2014)}]{LozanoDuran_Jimenez_2014}%
  \BibitemOpen
  \bibfield  {author} {\bibinfo {author} {\bibfnamefont {A.}~\bibnamefont {{Lozano-Dur\'an}}}\ and\ \bibinfo {author} {\bibfnamefont {J.}~\bibnamefont {Jim\'enez}},\ }\bibfield  {title} {\bibinfo {title} {Time-resolved evolution of coherent structures in turbulent channels: characterization of eddies and cascades},\ }\href {https://doi.org/doi.org/10.1017/jfm.2014.575} {\bibfield  {journal} {\bibinfo  {journal} {J. Fluid Mech.}\ }\textbf {\bibinfo {volume} {759}},\ \bibinfo {pages} {432} (\bibinfo {year} {2014})}\BibitemShut {NoStop}%
\bibitem [{\citenamefont {Herlina}\ and\ \citenamefont {Wissink}(2019)}]{herlina2019}%
  \BibitemOpen
  \bibfield  {author} {\bibinfo {author} {\bibnamefont {Herlina}}\ and\ \bibinfo {author} {\bibfnamefont {J.~G.}\ \bibnamefont {Wissink}},\ }\bibfield  {title} {\bibinfo {title} {Simulation of air--water interfacial mass transfer driven by high-intensity isotropic turbulence},\ }\href {https://doi.org/10.1017/jfm.2018.884} {\bibfield  {journal} {\bibinfo  {journal} {J. Fluid Mech.}\ }\textbf {\bibinfo {volume} {860}},\ \bibinfo {pages} {419} (\bibinfo {year} {2019})}\BibitemShut {NoStop}%
\bibitem [{\citenamefont {Bellani}\ \emph {et~al.}(2013)\citenamefont {Bellani}, \citenamefont {Nole},\ and\ \citenamefont {Variano}}]{bellani_turbulence_2013}%
  \BibitemOpen
  \bibfield  {author} {\bibinfo {author} {\bibfnamefont {G.}~\bibnamefont {Bellani}}, \bibinfo {author} {\bibfnamefont {M.~A.}\ \bibnamefont {Nole}},\ and\ \bibinfo {author} {\bibfnamefont {E.~A.}\ \bibnamefont {Variano}},\ }\bibfield  {title} {\bibinfo {title} {Turbulence modulation by large ellipsoidal particles: concentration effects},\ }\href {https://doi.org/10.1007/s00707-013-0925-z} {\bibfield  {journal} {\bibinfo  {journal} {Acta Mechanica}\ }\textbf {\bibinfo {volume} {224}},\ \bibinfo {pages} {2291} (\bibinfo {year} {2013})}\BibitemShut {NoStop}%
\bibitem [{\citenamefont {Esteban}\ \emph {et~al.}(2019)\citenamefont {Esteban}, \citenamefont {Shrimpton},\ and\ \citenamefont {Ganapathisubramani}}]{esteban_laboratory_2019}%
  \BibitemOpen
  \bibfield  {author} {\bibinfo {author} {\bibfnamefont {L.~B.}\ \bibnamefont {Esteban}}, \bibinfo {author} {\bibfnamefont {J.~S.}\ \bibnamefont {Shrimpton}},\ and\ \bibinfo {author} {\bibfnamefont {B.}~\bibnamefont {Ganapathisubramani}},\ }\bibfield  {title} {\bibinfo {title} {Laboratory experiments on the temporal decay of homogeneous anisotropic turbulence},\ }\href {https://doi.org/10.1017/jfm.2018.964} {\bibfield  {journal} {\bibinfo  {journal} {J. Fluid Mech.}\ }\textbf {\bibinfo {volume} {862}},\ \bibinfo {pages} {99} (\bibinfo {year} {2019})}\BibitemShut {NoStop}%
\bibitem [{\citenamefont {Nezami}\ \emph {et~al.}(2023)\citenamefont {Nezami}, \citenamefont {Byron},\ and\ \citenamefont {Johnson}}]{nezami2023laboratory}%
  \BibitemOpen
  \bibfield  {author} {\bibinfo {author} {\bibfnamefont {A.~G.}\ \bibnamefont {Nezami}}, \bibinfo {author} {\bibfnamefont {M.}~\bibnamefont {Byron}},\ and\ \bibinfo {author} {\bibfnamefont {B.~A.}\ \bibnamefont {Johnson}},\ }\bibfield  {title} {\bibinfo {title} {Laboratory generation of zero-mean-flow homogeneous isotropic turbulence: Non-grid approaches},\ }\href {https://doi.org/10.1017/flo.2023.36} {\bibfield  {journal} {\bibinfo  {journal} {Flow}\ }\textbf {\bibinfo {volume} {3}},\ \bibinfo {pages} {E42} (\bibinfo {year} {2023})}\BibitemShut {NoStop}%
\bibitem [{\citenamefont {Guo}\ and\ \citenamefont {Shen}(2010)}]{guoInteractionDeformableFree2010}%
  \BibitemOpen
  \bibfield  {author} {\bibinfo {author} {\bibfnamefont {X.}~\bibnamefont {Guo}}\ and\ \bibinfo {author} {\bibfnamefont {L.}~\bibnamefont {Shen}},\ }\bibfield  {title} {\bibinfo {title} {Interaction of a deformable free surface with statistically steady homogeneous turbulence},\ }\href {https://doi.org/10.1017/S0022112010001539} {\bibfield  {journal} {\bibinfo  {journal} {J. Fluid Mech.}\ }\textbf {\bibinfo {volume} {658}},\ \bibinfo {pages} {33} (\bibinfo {year} {2010})}\BibitemShut {NoStop}%
\bibitem [{\citenamefont {Westerweel}\ and\ \citenamefont {Scarano}(2005)}]{Westerweel2005}%
  \BibitemOpen
  \bibfield  {author} {\bibinfo {author} {\bibfnamefont {J.}~\bibnamefont {Westerweel}}\ and\ \bibinfo {author} {\bibfnamefont {F.}~\bibnamefont {Scarano}},\ }\bibfield  {title} {\bibinfo {title} {Universal outlier detection for {PIV} data},\ }\href {https://doi.org/10.1007/s00348-005-0016-6} {\bibfield  {journal} {\bibinfo  {journal} {Exp. Fluids}\ }\textbf {\bibinfo {volume} {39}},\ \bibinfo {pages} {1096} (\bibinfo {year} {2005})}\BibitemShut {NoStop}%
\bibitem [{\citenamefont {Wieneke}(2015)}]{Wieneke2015}%
  \BibitemOpen
  \bibfield  {author} {\bibinfo {author} {\bibfnamefont {B.}~\bibnamefont {Wieneke}},\ }\bibfield  {title} {\bibinfo {title} {{PIV} uncertainty quantification from correlation statistics},\ }\href {https://doi.org/10.1088/0957-0233/26/7/074002} {\bibfield  {journal} {\bibinfo  {journal} {Meas. Sci. Tech.}\ }\textbf {\bibinfo {volume} {26}},\ \bibinfo {pages} {074002} (\bibinfo {year} {2015})}\BibitemShut {NoStop}%
\bibitem [{\citenamefont {Cobelli}\ \emph {et~al.}(2009)\citenamefont {Cobelli}, \citenamefont {Maurel}, \citenamefont {Pagneux},\ and\ \citenamefont {Petitjeans}}]{cobelli2009}%
  \BibitemOpen
  \bibfield  {author} {\bibinfo {author} {\bibfnamefont {P.~J.}\ \bibnamefont {Cobelli}}, \bibinfo {author} {\bibfnamefont {A.}~\bibnamefont {Maurel}}, \bibinfo {author} {\bibfnamefont {V.}~\bibnamefont {Pagneux}},\ and\ \bibinfo {author} {\bibfnamefont {P.}~\bibnamefont {Petitjeans}},\ }\bibfield  {title} {\bibinfo {title} {Global measurement of water waves by {{Fourier}} transform profilometry},\ }\href {https://doi.org/10.1007/s00348-009-0611-z} {\bibfield  {journal} {\bibinfo  {journal} {Exp. Fluids}\ }\textbf {\bibinfo {volume} {46}},\ \bibinfo {pages} {1037} (\bibinfo {year} {2009})}\BibitemShut {NoStop}%
\bibitem [{\citenamefont {Xuan}\ and\ \citenamefont {Shen}(2019)}]{xuanConservativeSchemeSimulation2019}%
  \BibitemOpen
  \bibfield  {author} {\bibinfo {author} {\bibfnamefont {A.}~\bibnamefont {Xuan}}\ and\ \bibinfo {author} {\bibfnamefont {L.}~\bibnamefont {Shen}},\ }\bibfield  {title} {\bibinfo {title} {A conservative scheme for simulation of free-surface turbulent and wave flows},\ }\href {https://doi.org/10.1016/j.jcp.2018.10.046} {\bibfield  {journal} {\bibinfo  {journal} {J. Comput. Phys.}\ }\textbf {\bibinfo {volume} {378}},\ \bibinfo {pages} {18} (\bibinfo {year} {2019})}\BibitemShut {NoStop}%
\bibitem [{\citenamefont {Rosales}\ and\ \citenamefont {Meneveau}(2005)}]{rosalesLinearForcingNumerical2005}%
  \BibitemOpen
  \bibfield  {author} {\bibinfo {author} {\bibfnamefont {C.}~\bibnamefont {Rosales}}\ and\ \bibinfo {author} {\bibfnamefont {C.}~\bibnamefont {Meneveau}},\ }\bibfield  {title} {\bibinfo {title} {Linear forcing in numerical simulations of isotropic turbulence: {{Physical}} space implementations and convergence properties},\ }\href {https://doi.org/10.1063/1.2047568} {\bibfield  {journal} {\bibinfo  {journal} {Phys. Fluids}\ }\textbf {\bibinfo {volume} {17}},\ \bibinfo {pages} {095106} (\bibinfo {year} {2005})}\BibitemShut {NoStop}%
\bibitem [{\citenamefont {Tennekes}\ and\ \citenamefont {Lumley}(1972)}]{tennekes1972}%
  \BibitemOpen
  \bibfield  {author} {\bibinfo {author} {\bibfnamefont {H.}~\bibnamefont {Tennekes}}\ and\ \bibinfo {author} {\bibfnamefont {J.~L.}\ \bibnamefont {Lumley}},\ }{\emph {\bibinfo {title} {A first course in turbulence}}}\ (\bibinfo  {publisher} {MIT Press},\ \bibinfo {address} {Cambridge, Mass},\ \bibinfo {year} {1972})\BibitemShut {NoStop}%
\bibitem [{\citenamefont {Hunt}\ and\ \citenamefont {Graham}(1978)}]{hunt1978}%
  \BibitemOpen
  \bibfield  {author} {\bibinfo {author} {\bibfnamefont {J.~C.~R.}\ \bibnamefont {Hunt}}\ and\ \bibinfo {author} {\bibfnamefont {J.~M.~R.}\ \bibnamefont {Graham}},\ }\bibfield  {title} {\bibinfo {title} {Free-stream turbulence near plane boundaries},\ }\href {https://doi.org/10.1017/S0022112078000130} {\bibfield  {journal} {\bibinfo  {journal} {J. Fluid Mech.}\ }\textbf {\bibinfo {volume} {84}},\ \bibinfo {pages} {209} (\bibinfo {year} {1978})}\BibitemShut {NoStop}%
\bibitem [{\citenamefont {Hunt}(1984)}]{hunt1984}%
  \BibitemOpen
  \bibfield  {author} {\bibinfo {author} {\bibfnamefont {J.~C.~R.}\ \bibnamefont {Hunt}},\ }\bibfield  {title} {\bibinfo {title} {Turbulence structure in thermal convection and shear-free boundary layers},\ }\href {https://doi.org/10.1017/S0022112084000070} {\bibfield  {journal} {\bibinfo  {journal} {J. Fluid Mech}\ }\textbf {\bibinfo {volume} {138}},\ \bibinfo {pages} {161–184} (\bibinfo {year} {1984})}\BibitemShut {NoStop}%
\bibitem [{\citenamefont {Shen}\ \emph {et~al.}(1999)\citenamefont {Shen}, \citenamefont {Zhang}, \citenamefont {Yue},\ and\ \citenamefont {Triantafyllou}}]{shenSurfaceLayerFreesurface1999}%
  \BibitemOpen
  \bibfield  {author} {\bibinfo {author} {\bibfnamefont {L.}~\bibnamefont {Shen}}, \bibinfo {author} {\bibfnamefont {X.}~\bibnamefont {Zhang}}, \bibinfo {author} {\bibfnamefont {D.~K.~P.}\ \bibnamefont {Yue}},\ and\ \bibinfo {author} {\bibfnamefont {G.~S.}\ \bibnamefont {Triantafyllou}},\ }\bibfield  {title} {\bibinfo {title} {The surface layer for free-surface turbulent flows},\ }\href {https://doi.org/10.1017/S0022112099004590} {\bibfield  {journal} {\bibinfo  {journal} {J. Fluid Mech.}\ }\textbf {\bibinfo {volume} {386}},\ \bibinfo {pages} {167} (\bibinfo {year} {1999})}\BibitemShut {NoStop}%
\bibitem [{\citenamefont {Calmet}\ and\ \citenamefont {Magnaudet}(2003)}]{calmet2003}%
  \BibitemOpen
  \bibfield  {author} {\bibinfo {author} {\bibfnamefont {I.}~\bibnamefont {Calmet}}\ and\ \bibinfo {author} {\bibfnamefont {J.}~\bibnamefont {Magnaudet}},\ }\bibfield  {title} {\bibinfo {title} {Statistical structure of high-{R}eynolds-number turbulence close to the free surface of an open-channel flow},\ }\href {https://doi.org/10.1017/S0022112002002793} {\bibfield  {journal} {\bibinfo  {journal} {J. Fluid Mech.}\ }\textbf {\bibinfo {volume} {474}},\ \bibinfo {pages} {355} (\bibinfo {year} {2003})}\BibitemShut {NoStop}%
\bibitem [{\citenamefont {Magnaudet}(2003)}]{magnaudet2003}%
  \BibitemOpen
  \bibfield  {author} {\bibinfo {author} {\bibfnamefont {J.}~\bibnamefont {Magnaudet}},\ }\bibfield  {title} {\bibinfo {title} {High-{Reynolds-number} turbulence in a shear-free boundary layer: Revisiting the {Hunt}\textendash{Graham} theory},\ }\href {https://doi.org/10.1017/S0022112003004245} {\bibfield  {journal} {\bibinfo  {journal} {J. Fluid Mech.}\ }\textbf {\bibinfo {volume} {484}},\ \bibinfo {pages} {167} (\bibinfo {year} {2003})}\BibitemShut {NoStop}%
\bibitem [{\citenamefont {Fuchs}\ \emph {et~al.}(2022)\citenamefont {Fuchs}, \citenamefont {Kharche}, \citenamefont {Patil}, \citenamefont {Friedrich}, \citenamefont {W\"{a}chter},\ and\ \citenamefont {Peinke}}]{fuchs2022}%
  \BibitemOpen
  \bibfield  {author} {\bibinfo {author} {\bibfnamefont {A.}~\bibnamefont {Fuchs}}, \bibinfo {author} {\bibfnamefont {S.}~\bibnamefont {Kharche}}, \bibinfo {author} {\bibfnamefont {A.}~\bibnamefont {Patil}}, \bibinfo {author} {\bibfnamefont {J.}~\bibnamefont {Friedrich}}, \bibinfo {author} {\bibfnamefont {M.}~\bibnamefont {W\"{a}chter}},\ and\ \bibinfo {author} {\bibfnamefont {J.}~\bibnamefont {Peinke}},\ }\bibfield  {title} {\bibinfo {title} {An open source package to perform basic and advanced statistical analysis of turbulence data and other complex systems},\ }\bibfield  {journal} {\bibinfo  {journal} {Phys. fluids}\ }\textbf {\bibinfo {volume} {34}},\ \href {https://doi.org/10.1063/5.0107974} {10.1063/5.0107974} (\bibinfo {year} {2022})\BibitemShut {NoStop}%
\bibitem [{\citenamefont {Turney}\ and\ \citenamefont {Banerjee}(2013)}]{turney2013}%
  \BibitemOpen
  \bibfield  {author} {\bibinfo {author} {\bibfnamefont {D.~E.}\ \bibnamefont {Turney}}\ and\ \bibinfo {author} {\bibfnamefont {S.}~\bibnamefont {Banerjee}},\ }\bibfield  {title} {\bibinfo {title} {Air\textendash water gas transfer and near-surface motions},\ }\href {https://doi.org/10.1017/jfm.2013.435} {\bibfield  {journal} {\bibinfo  {journal} {J. Fluid Mech.}\ }\textbf {\bibinfo {volume} {733}},\ \bibinfo {pages} {588} (\bibinfo {year} {2013})}\BibitemShut {NoStop}%
\bibitem [{\citenamefont {McCready}\ \emph {et~al.}(1986)\citenamefont {McCready}, \citenamefont {Vassiliadou},\ and\ \citenamefont {Hanratty}}]{Mcready1986}%
  \BibitemOpen
  \bibfield  {author} {\bibinfo {author} {\bibfnamefont {M.}~\bibnamefont {McCready}}, \bibinfo {author} {\bibfnamefont {E.}~\bibnamefont {Vassiliadou}},\ and\ \bibinfo {author} {\bibfnamefont {T.}~\bibnamefont {Hanratty}},\ }\bibfield  {title} {\bibinfo {title} {Computer simulation of turbulent mass transfer at a mobile interface},\ }\href {https://doi.org/10.1002/aic.690320707} {\bibfield  {journal} {\bibinfo  {journal} {AIChE J.}\ }\textbf {\bibinfo {volume} {32}},\ \bibinfo {pages} {1108} (\bibinfo {year} {1986})}\BibitemShut {NoStop}%
\bibitem [{\citenamefont {Turney}\ \emph {et~al.}(2005)\citenamefont {Turney}, \citenamefont {Smith},\ and\ \citenamefont {Banerjee}}]{turney2005}%
  \BibitemOpen
  \bibfield  {author} {\bibinfo {author} {\bibfnamefont {D.~E.}\ \bibnamefont {Turney}}, \bibinfo {author} {\bibfnamefont {W.~C.}\ \bibnamefont {Smith}},\ and\ \bibinfo {author} {\bibfnamefont {S.}~\bibnamefont {Banerjee}},\ }\bibfield  {title} {\bibinfo {title} {A measure of near-surface fluid motions that predicts air--water gas transfer in a wide range of conditions},\ }\href {https://doi.org/10.1029/2004GL021671} {\bibfield  {journal} {\bibinfo  {journal} {Geophys. Res. Lett.}\ }\textbf {\bibinfo {volume} {32}},\ \bibinfo {pages} {L04607} (\bibinfo {year} {2005})}\BibitemShut {NoStop}%
\bibitem [{\citenamefont {Nagaosa}(1999)}]{nagaosa1999}%
  \BibitemOpen
  \bibfield  {author} {\bibinfo {author} {\bibfnamefont {R.}~\bibnamefont {Nagaosa}},\ }\bibfield  {title} {\bibinfo {title} {Direct numerical simulation of vortex structures and turbulent scalar transfer across a free surface in a fully developed turbulence},\ }\href {https://doi.org/10.1063/1.870020} {\bibfield  {journal} {\bibinfo  {journal} {Phys. Fluids}\ }\textbf {\bibinfo {volume} {11}},\ \bibinfo {pages} {1581} (\bibinfo {year} {1999})}\BibitemShut {NoStop}%
\bibitem [{\citenamefont {Aarnes}\ \emph {et~al.}(2025{\natexlab{b}})\citenamefont {Aarnes}, \citenamefont {Babiker}, \citenamefont {Xuan}, \citenamefont {Shen},\ and\ \citenamefont {Ellingsen}}]{aarnes25data}%
  \BibitemOpen
  \bibfield  {author} {\bibinfo {author} {\bibfnamefont {J.~R.}\ \bibnamefont {Aarnes}}, \bibinfo {author} {\bibfnamefont {O.~M.}\ \bibnamefont {Babiker}}, \bibinfo {author} {\bibfnamefont {A.}~\bibnamefont {Xuan}}, \bibinfo {author} {\bibfnamefont {L.}~\bibnamefont {Shen}},\ and\ \bibinfo {author} {\bibfnamefont {S.~{\AA}.}\ \bibnamefont {Ellingsen}},\ }\href {https://doi.org/10.18710/XQ81WH} {\bibinfo {title} {Replication data for: ``vortex structures under dimples and scars in turbulent free-surface flows`` dataverse, part 1}} (\bibinfo {year} {2025}{\natexlab{b}})\BibitemShut {NoStop}%
\bibitem [{\citenamefont {Babiker}\ \emph {et~al.}(2026)\citenamefont {Babiker}, \citenamefont {Aarnes}, \citenamefont {Semati}, \citenamefont {Ferran}, \citenamefont {Tee}, \citenamefont {Hearst},\ and\ \citenamefont {Ellingsen}}]{babiker26data}%
  \BibitemOpen
  \bibfield  {author} {\bibinfo {author} {\bibfnamefont {O.~M.}\ \bibnamefont {Babiker}}, \bibinfo {author} {\bibfnamefont {J.~R.}\ \bibnamefont {Aarnes}}, \bibinfo {author} {\bibfnamefont {A.}~\bibnamefont {Semati}}, \bibinfo {author} {\bibfnamefont {A.}~\bibnamefont {Ferran}}, \bibinfo {author} {\bibfnamefont {Y.~H.}\ \bibnamefont {Tee}}, \bibinfo {author} {\bibfnamefont {R.~J.}\ \bibnamefont {Hearst}},\ and\ \bibinfo {author} {\bibfnamefont {S.~{\AA}.}\ \bibnamefont {Ellingsen}},\ }\href {https://doi.org/10.18710/TSIZCQ} {\bibinfo {title} {Replication data for: Experimental investigation relating free-surface features to sub-surface turbulence.}} (\bibinfo {year} {2026})\BibitemShut {NoStop}%
\bibitem [{\citenamefont {Poelma}\ \emph {et~al.}(2006)\citenamefont {Poelma}, \citenamefont {Westerweel},\ and\ \citenamefont {Ooms}}]{poelma2006}%
  \BibitemOpen
  \bibfield  {author} {\bibinfo {author} {\bibfnamefont {C.}~\bibnamefont {Poelma}}, \bibinfo {author} {\bibfnamefont {J.}~\bibnamefont {Westerweel}},\ and\ \bibinfo {author} {\bibfnamefont {G.}~\bibnamefont {Ooms}},\ }\bibfield  {title} {\bibinfo {title} {Turbulence statistics from optical whole-field measurements in particle-laden turbulence},\ }\href {https://doi.org/10.1007/s00348-005-0072-y} {\bibfield  {journal} {\bibinfo  {journal} {Exp. Fluids}\ }\textbf {\bibinfo {volume} {40}},\ \bibinfo {pages} {347–363} (\bibinfo {year} {2006})}\BibitemShut {NoStop}%
\bibitem [{\citenamefont {Lavoie}\ \emph {et~al.}(2007)\citenamefont {Lavoie}, \citenamefont {Avallone}, \citenamefont {De~Gregorio}, \citenamefont {Romano},\ and\ \citenamefont {Antonia}}]{lavoie07}%
  \BibitemOpen
  \bibfield  {author} {\bibinfo {author} {\bibfnamefont {P.}~\bibnamefont {Lavoie}}, \bibinfo {author} {\bibfnamefont {G.}~\bibnamefont {Avallone}}, \bibinfo {author} {\bibfnamefont {F.}~\bibnamefont {De~Gregorio}}, \bibinfo {author} {\bibfnamefont {G.~P.}\ \bibnamefont {Romano}},\ and\ \bibinfo {author} {\bibfnamefont {R.~A.}\ \bibnamefont {Antonia}},\ }\bibfield  {title} {\bibinfo {title} {Spatial resolution of {PIV} for the measurement of turbulence},\ }\href {https://doi.org/10.1007/s00348-007-0319-x} {\bibfield  {journal} {\bibinfo  {journal} {Exp. Fluids}\ }\textbf {\bibinfo {volume} {43}},\ \bibinfo {pages} {39} (\bibinfo {year} {2007})}\BibitemShut {NoStop}%
\bibitem [{\citenamefont {Hearst}\ and\ \citenamefont {Lavoie}(2014)}]{hearst2014}%
  \BibitemOpen
  \bibfield  {author} {\bibinfo {author} {\bibfnamefont {R.~J.}\ \bibnamefont {Hearst}}\ and\ \bibinfo {author} {\bibfnamefont {P.}~\bibnamefont {Lavoie}},\ }\bibfield  {title} {\bibinfo {title} {Decay of turbulence generated by a square-fractal-element grid},\ }\href {https://doi.org/10.1017/jfm.2013.684} {\bibfield  {journal} {\bibinfo  {journal} {J. Fluid Mech}\ }\textbf {\bibinfo {volume} {741}},\ \bibinfo {pages} {567} (\bibinfo {year} {2014})}\BibitemShut {NoStop}%
\bibitem [{\citenamefont {Vassilicos}(2015)}]{vassillicos15}%
  \BibitemOpen
  \bibfield  {author} {\bibinfo {author} {\bibfnamefont {J.~C.}\ \bibnamefont {Vassilicos}},\ }\bibfield  {title} {\bibinfo {title} {Dissipation in turbulent flows},\ }\href {https://doi.org/10.1146/annurev-fluid-010814-014637} {\bibfield  {journal} {\bibinfo  {journal} {Annu. Rev. Fluid Mech.}\ }\textbf {\bibinfo {volume} {47}},\ \bibinfo {pages} {95} (\bibinfo {year} {2015})}\BibitemShut {NoStop}%
\bibitem [{\citenamefont {Kolmogorov}(1941)}]{kolmogorov41a}%
  \BibitemOpen
  \bibfield  {author} {\bibinfo {author} {\bibfnamefont {A.~N.}\ \bibnamefont {Kolmogorov}},\ }\bibfield  {title} {\bibinfo {title} {The local structure of turbulence in incompressible viscous fluid for very large {R}eynolds numbers},\ }\href {https://doi.org/10.1098/rspa.1991.0075} {\bibfield  {journal} {\bibinfo  {journal} {Dokl. Akad. Nauk SSSR}\ }\textbf {\bibinfo {volume} {30}},\ \bibinfo {pages} {301} (\bibinfo {year} {1941})}\BibitemShut {NoStop}%
\bibitem [{\citenamefont {Mora}\ \emph {et~al.}(2019)\citenamefont {Mora}, \citenamefont {Mu\~{n}iz Pladellorens}, \citenamefont {Riera~Turr\'{o}}, \citenamefont {Lagauzere},\ and\ \citenamefont {Obligado}}]{Mora2019}%
  \BibitemOpen
  \bibfield  {author} {\bibinfo {author} {\bibfnamefont {D.}~\bibnamefont {Mora}}, \bibinfo {author} {\bibfnamefont {E.}~\bibnamefont {Mu\~{n}iz Pladellorens}}, \bibinfo {author} {\bibfnamefont {P.}~\bibnamefont {Riera~Turr\'{o}}}, \bibinfo {author} {\bibfnamefont {M.}~\bibnamefont {Lagauzere}},\ and\ \bibinfo {author} {\bibfnamefont {M.}~\bibnamefont {Obligado}},\ }\bibfield  {title} {\bibinfo {title} {Energy cascades in active-grid-generated turbulent flows},\ }\href {https://doi.org/10.1103/PhysRevFluids.4.104601} {\bibfield  {journal} {\bibinfo  {journal} {Phys. Rev. Fluids}\ }\textbf {\bibinfo {volume} {4}},\ \bibinfo {pages} {104601} (\bibinfo {year} {2019})}\BibitemShut {NoStop}%
\bibitem [{\citenamefont {Herlina}\ and\ \citenamefont {Wissink}(2014)}]{herlina2014}%
  \BibitemOpen
  \bibfield  {author} {\bibinfo {author} {\bibnamefont {Herlina}}\ and\ \bibinfo {author} {\bibfnamefont {J.~G.}\ \bibnamefont {Wissink}},\ }\bibfield  {title} {\bibinfo {title} {Direct numerical simulation of turbulent scalar transport across a flat surface},\ }\href {https://doi.org/10.1017/jfm.2014.68} {\bibfield  {journal} {\bibinfo  {journal} {J. Fluid Mech.}\ }\textbf {\bibinfo {volume} {744}},\ \bibinfo {pages} {217} (\bibinfo {year} {2014})}\BibitemShut {NoStop}%
\end{thebibliography}

%

\end{document}